\documentclass{MyJHEP3}
\usepackage{amsmath}
\usepackage{amssymb}
\usepackage{graphicx}

\newcommand{\OMo}{\Omega_{M}^0}

\newcommand{\OLo}{\Omega_{\Lambda}^0}

\newcommand{\rc}{\rho_c}
\newcommand{\rco}{\rho_{c}^0}
\newcommand{\rmo}{\rho_{m}^0}

\newcommand{\rM}{\rho_M}
\newcommand{\rmr}{\rho_m}
\newcommand{\pmr}{p_m}
\newcommand{\rMo}{\rho_{M}^0}

\newcommand{\rR}{\rho_R}

\newcommand{\rX}{\rho_X}
\newcommand{\rB}{\rho_B}
\newcommand{\rLep}{\rho_L}

\newcommand{\wm}{\omega_m}

\newcommand{\weff}{\omega_{\rm eff}}

\newcommand{\rL}{\rho_{\CC}}

\newcommand{\rLo}{\rho_{\CC}^0}

\newcommand{\CC}{\Lambda}

\newcommand{\rLi}{\rho_{\Lambda}^{i}}

\newcommand{\G}{{\cal G}}
\newcommand{\F}{{\cal F}}

\newcommand{\HB}{{\cal H}}
\newcommand{\CUV}{\Lambda_{\rm UV}}
\newcommand{\E}{{\cal E}}
\newcommand{\ZPE}{V_{\rm ZPE}}
\newcommand{\rLV}{\rho_{\CC\rm vac}}
\newcommand{\rLb}{\rho_{\CC}^{({\rm b})}}

\newcommand{\rVu}{\rho_{\rm vac}^{(1)}}
\newcommand{\Gb}{G^{({\rm b})}}

\newcommand{\Gbu}{G^{(1)}}
\newcommand{\BCC}{\Lambda^{({\rm b})}}
\newcommand{\pc}{\phi_c}

\newcommand{\gmnu}{g^{\mu\nu}}
\newcommand{\rV}{\rho_{\rm vac}}
\newcommand{\CP}{{\cal CP}}

\newcommand{\LQCD}{\Lambda_{\rm QCD}}
\newcommand{\mupe}{\mu_{\rm pe}}
\newcommand{\OMB}{\Omega_B}
\newcommand{\OMBo}{\Omega^0_B}
\newcommand{\ODM}{\Omega_{\rm DM}}
\newcommand{\ODMo}{\Omega^0_{\rm DM}}
\newcommand{\nuQCD}{\nu_{\rm QCD}}
\newcommand{\nuX}{\nu_{X}}
\newcommand{\nuB}{\nu_{B}}
\newcommand{\nueff}{\nu_{\rm eff}}
\newcommand{\K}{{\cal K}}
\newcommand{\VP}{\langle V(\phi)\rangle}
\newcommand{\rLI}{\rho_{\CC\rm ind}}
\newcommand{\Hp}{\cal H}
\newcommand{\rLP}{\rho_{\CC{\rm ph}}}

\newcommand{\EP}{V_{\rm eff}}

\newcommand{\EPR}{V_{\rm eff}}

\newcommand{\tEP}{{V}_{\rm scal}}

\newcommand{\newtext}[1]{\text{#1}}

\title{\bf Cosmological constant and vacuum energy: \\ old and new
ideas}

\author{%
Joan Sol\`a\\
High Energy Physics Group, Dept. Estructura i Constituents de la Mat\`eria\\
and  Institut de Ci{\`e}ncies del Cosmos\\
Univ. de Barcelona, Av. Diagonal 647, E-08028 Barcelona, Catalonia, Spain\\

\email{sola@ecm.ub.edu}}

\abstract{The cosmological constant (CC) term in Einstein's equations,
$\CC$, was first associated to the idea of vacuum energy density.
Notwithstanding, it is well-known that there is a huge, in fact appalling,
discrepancy between the theoretical prediction and the observed value
picked from the modern cosmological data. This is the famous, and
extremely difficult, ``CC problem''. Paradoxically, the recent observation
at the CERN Large Hadron Collider of a Higgs-like particle, should
actually be considered ambivalent: on the one hand it appears as a likely
great triumph of particle physics, but on the other hand it wide opens
Pandora's box of the cosmological uproar, for it may provide (alas!) the
experimental certification of the existence of the electroweak (EW) vacuum
energy, and thus of the intriguing reality of the CC problem. Even if only
counting on this contribution to the inventory of vacuum energies in the
universe, the discrepancy with the cosmologically observed value is
already of 55 orders of magnitude. This is the (hitherto) ``real''
magnitude of the CC problem, rather than the (too often) brandished $123$
ones from the upper (but fully unexplored!) ultrahigh energy scales. Such
is the baffling situation after 96 years of introducing the $\CC$-term by
Einstein. In the following I will briefly (and hopefully pedagogically)
fly over some of the old and new ideas on the CC problem. Since, however,
the Higgs boson just knocked our door and recalled us that the vacuum
energy may be a fully tangible concept in real phenomenology, I will
exclusively address the CC problem from the original notion of vacuum
energy, and its possible ``running'' with the expansion of the universe,
rather than venturing into the numberless attempts to replace the CC by
the multifarious concept of dark energy.}

\keywords{Cosmological constant, Vacuum Energy, Quantum Field
Theory}

\begin{document}

\section{Introduction}
\label{sec:introduction}

Undoubtedly the most prominent performance of modern cosmology has been to
provide observational evidence for the accelerated expansion of the
universe\,\cite{SNIa,WMAP,Kom11,PLANCK2013} and for the existence of
(other) large scale dynamical phenomena possibly caused by forms of matter
beyond the usual baryonic component. To ``explain'' the accelerated
evolution the name ``dark energy'' (DE) was
coined\,\cite{NameDEHutererTurner99}; it refers to some mysterious form of
diffuse (i.e. non-clustering) energy presumably permeating all corners of
the universe, possessing negative pressure and thus being capable of
boosting the expansion of the universe as a whole. Similarly, to
``explain'' the anomalous dynamics of galaxies and of galaxy clusters, we
have imagined that there is a large deficit of matter at different
astronomical scales in the form of unknown stable particles, which are
neither electrons nor protons, not even neutrinos, but some form of
electrically neutral heavy stuff beyond the spectrum of the Standard Model
(SM) of the strong and electroweak interactions, and referred to as ``dark
matter'' (DM) particles. We do not yet know if any of these hypotheses is
true at all, although new hints (not realities, yet) might be around
recently; the only thing we know for sure is the reality of the observed
physical phenomena that we are trying to explain. In the following I will
not elaborate on the whereabouts of the hypothetical DM particles, I will
rather focus on a few aspects of the DE problem, or more specifically the
cosmological constant (CC)
problem\,\cite{CCPWeinberg,CCproblem2,PeeblesRatra03}, which is perhaps
the most intriguing of all cosmological puzzles.

It is often stated in the literature that the CC term, and its association
with the notion of vacuum energy, cannot be a valid theoretical
explanation for the accelerated expansion of the universe, and that we
necessarily have to ``go beyond $\CC$''. The adduced reasons are manyfold,
but perhaps the most brandished one is that the various contributions to
the vacuum energy cannot possibly be successfully fine tuned to the
measured value by any known mechanism, and therefore the idea of the
vacuum energy and its connection with the CC is viewed as completely
unnatural. Barring the fact that this need not be true, since the whole
issue is quite debatable (as we will try to show) and moreover there are
dynamical mechanisms within modified gravity that could efficiently help
to cure the fine tuning decease  -- if only from a mere technical point of
view $\CC$ (cf. Ref.\,\cite{RelaxedUniverse}) -- it is nevertheless a
curious ``reason'' to wield, as usually nothing more fundamental is
offered as an alternative, except a defense (tooth and nail) of some
particular form of DE, say from quintessence to string landscape.
Unfortunately, as we know, none of these alternatives seem to improve in
any practical way the fine-tuning illness\,\cite{CCPWeinberg} -- a very
serious matter, in principle, which for some (no less) mysterious reason
is unjustly blamed to the CC option almost exclusively. In fact, the fine
tuning problem in such new frameworks not only does not become milder but
it gets even worse than in the CC case, simply because the traditional
vacuum energy of the SM is still there, and so one has to cope with its
fine tuning, plus the (no less severe) one associated to the field or
string object (usually linked to some form of high energy physics)
purportedly replacing the CC term. As a result the two fine tunings make
the overall job even more bizarre. Let alone that, quite often, in these
frameworks an extremely light new particle is predicted in the ballpark of
$\sim 10^{-33}$ eV. However, we should seriously worry about the fact that
such (incommensurably tiny!) mass scale is some 30 orders of magnitude
smaller than the mass scale which these models aim to explain -- namely
the millielectronvolt ($\sim 10^{-3}$ eV) mass scale associated to the CC
term! Why such strategy is not perceived as trying to solve a big problem
by creating an even major one?  The answer is perhaps another profound
mystery of Nature; quite likely it must be that the CC problem is such a
disproportionately big problem that we are -- too soon -- ready to
redefine dramatically the scope and limits of our physical perceptions.

The  ``instinctive'' tendency to replace the vacuum energy by alternative
theoretical constructs can be counterproductive, though, because in doing
so many people (consciously or unconsciously) may give up the duty of
explaining why the usual vacuum energy of QFT does not participate at all
in accounting for the value of the CC. Please notice that after the likely
discovery of a Higgs boson of mass $M_{\cal H}=125.7\pm 0.4$ GeV at the
LHC collider\,\cite{HiggsDiscovery1}, and corresponding precise
determination of all the relevant SM
parameters\,\cite{RPP2012,Baak12,Sirlin2013}, the reality of the vacuum
energy density associated to the spontaneous symmetry breaking mechanism
of the electroweak (EW) theory starts to acquire a palpable reality, its
value being directly dependent on the product squared of the Higgs mass
and Fermi's constant: $\langle V\rangle_{\rm EW}\sim M_{\cal H}^2/G_F\sim
10^8$ GeV$^4$.  If we wish to face the CC problem in earnest, we should
somehow move on and stop leaving the vacuum energy of the SM in the most
complete oblivion, literally as if the mere fact of not thinking or
talking about it would make it disappear from our world! If we think
seriously about it, wouldn't this attitude be more typical of an
inhabitant of some ``Ostrichland''?

In the following I shall dwell on some properties of the CC term in
Einstein's equations and generalizations thereof. Our main aim here is to
consider models where the prime driving force accelerating the universe is
dynamical vacuum energy and hence a time variable CC. I will also discuss
the CC fine tuning problem (the ``old CC problem''\,\cite{CCPWeinberg})
from several perspectives, always within the notion of vacuum energy.
While a climax of mystery and severity usually hovers over the CC fine
tuning problem, I will try to relax it by introducing a number of remarks
and reflections on how to possibly interpret this fantastic conundrum in
more paused and balanced terms.

Finally, I will discuss some intriguing phenomenological implications of
the dynamical vacuum framework as a potential source for a mild
variability of the fundamental ``constants'' of Nature. This could help in
effectively testing these ideas using present day tech facilities in the
ground lab and in the sky. It is well-known that some fundamental
``constants'', such as the fine structure constant and the proton mass
could show some cosmic time variation. There is no firm experimental
evidence yet, but there are some indications. Since the proton mass is
intimately connected with the QCD scale of the strong interactions,
$\LQCD$, the latter could ultimately show a cosmic time evolution.
However, covariance of General Relativity (GR) demands that this should be
accompanied by the corresponding time variation of another quantity. We
claim that such quantity is the vacuum energy
density\,\cite{FritzschSola2012}. This could provide a \textit{raison
d'\^etre} for dynamical dark energy.

Throughout the exposition I'll strive for being accessible and
(hopefully) pedagogical. For a more detailed and technical
presentation, see e.g. \cite{SolaReview2013b}; and for a summarized
introduction to time evolving vacuum models along these lines,
see\,\cite{SolaReview2011,SolaReview05}.

%\newpage

\section{Dark Energy and Einstein's original ``constant cosmological constant''}
\label{sect:EinsteinCC}

Historically, the $\CC$-term in the field equations was introduced by A.
Einstein 96 years ago\,\cite{Einstein1917}, but the ``CC problem'' as such
was formulated 50 years later by Y. B. Zeldovich\,\cite{Zeldovich67}. The
latter is the realization that the quantum theory applied to the world of
the elementary particles seems to predict an effective value for $\CC$
which is much larger than the critical density of the universe (to which
the vacuum energy density is found to be comparable) -- see the reviews
\cite{CCPWeinberg,CCproblem2,PeeblesRatra03}. The first models trying to
circumvent this tough difficulty from a fundamental quantum field theory
(QFT) point of view -- confer
e.g.\,\cite{Dolgov82,Abbott85,PSW87,BarrFord87,Sola89} -- tried to use
dynamical scalar fields and sought to explain the small value of the CC
density $\rL=\CC/(8\pi\,G)$, assuming that the latter was actually the
energy density of a cosmic scalar field (called ``cosmon'' in one of the
formulations\,\cite{PSW87}) starting with a huge value in the early
universe and then eventually settling down dynamically (hence without fine
tuning) to the present current value\,\cite{PLANCK2013} $\rLo\simeq
2.5\times 10^{-47}$ GeV$^4$.

No model of this kind ever succeeded in achieving that main aim, and the
cosmic scalar field models were basically used to endow the vacuum energy
with a dynamical character\,\cite{Wetterich88,RatraPeebles88} -- still not
being called quintessence, but playing this role. Ten years later the
first models bearing this name appeared in the market\,\cite{Caldwell98},
although again with a much more modest aim: they did not try to explain
the value of $\rLo$, but just the cosmic coincidence problem, i.e. the
reason why the current value of the CC density is so close to the present
matter density -- see \cite{PeeblesRatra03} for a comprehensive exposition
and more references. However, not even this lower target was ever reached
satisfactorily. Nor could be reached either by the many phenomenological
proposals for a time dependent cosmological term or other alternative
notions. However, a great deal of attention was dedicated to these models,
motivated basically by the age of the universe and CC
problems\,\cite{OzerTaha87,Bertolami86,FreeseET87,
CarvalhoET92,ArcuriWaga94,Waga93,LM93,LM94,LT96,SalimWaga93,Arbab97} --
see also the reviews \cite{Overduin98} and \cite{Lima04}. Subsequently (in
the years after the discovery of the accelerated expansion of the
universe\,\cite{SNIa}) a new and powerful wave of proposals invaded all
corners of the cosmological literature: the multifarious notion of dark
energy (DE) was born\,\cite{CCproblem2,PeeblesRatra03,DEbook2010}, its
main purpose being to replace in infinitely many different disguises (the
last one being usually more bewildering than the previous) the very
function made by the $\CC$-term in Einstein's equations. If that is not
enough, issues on the interpretation of what we have really measured are
still not settled either\,\cite{Durrer2011}. And here we are, still
fighting with the phenomenal cosmological problems triggered by Einstein's
first idea to introduce the $\CC$ term in his almost centennial General
Relativity framework! Somehow Einstein knew he was going to raise such a
pandemonium, as he felt that by introducing $\CC$ in the field equations
he was in danger of being interned in a madhouse: \textit{``Ich habe
wieder etwas verbrochen in der Gravitationstheorie, was mich ein wenig in
Gefahr bringt, in ein Tollhaus interniert zu werden''} (A. Einstein,
Letter to P. Ehrenfest, February 4th 1917). We have to admit that at least
Einstein was quite honest with his CC idea (despite it being his ``biggest
blunder''\,\cite{Blunder}), but we never heard what the defenders of the
many ersatz $\CC$ notions have in mind about the possible craziness of
their own proposals. After all the $\CC$ term is no blunder at all.

Almost a century ago (98 years to be precise) the gravitational field
equations (\textit{Die Feldgleichungen der Gravitation}) were first
introduced by A. Einstein in 1915, with no cosmological term at
all\,\cite{Einstein1915}:
\begin{equation}
G_{\mu\nu}\equiv R_{\mu\nu}-\frac12\,\,g_{\mu\nu}\,R=8\pi G\,T_{\mu\nu}\,. \label{EEnoCC1}
\end{equation}
It was only two years later when Einstein, in an attempt to describe the
largely accepted idea, at that time, of a finite, static and closed
universe hypothetical fulfilling Mach's principle, introduced the $\CC$
term\,\cite{Einstein1917} and modified his field equations in the form we
still write them today:
\begin{equation}
G_{\mu\nu}-g_{\mu\nu}\,\CC=8\pi G\,T_{\mu\nu}\,. \label{EEnoCC2}
\end{equation}
The reason why he was able to introduce that term is because it is
perfectly consistent with the idea of general covariance, which was the
building principle of GR. This is mathematically expressed by the fact
that the covariant derivative on both sides of the original field
equations (\ref{EEnoCC1}) gives zero: indeed $\nabla^{\mu}G_{\mu\nu}=0$ is
always (automatically) satisfied on account of the Bianchi identity of the
Riemann tensor; whereas the covariant derivative on the right-hand-side
can be satisfied in different ways, the simplest one being perhaps to
assume that $G$ (Newton's gravitational coupling) is a fundamental
constant, and that matter is covariantly conserved (i.e.
$\nabla^{\mu}T_{\mu\nu}=0$). Under these conditions, if we compute once
more the covariant derivative on both sides, but this time on the modified
field equations (\ref{EEnoCC2}), and under the same set of assumptions (on
fixed $G$ and matter conservation), we are immediately led to

\begin{equation}\label{eq:CCconstant}
\nabla_{\mu}\CC=\partial_{\mu}\CC=0\ \  \ \ \Rightarrow \ \  \ \ \CC={\rm const}\,.
\end{equation}
This is of course what justifies the name ``cosmological constant'' to the
parameter $\CC$ introduced by Einstein in Eq.\,(\ref{EEnoCC2}). But in
point of fact the justification is only partial, as it should be clear
from our carefully stating the conditions under which it has been derived.
If $G$ is not constant and/or matter is not covariantly conserved (both of
them being assumptions which should not be rejected too fast) then the
canonical conclusion (\ref{eq:CCconstant}) is not guaranteed at all. A
time dependent $\CC$, or more precisely, a spatially homogeneous function
of the cosmic time, $\CC=\CC(t)$, would still be perfectly compatible with
the Cosmological Principle. However, in order to still fulfill the Bianchi
identity we would need either a time dependent gravitational coupling,
$G=G(t)$, or to admit the possibility that matter exchanges energy with
vacuum (hence that matter is not self-conserved, in a locally covariant
sense), or a combination of the two possibilities. While some of these
possibilities may look bizarre at first sight, it is no less bizarre than
the possibility that the fundamental ``constants'' of Nature might not be
constant after all. There are actually some hints that this could be the
case -- and also that the two ``bizarre stories'' could in fact be deeply
related\,\cite{FritzschSola2012} -- see
Sect.\,\ref{sect:DynamicalVacFundConst}.

The Cosmological Principle is based on the homogeneous and isotropic
Friedmann-Lema\^\i tre-Robertson-Walker (FLRW)
metric\,\cite{CosmologyBooks}
\begin{equation}\label{eq:FLRWmetric}
ds^2=c^2dt^2-a^2(t)\left[\frac{dr^2}{1-K\,r^2}+r^2\,d\Omega^2\right]\,,
\end{equation}
with $d\,\Omega^2=r^2\,d\theta^2+r^2\sin^2\theta\,d\phi^2$. The basic
cosmological equations emerging from Einstein's field equations with
$\CC$-term (\ref{EEnoCC2}) in the FLRW metric are well-known. Adopting for
matter the energy-momentum tensor for a perfect cosmic fluid,
\begin{equation}\label{eq:Tmunu}
T_{\mu\nu}=(\rho_m+p_m)U_{\mu}U_{\nu}-p_m\, g_{\mu\nu}\,,
\end{equation}
and computing the various components of the geometric tensors in
(\ref{EEnoCC2}), we find the desired result, which is summarized in two
fundamental equations. On the one hand we have
\begin{equation}\label{eq:FriedmannK}
H^2\equiv\left(\frac{\dot{a}}{a}\right)^2=\frac{8\pi
G}{3}\rmr+\frac{\CC}{3}-\frac{K}{a^2}\,,
\end{equation}
where the constant $K$ is the spatial curvature parameter appearing in
(\ref{eq:FLRWmetric}), and on the other
\begin{equation}\label{eq:acceleration}
\frac{\ddot{a}}{a}=-\frac{4\pi\,G}{3}\,(\rmr+3\pmr)+\frac{\CC}{3}\,.
\end{equation}
Here $a=a(t)$ is the scale factor of the FLRW metric
(\ref{eq:FLRWmetric}). The first equation (\ref{eq:FriedmannK}) is called
the Friedmann-Lema\^\i tre equation, whereas equation
(\ref{eq:acceleration}) is the acceleration equation.

If we would take the ``simplest'' possibility conceived by Einstein,
namely a strictly constant $\CC$ and spherical symmetry of the
three-dimensional space -- entailing  $K=+1$ in Eq. (\ref{eq:FriedmannK})
--  one can easily derive the explicit form of the field equations in
terms of the constant matter density $\rmr$, the newly introduced
$\CC$-term and the scale factor $a$ of the metric. We can, of course,
recover Einstein's universe as a very particular case of the above
dynamical equations (\ref{eq:FriedmannK})-(\ref{eq:acceleration}). Indeed,
assuming a universe made of dust (hence zero matter pressure, $p_m=0$) and
imposing static equilibrium between the gravity force and the negative
vacuum pressure (thus insuring a strictly static universe), we have $H=0$
and $\ddot{a}=0$, and then a simple relation ensues immediately between
these three quantities: $4\pi G\rmr=1/a^2=\CC$, or equivalently
\begin{equation}\label{eq:EinsteinUniverse}
\rmr=\frac{1}{4\pi G\,a^2}=2\,\rL\,,
\end{equation}
where we recall that $\CC=8\pi G\rL$. Clearly, it is thanks to the assumed
nonvanishing, and positive, $\CC$-term that such relation is possible.
This made Einstein happy, but his happiness was ephemeral. It is not only
that E. Hubble soon provided evidence that our universe is actually
expanding, but the fact (noticed by A.S. Eddington\,\cite{Eddington1931})
that even in the absence of this information the relation
(\ref{eq:EinsteinUniverse}) is completely unrealistic, for it corresponds
to an unstable position of equilibrium! It means that the slightest
perturbation in the value of $\rmr$ around the one satisfying
(\ref{eq:EinsteinUniverse}) triggers a runaway solution that kicks forever
the ``marble off the hill's top''! One can easily check it by considering
the  resulting Eq.\,(\ref{eq:acceleration}) for $a\to a_{\rm eq}+\delta a$
with $\rmr\sim 1/a^3$, where $a_{\rm eq}$ is the purported equilibrium
position. One immediately gets $\ddot{a}/{a}\sim 4\pi\,G\delta a$. Thus,
having the acceleration the same sign as the perturbation, the point
$a_{\rm eq}$ is unstable. The result is perfectly intuitive, and can also
be understood as follows. The gravitational law corrected with the
$\CC$-term reads
\begin{equation}\label{eq:NewtonLambda}
\bold{g}(r)=-G\,\frac{M}{r^2}\bold{u}_r+\frac13\,\,\Lambda\,r\,\bold{u}_r\,,
\end{equation}
where $\bold{u}_r$ is a unit vector directed radially outwards with
respect to the position of the body of mass $M$ creating the field. Thus,
if for some reason this universe would expand slightly ($r\to r+\delta
r$), this would diminish the gravitational attraction but at the same time
would enhance the repulsive $\CC$-force because the latter is larger the
larger is the separation between particles, if $\CC> 0$ -- as implicit in
Eq.\,(\ref{eq:EinsteinUniverse}). So there is no possible compensation
between the two. As a result the original "seed expansion", no matter how
small it is, would destabilize the universe into a "runaway expansion".
Similarly, an initial "seed contraction" ($r\to r-\delta r$) would cause
the universe to shrink indefinitely into the ``Big Crunch''.

Paradoxically, ten years later after $\CC$ was introduced by Einstein to
insure a static non-evolving universe, G. Lema\^\i
tre\,\cite{Lemaitre1927} used the form (\ref{EEnoCC2}) with nonvanishing
$\CC$ to discuss his dynamical models of the expansion of the universe,
strongly motivated by E. Hubble's observations prior their publication in
1929. The same models had already been independently discussed (without
$\CC$) a few years before by A. Friedmann\,\cite{Friedmann1922and34} on
pure mathematical grounds and with no connection whatsoever with
observations. In 1931, fourteen years after Einstein had introduced the
$\CC$-term, he finally rejected it\,\cite{Einstein1931}, as if the
observational evidence collected by Hubble against his original static
model of the universe should have any impact at all on the theoretical
status of the $\CC$ term. As we know, 96 years after its introduction in
the gravitational field equations, $\CC$ is still there, ``alive and
kicking'', we like it or not!  \newtext{The experimental confirmation} for
a nonvanishing (and positive) $\CC$-term is only from $15$ years
ago\,\cite{SNIa98}, but nowadays we have plenty of additional evidence
from independent data sets derived from the observation of distant
supernovae, the anisotropies of the CMB, the lensing effects on the
propagation of light through weak gravitational fields, and the inventory
of cosmic matter from the large scale structures (LSS) of the
Universe.\,\cite{SNIa,WMAP,Kom11,PLANCK2013,Lensing,LSS}.

\section{The electroweak Higgs vacuum in classical field theory}
\label{sect:HiggsVacuum}

In the following we summarize the old CC problem and perform a preliminary
discussion of the associated fine-tuning problem, leaving the more
sophisticated effects for the subsequent sections. We wish to illustrate
the problem within the context of the standard model (SM) of particle
physics, and more specifically within the Glashow-Weinberg-Salam model of
electroweak interactions\,\cite{GWS67}. This is the most successful QFT we
have at present (together with the QCD theory of strong
interactions\,\cite{GellMannFritzsch72}), both theoretically and
phenomenologically, and therefore it is the ideal scenario where to
formulate the origin of the problem. As is well-known, the unification of
weak and electromagnetic interactions into a renormalizable theory
requires to use the principle of local gauge symmetry in combination with
the phenomenon of spontaneous symmetry breaking
(SSB)\,\cite{HiggsMechanism64}. {It is} indeed the only known way to
generate all the particle masses by preserving the underlying gauge
symmetry. {In the SM}, one must introduce a fundamental complex doublet of
scalar fields. However, in order to simplify the discussion, let us
consider a field theory with a real single scalar field $\phi$, as this
does not alter at all the nature of the problem under discussion. To
trigger SSB, one must introduce a potential for the field $\phi$, which in
renormalizable QFT takes the form (the {tree-level} Higgs potential):
\begin{equation}\label{Poten}
V(\phi)=\frac12\,m^2\,\phi^2+\frac{1}{4!}\,\lambda\,\phi^4\ \
\ \ \ \ (\lambda>0)\,.
\end{equation}
Since we are dealing with a problem related with the CC, we must
inexcusably consider the influence of gravity. To this effect, we
shall conduct our investigation of the CC problem within the
semiclassical context, i.e.\ from the point of view of quantum field
theory (QFT) in curved
space-time\,{\cite{QFTBook1,QFTBook2,QFTBook3}. It means that we
address the CC problem in a framework where gravity is an external
gravitational field and we quantize matter fields
only\,\cite{ShapSol09,ShapSol09b}. The potential in equation
(\ref{Poten}) is given at the moment only at the classical level,
but it will eventually acquire quantum effects generated by the
matter fields themselves (cf. Sect.\ref{sect:effHiggsFLAT}). In this
context, we need to study what impact the presence of such potential
may have on Einstein's equations both at the classical and at the
quantum level.

Einstein's field equations for the classical metric in vacuo are derived
from the Einstein-Hilbert (EH) action with a cosmological term  $\BCC$
({hereafter} the CC \textit{vacuum term}). The EH action in vacuo
reads\,\cite{CosmologyBooks}:
\begin{eqnarray} S_{\rm EH} &=& \frac{-1}{16\pi\,G}\,\int d^4
x\sqrt{-g}\, \left(\,R + 2\,\CC\,\right)=-\,\int d^4
x\sqrt{-g}\,\left(\frac{1}{16\pi\,G} \,R+\rLV\right).\nonumber\\
\label{EH}
\end{eqnarray}
Here we have defined $\rLV$, the energy density associated to the CC
vacuum term:
\begin{equation}\label{rLV}
\rLV=\frac{\CC}{8\pi\,G_N}\,.
\end{equation}
The classical action including the scalar field $\phi$ with its potential
(\ref{Poten}) is
\begin{eqnarray}
S = S_{EH} +\int
d^4x\,\sqrt{|g|}\,\left[\frac12\,g^{\mu\nu}\,\partial_{\mu}\phi\,\partial_{\nu}\phi-V(\phi)\right]\,.
\label{Stotal}
\end{eqnarray}
Due to the usual interpretation of Einstein's equations as an equality
between geometry and a matter-energy source, it is convenient to place the
$\rLV$ term as a part of the matter action, $S[\phi]$. Then the total
action (\ref{Stotal}) can be reorganized as
\begin{eqnarray}
S = \,\frac{1}{16\pi\,G_N}\,\int d^4 x\sqrt{|g|}\,R+S[\phi]\,,
\label{Stotal2}
\end{eqnarray}
with
\begin{eqnarray}\label{Sphi}
    && S[\phi]=\int
    d^4x\,\sqrt{|g|}\,\left[\frac12\,g^{\mu\nu}\,\partial_{\mu}\phi\,\partial_{\nu}\phi-\rLV-{V}(\phi)\right]
    \equiv \int
    d^4x\,\sqrt{|g|}\ \mathcal{L}_{\phi}\,,
\end{eqnarray}
where $ \mathcal{L}_{\phi}$ is the matter Lagrangian for $\phi$. For the
moment, we will treat the matter fields contained in $\mathcal{L}_{\phi}$
as classical fields, and in particular the potential $V$ is supposed to
take the classical form (\ref{Poten}) with no quantum corrections. If we
compute the energy-momentum tensor of the scalar field $\phi$ in the
presence of the vacuum term $\rLV$, let us call it
$\tilde{T}_{\mu\nu}^{\phi}$, we obtain
\begin{eqnarray}\label{Tmunu}
    &&\tilde{T}_{\mu\nu}^{\phi}=\frac{2}{\sqrt{|g|}}\,{\delta\,S[\phi]\over\delta\,g^{\mu\nu}}
=2\,\frac{\partial\mathcal{L}_{\phi}}{\partial\,g^{\mu\nu}}-
g_{\mu\nu}\,\mathcal{L}_{\phi} =
g_{\mu\nu}\,\rLV+{T}_{\mu\nu}^{\phi}\,,
\end{eqnarray}
where we have used $\partial\sqrt{|g|}/\partial  g^{\mu\nu} =
-(1/2)\sqrt{|g|}\,g_{\mu\nu}$. Here
\begin{eqnarray}\label{Tmunu2}
    &&{T}_{\mu\nu}^{\phi} = \left[\,\partial_{\mu}\phi\,\partial_{\nu}\phi-\frac12\,g_{\mu\nu}\
\partial_{\sigma}\phi\,\partial^{\sigma}\phi
\right]+g_{\mu\nu}\,V(\phi)
\end{eqnarray}
is the ordinary energy-momentum tensor of the scalar field $\phi$.

In the vacuum (i.e.\ in the ground state of $\phi$) there is no kinetic
energy, so that the first term on the \textit{r.h.s} of (\ref{Tmunu2})
does not contribute {in that state}. Only the potential can take a
non-vanishing vacuum {expectation} value, which we may call $\VP$. Thus,
the ground state value of (\ref{Tmunu}) is
\begin{equation}\label{vacTmunu}
    \langle \tilde{T}_{\mu\nu}^{\phi}\rangle=g_{\mu\nu}\,\rLV+\langle
    {T}_{\mu\nu}^{\phi}\rangle=g_{\mu\nu}\,(\rLV+\VP)\equiv\rV^{\rm
    cl}\ g_{\mu\nu}\ \ \ \ \ \ ( m^2<0)\,,
\end{equation}
where $\rV^{\rm cl}$ is the \textit{classical vacuum energy} in the
presence of the field $\phi$.

If $m^2>0$ in equation (\ref{Poten}), then $\langle\phi\rangle=0\
\Rightarrow$ $\VP=0$, and there is no SSB. The classical vacuum energy is
just the original $\rLV$ term,
\begin{equation}\label{TmunuNoSSB}
    \langle \tilde{T}_{\mu\nu}^{\phi}\rangle=g_{\mu\nu}\,\rLV\ \ \ \ \ \ \ \ \ (
    m^2>0)\,.
\end{equation}
This result also applies in the free field theory case.  However, if the
phenomenon of SSB is active, which occurs when $m^2<0$, we have a
non-trivial ground-state value for $\phi$, or vacuum expectation value
(VEV):
\begin{equation}
v\equiv\langle\phi\rangle =\sqrt{\frac{-6\,m^{2}}{\lambda}}\,.
\label{5N}
\end{equation}
In this case, there is an \textit{induced part} of the vacuum energy at
the classical level owing to the electroweak phase transition generated by
the Higgs potential. This transition induces a non-vanishing contribution
to the cosmological term which is usually called the ``induced CC''. At
the classical level, {it is} given by
\begin{equation}
\rLI\equiv\VP=-\frac{3\,m^{4}}{2\lambda}=\frac14\,m^2\,v^2=
-\frac18\,M_{\Hp}^2\,v^2= -\frac{1}{8\sqrt{2}}\,M_{\Hp}^2\,M_F^2\,,
\label{eq:Higgstree}
\end{equation}
In the last equation we have used the physical Higgs mass squared:
\begin{equation}\label{eq:HiggsMass}
M_{\Hp}^2=\left.\frac{\partial^2\,V(\phi)}{\partial\phi^2}\right|_{\phi=v}=m^2+\frac12\,\lambda\,v^2=-2m^2>0\,.
\end{equation}
Indeed, if we redefine the Higgs field as ${\Hp}=\phi-v$ then its value at
the minimum will obviously be zero. {This is} the standard position for
the ground state of the field before doing perturbation theory. The
physical mass is determined by the oscillations of $\Hp$ around this
minimum, i.e.\ it follows from the second derivative of $V$ at $\phi=v$,
as in Eq.\,(\ref{eq:HiggsMass}). In the last equality of that equation we
have also introduced the so-called Fermi's scale $M_F\equiv
G_F^{-1/2}\simeq 293$ GeV, which is defined from Fermi's constant obtained
from muon decay, $G_F\simeq {1.166}\times 10^{-5}$ GeV$^{-2}$. The
relation of $G_F$ with the $W^{\pm}$ gauge boson mass and  the $SU(2)_L$
weak gauge coupling, $g$, reads (at the lowest order):
\begin{equation}\label{eq:GF}
\frac{G_F}{\sqrt{2}}=\frac{g^2}{8M_W^2}=\frac{1}{2\,v^2}\,,
\end{equation}
where in the second equality we have used the formula for the $W^{\pm}$
mass in the SM, namely $M_W^2=(1/4)\,g^2\,v^2$. For completeness let us
recall that $M_Z^2=(1/4)\,\left(g^2+g'^2\right)\,v^2$, where $g'$ is the
$U_Y(1)$ gauge coupling. In this way, Eq.\,(\ref{eq:GF}) provides a direct
determination of the Higgs VEV in terms of Fermi's constant:
\begin{equation}\label{eq:GFv}
v=\left(\sqrt{2}\,G_F\right)^{-1/2}\simeq 246\,{\rm GeV}\,,
\end{equation}
and this relation has been used in the last equality of
Eq.\,(\ref{eq:Higgstree}).

In view of the SSB phenomenon, it is clear that we must replace
$g_{\mu\nu}\,\rLV\to g_{\mu\nu}\,\rV^{\rm  cl}\ $, given by
Eq.\,(\ref{vacTmunu}), in the expression of Einstein's equations in vacuo.
This modifies the effective cosmological constant contribution in
Einstein's equations. Furthermore, in the presence of incoherent matter
contributions (e.g.\ from dust and radiation) described by a perfect fluid
we have the additional contribution (\ref{eq:Tmunu}). Therefore, the final
Einstein's equations in terms of coherent and incoherent contributions of
matter, plus the vacuum energy of the fields, finally read
\begin{equation}\label{EEvac}
R_{\mu\nu}-\frac{1}{2}g_{ab}R=8\pi\,G_N\,\left(\langle\tilde{T}_{\mu\nu}^{\phi}\rangle+
T_{\mu\nu}\right)=8\pi\,G_N\,\left[g_{\mu\nu}\left(\rLV+\rLI\right)+T_{\mu\nu}\right]\,.
\end{equation}
We conclude that the ``physical value'' of the CC, at this stage, is not
simply the original term  $\rLV$, but
\begin{equation}\label{rLphclass}
\rLP=\rLV+\rLI\,,
\end{equation}
where the induced part is given by (\ref{eq:Higgstree}). On the face
of this result, it is pretty obvious that when we compare theory and
experiment a severe fine tuning problem is conjured in equation
(\ref{rLphclass}). Indeed, the lowest order contribution from the
Higgs potential, as given by equation (\ref{eq:Higgstree}), is
already much larger than the observational value of the CC. Using
the recent LHC measurement of the mass of the Higgs-like particle,
suggesting the value $M_{\Hp}\simeq 125$
GeV\,\cite{HiggsDiscovery1}, equation (\ref{eq:Higgstree}) yields
$\rLI\simeq -1.2\times 10^8$ GeV$^4$. Thus, being the CC observed
value of order $\rLo\sim 10^{-47}$ GeV$^4$, the electroweak vacuum
energy density is predicted to be $55$ orders of magnitude larger
than the currently measured $\rLo$:
\begin{equation}\label{eq:ratioCCindCCzero}
\left|\frac{\rLI^{\rm EW}}{\rLo}\right|={\cal O}(10^{55})\,.
\end{equation}
Suppose that the induced result would exactly be $\rLI= -10^8$ {GeV}$^4$
and that the vacuum density would exactly be $\rLo=+10^{-47}$ {GeV}$^4$.
In such case one would have to choose the vacuum term $\rLV$ in equation
(\ref{rLphclass}) with the rather bizarre precision of $55$ decimal places
in order to fulfill the equation
\begin{equation}\label{finetuningclassical}
10^{-47}\,{\rm GeV}^4=\rLV+\rLI=\rLV-10^8\,{\rm GeV}^4\,.
\end{equation}
This is of course the famous fine-tuning problem.

Let us note that this problem is in no way privative of the cosmological
constant approach to the DE, but is virtually present in \textit{any}
known model of the DE, in particular also in the quintessence
approach\,\cite{PeeblesRatra03}. Indeed, the quintessence scalar field
potential $V(\varphi)$ is supposed to precisely match the value of the
measured DE density at present, $\rLo$, starting from a high energy scale,
usually some grand unified theory (GUT) scale $\varphi=M_X$ between $\sim
10^{16}$ GeV and $M_P\sim 10^{19}$ GeV. Even in the simplest case
$V(\varphi)\sim m_{\varphi}^2\,\varphi^2$, one finds
$m_{\varphi}^2\sim\rLo/M_X^2$. Defining the mass scale associated to the
current CC value, $m_{\CC}\equiv({\rLo})^{1/4}={\cal O}(10^{-3})$ eV, we
have the ratio
\begin{equation}\label{eq:quintessversusCC}
\frac{m_{\varphi}}{m_{\CC}}\sim \frac{m_{\CC}}{M_X}\sim
10^{-30}\,,
\end{equation}
which tells us that the mass of the quintessence field should be some
thirty orders of magnitude smaller than the CC mass scale that one tries
to explain! Apart from the numerical mass value that this implies
($m_{\varphi}\sim 10^{-33}$ eV), such situation is of course preposterous.
Therefore, the quintessence approaches, in addition from being plagued
with fine-tuning problems in no lesser degree than the original CC
problem, do introduce extremely unnatural small mass scales.

However, this is not quite the end of the story yet. In QFT the induced
value of the vacuum energy is much more complicated than the simple result
(\ref{eq:Higgstree}), and the fine-tuning problem is much more cumbersome
than the one expressed in equation (\ref{finetuningclassical}) -- see
Sect.\ref{sect:effHiggsFLAT} for some clues about the seemingly
devastating, new complications.

\section{Vacuum energy: zero-point energy and some cosmic numerology} \label{sect:ZPE1}

As we know, a nonvanishing $\CC$ leads to a nonvanishing value of
the vacuum energy density, or CC density, $\rL=\CC/(8\pi G)$. In the
previous section we have seen that even at the classical level we
can get a large contribution to $\rL$ from the spontaneous symmetry
breaking (SSB) of the electroweak symmetry, i.e. from the ground
state of the Higgs potential. In addition, quantum corrections to
the this potential can be quite significant, as we will discuss
later. The possible confirmation of the Higgs finding at the LHC
collider at CERN certainly strengthens the case for the vacuum
energy in QFT, but the quantum effects which we are going to refer
now do not depend on the existence of the Higgs potential as they
have a very generic character. They were noted much before the Higgs
potential and the SSB phenomenon entered the stage. We pay now some
preliminary attention to this important (and generic) phenomenon of
the quantum vacuum, only to come back to it in subsequent sections
from a more rigorous point of view.

\subsection{Zero-point energy in the old days} \label{sect:ZPEold}

The quantum effects we are now referring to are those emerging from the
vacuum-to-vacuum fluctuations of the quantum matter fields. They occur
already in the free field theory, in contrast to the SSB phenomenon. These
quantum fluctuations correspond to closed loop diagrams without external
tails (``blobs''). They describe the infinite number of oscillators with
all possible frequencies $\omega_{\bf k}$ to which we attach to any free
quantum matter field. The sum of all the (nonvanishing) ground state
energies of these oscillators constitutes the zero-point energy (ZPE) of
the corresponding quantum field: $E_0=(1/2)\sum_{\bf k} \hbar\omega_{\bf
k}$. Let us recall that the historical origin of the ZPE emerges from
Planck's theory for the black-body radiation in 1900. For the average
energy of an oscillator of frequency $\omega$ in equilibrium with
radiation at temperature $T$, Planck obtained
\begin{equation}\label{eq:PlanckBlackbody}
E_{\omega}=\frac{\hbar\omega}{e^{\hbar\omega/kT}-1}+\frac12\,\hbar\omega\,.
\end{equation}
He realized immediately that $E_{\omega}\to (1/2)\,\,\hbar\omega$
for $T\to 0$. This is the \textit{Nullpunktsenergie} or zero-point
energy of the oscillator, therefore a nonvanishing one. Despite he
soon tried to normalize it away treating it as an unphysical effect,
thirteen years later Einstein and Stern paid clever attention to the
fact that if one expands the formula (\ref{eq:PlanckBlackbody}) in
the classical limit $kT\gg\hbar\omega$ one finds $E_{\omega}\simeq
kT-(1/2)\,\,\hbar\omega+(1/2)\,\,\hbar\omega=kT$, as easily checked.
In other words, thanks to the isolated ZPE term on the
\textit{r.h.s.} of (\ref{eq:PlanckBlackbody}) one can recover the
expected classical limit for the average energy of an oscillator in
thermal equilibrium at temperature $T$. This convinced Einstein that
the ZPE could have some physical meaning after all.

But things are not so easy. The electromagnetic field oscillates with all
frequencies, and therefore one has to compute the infinite sum
$E_0=(1/2)\sum_{\bf k} \hbar\omega_{\bf k}$. The unrenormalized result is
of course infinite, but the renormalized quantity is perfectly finite (cf.
Sect.\,\ref{sect:ZPE2}) and its value may be of concern. The first
considerations on the ZPE of the quantum vacuum dates back to the
theoretical discussions made by W. Nernst as early as 1916 and by W. Pauli
in the 1920s (and again in a more formal way in 1933, see
Sect.\,\ref{sect:ZPE2}). However, all these discussions were mainly
focused on the electromagnetic field. As, however, the renormalized form
of the ZPE is proportional to the quartic power of the renormalized mass
of the corresponding field quantum ($\rL\propto m^4$, see below), in the
electromagnetic case we have no such physical contribution to the ZPE (at
least in the way it was originally conceived by the first pioneers
discussing these matters) owing to the massless nature of the photon.

It was probably Y.B. Zeldovich who first raised a serious concern on the
contribution to the ZPE from a ``typical'' massive
particle\,\cite{Zeldovich67} and expressed the necessity to subtract its
leading -- or ``first order''-- effect in the computation of the physical
value of the CC\,\footnote{It is not my intention to make full justice, in
this summarized account, to the extensive historical literature on the CC
problem -- see e.g. \cite{CCPWeinberg,CCproblem2} and
\cite{Rugh2002,Straumann2000}, and references therein. It will suffice to
say that the connection between vacuum energy and CC had already been
glimpsed by Lema\^\i tre in 1934 a few years after he found the
cosmological expanding solution in the presence of a $\CC$-term
\,\cite{Lemaitre1934}.}. In particular he noticed that the leading proton
contribution to the ZPE ($\rL\propto m_p^4\sim 1$GeV$^4$) is
overwhelmingly large as compared to any cosmic density, say the current
critical density $\rc^0$ expressed in typical particle physics units:
\begin{equation}\label{eq:criticaldensity}
\rc^0=\frac{3\,H_0^2}{8\pi\,G}=\frac{3}{8\pi}\,H_0^2\,M_P^2\sim 10^{-47}\,{\rm GeV}^4\,,
\end{equation}
where $H_0\sim 10^{-42}$ GeV is the current value of the Hubble rate and
$M_P=1/\sqrt{G}\sim 10^{19}$ GeV is Planck's mass (both expressed in
natural units). The current matter and vacuum energy densities ($\rmo$ and
$\rLo$) are proportional to (\ref{eq:criticaldensity}) up to factors of
order one, i.e. $\rmo=\OMo\,\rco$ and $\rLo=\OLo\,\rco$, with
$\OMo\sim\OLo={\cal O}(1)$. After subtracting the exceedingly large
leading term in the theoretical ZPE estimate, Zeldovich realized that
starting once more from $m_p$ as the ``typical'' mass scale of particle
physics one could produce a much more reasonable order of magnitude
estimate of the CC density  by means of a ``second order'' (or
``next-to-leading'') formula involving the natural presence of the
gravitational coupling, $G$. To this purpose he concocted the
dimensionally consistent ``construct''
\begin{equation}\label{eq:Zeldovich}
\rL\simeq G\,m_p^6=\frac{m_p^6}{M_P^2}\sim 10^{-38}\,{\rm GeV}^4\,.
\end{equation}
The previous estimate still strays off the modern value, but by ``only''
nine orders of magnitude -- in contrast to the demolishing 47 orders of
magnitude that one has to face if keeping the leading term $\rL\sim
m_p^4$. Needless to say at the time of Zeldovich there was no real
measurement of $\CC$, although of course an upper bound estimate of the
order of the critical density was in force. Therefore the order of
magnitude discrepancy was not that different from the present one. Even
so, formula (\ref{eq:Zeldovich}) could somehow still be considered as a
respectable estimate. An even more intriguing result obtains if one
replaces the proton by the pion (whose mass $m_{\pi}\simeq 0.1$ GeV is
roughly ten times smaller than that of the proton). Then one gets a better
approximation which differs now by ``only'' three orders of magnitude from
the correct order of magnitude result (\ref{eq:criticaldensity}). It turns
out that a kind of ``cosmic prediction'' of the pion mass was proposed by
Weinberg in 1972\,\cite{Weinberg1972} through the curious numerical
relation
\begin{equation}\label{eq:Weinberg}
m_{\pi}^3\sim\frac{H_0}{G}=H_0\,M_P^2\sim 10^{-4}\,{\rm GeV}^3\,,
\end{equation}
which indeed leads to $m_{\pi}={\cal O}(0.1)$ GeV. Amusingly, if we now
substitute Eq.\,(\ref{eq:Weinberg}) into (\ref{eq:Zeldovich})
--- after first replacing $m_p\to m_{\pi}$, according to our prescription -- we find
\begin{equation}\label{eq:geometric}
\rL\sim H_0^2\,M_P^2\sim 10^{-46}\,{\rm GeV}^4\,,
\end{equation}
which, according to (\ref{eq:criticaldensity}), is very close to the
correct order of magnitude of the current CC density, since
$\rLo=\OLo\,\rco$, with $\OLo\simeq 0.7$. From (\ref{eq:Weinberg}) and
(\ref{eq:geometric}) it also ensues the (no less ``cabalistic'') relation
$\rL\sim m_{\pi}^3\,H_0$, somehow suggesting a possible link between the
meson world of particle physics and cosmology. Unfortunately such relation
is untenable within GR, and therefore such link is impossible if following
that pathway. Equally problematic is a (subtlety disguised) form of the
previous relation, which has iteratively appeared in the literature in
more recent times, to wit:
\begin{equation}\label{eq:Schutzhold}
\rL\sim H_0\,\Lambda_{\rm QCD}^3\,,
\end{equation}
where $\Lambda_{\rm QCD}\simeq 200$ MeV is the QCD scale of the
strong interactions (not far away from the pion mass). Some people
has tried hard to seek for a fundamental reason behind a formula
like (\ref{eq:Schutzhold})\,\cite{Schutzhold02,OddHpowers} and
others explore the phenomenological consequences\,\cite{Carneiro09}.
Numerically it is much worse than (\ref{eq:geometric}), as it yields
$\rL\sim 10^{-44}$ GeV$^4$, thus failing by ``only'' three orders of
magnitude. But it is not this numerical failure which is most
disturbing (as numbers are comparable to the situation with the
previous pion formula -- thought of as an improved form of
Zeldovich's one); the problem here is of theoretical nature. As has
been emphasized in \cite{ShapSol09,ShapSol09b}, an equation like
(\ref{eq:Schutzhold}) -- or, for that matter, any other relation
where $\rL$ is extracted from an odd power of the Hubble rate -- is,
in principle, incompatible with the general covariance of the
effective action in QFT in curved spacetime, see the next section
for more details\footnote{In Sect.\,\ref{sect:DynamicalVacFundConst}
we will see that the $\Lambda_{\rm QCD}$ scale could play a relevant
cosmological role, but for quite a different reason, namely on
account of its potential cosmic time dependence linked to that of
the vacuum energy, and within a fully covariant
formulation\,\cite{FritzschSola2012}.}.

There are actually a number of peculiar numerical coincidences around the
value of the cosmological constant. Let me display a final
example\,\cite{Bogan13}. The following proposal can be argued in certain
axiomatic approaches\,\cite{Beck2008}:
$\CC=\left(1/\hbar^4\right)\,G^2\,\left({m_e}/{\alpha}\right)^6$, where
$m_e$ is the electron's mass and $\alpha$ the electromagnetic fine
structure constant. In terms of vacuum energy density, it renders the
correct order of magnitude estimate in natural units:
\begin{equation}\label{eq:CCBeck}
\rL=\frac{1}{8\pi\,M_P^2}\,\left(\frac{m_e}{\alpha}\right)^6\sim 10^{-47}\ {\rm GeV}^4\,.
\end{equation}
As we can see, it is a kind of improved Zeldovich's type of second order
relation (\ref{eq:Zeldovich}), but using the electron mass rather than the
proton mass, and involving the QED coupling as well.

In principle, we may have (or not) a good reason behind any of these
numerical games. The only ``fundamental'' thing to be done here is to make
sure that we subtract the leading (``first order'') contribution from the
ZPE, which for all kinds of known elementary particles (except perhaps for
a light neutrino) is much larger than the measured $\rL$\footnote{The
maximum wildness (or should I say madness) of the $\sim m^4$ contributions
to $\rL$ is achieved for the Planck mass, for which the discrepancy with
the current value is $M_P^4/\rLo=\left(M_P/m_{\CC}\right)^4\sim
\left({\cal O}(10^{19}{\rm GeV})/{\cal O}(10^{-3} {\rm eV})\right)^4\sim
10^{123}$. This is the ultimate state of ``paroxysm'' of the CC problem.
But, as we have seen in Sect.\,\ref{sect:HiggsVacuum} (and as we shall
further emphasize later on), those $123$ orders of magnitude should not be
considered for the time being as the most obvious and worrisome aspect of
the ``real'' CC problem!}. For a hypothetical neutrino of a few
meV$=10^{-3}$ eV, we have the suggestive result\,\cite{ShapSol99}
\begin{equation}\label{eq:neutrino}
\rL\sim m_{\nu}^4\sim 10^{-11}\,{\rm meV}^4\sim 10^{-47}\,{\rm GeV}^4\,,
\end{equation}
that falls in the right ballpark of the cosmic densities
(\ref{eq:criticaldensity}). In the next sections we will go in a
summarized way thorough some  interesting models based on specific
evolution laws for the cosmological term. But before that, let us pause
and meditate a bit more on the power structure in $H$ of equations
(\ref{eq:geometric}) and (\ref{eq:Schutzhold}), and possible
generalizations thereof.

\subsection{Odd versus even powers of $H$: possible structures for
$\rL(H)$} \label{sect:oddH}

\newtext{The aforementioned} lack of covariance which an odd power of $H$
would imply, and therefore the apparent necessity for even powers of $H$
in the structure of $\rL(H)$, can be understood from an analogy with QED
and QCD. Let us focus at the moment on QED. The effective action and
effective Lagrangian of QED (related at any order through $\Gamma_{\rm
eff}=\int d^4x\,{\cal L}_{\rm eff}$) can be written in  terms of two gauge
and Lorentz invariants of the Maxwell field $A_{\mu}$,
\begin{eqnarray}\label{eq:FandG}
{\cal F}&\equiv &\frac14\,F_{\mu\nu}\,F^{\mu\nu}=\frac12\,\left({\bf B}^2-{\bf E}^2\right)\nonumber\\
{\cal G}^2&\equiv &\left(\frac14\,F_{\mu\nu}\,^{*}F^{\mu\nu}\right)^2=\left({\bf E}\cdot{\bf B}\right)^2\ \ \ \ ({\rm with}\ ^{*}F^{\mu\nu}=\frac12\,\epsilon_{\mu\nu\alpha\beta}\,F^{\alpha\beta})\,,
\end{eqnarray}
where $F_{\mu\nu}=\partial_{\mu}A_{\nu}-\partial_{\nu}A_{\mu}$ is the
standard field strength tensor for QED. For constant and homogeneous
electric and magnetic fields, these are the only possible Lorentz scalars
which are gauge invariant. Therefore ${\cal L}_{\rm eff}$ can only be a
function of them, and this leads to the well-known
Heisenberg-Euler-Weisskopf Lagrangian\,\cite{HEW1936} (see e.g.
\,\cite{Dunne04} for a review), which describes the effective theory of
photons in the weak field limit. It corresponds to the Lagrangian where
the electrons of QED are integrated out, i.e. eliminated as main d.o.f. of
the effective theory. The Lagrangian involves only photons, but with small
interactions induced by the ``heavy'' electrons. The influence of the
latter is seen only through mere decoupling effects, hence suppressed by
inverse powers of the electron mass $m$. In terms of the above invariants,
the effective Heisenberg-Euler Lagrangian for low energy QED takes on the
form
\begin{equation}\label{eq:EHLagrangian}
{\cal L}_{\rm eff}({\cal F},{\cal G}^2)=-{\cal F}+\frac{2\alpha^2}{45\,m^4}\left(4\,{\cal F}^2+7\,{\cal G}^2\right)+{\cal O}\left(\frac{{\cal F}^3}{m^8}\right)+...
\end{equation}
where $\alpha=e^2/4\pi$ is the fine structure constant. For weak fields,
both ${\cal F}^2$ and ${\cal G}^2$ are smaller than $m^4$, and moreover
there is the $\alpha^2$ coupling suppression. Therefore, the second (and
subsequent) term(s) on the \textit{r.h.s.} of (\ref{eq:EHLagrangian})
represents a tiny correction to the free Maxwell Lagrangian $-{\cal F}$.
Despite their smallness, the qualitative importance of the $\alpha^2$
non-linear correction in (\ref{eq:EHLagrangian}) is that it triggers
electric and magnetic polarization vectors of the vacuum. It all happens
as though the dielectric constant of the vacuum would be slightly shifted
from one and hence also changing a little bit the corresponding refraction
index. As a result the QED vacuum behaves as a polarizable medium with
dielectric properties.

For nonconstant/nonhomogeneous electric and magnetic fields, new gauge
invariant Lorentz scalars enter the effective Lagrangian. Using
dimensional analysis, we expect new terms that may involving insertions of
the $\Box\equiv\partial_{\mu}\partial^{\mu}$ operator. For example,
structures of the form
\begin{equation}\label{eq:additionalTerms}
F_{\mu\nu}\frac{\Box}{m^2}F^{\mu\nu}\,,\ \ \ F_{\mu\nu}\frac{\Box^2}{m^4}F^{\mu\nu}\,,\ \ \ F_{\mu\nu}\frac{\Box}{m^6}F^{\mu\nu}\,{\cal F}\,, \ \ \ \ {\rm etc}.
\end{equation}
Thus, the simplest extension of (\ref{eq:EHLagrangian}) for non-constant
electromagnetic fields would be
\begin{equation}\label{eq:EHLagrangian2}
{\cal L}_{\rm eff}({\cal F},{\cal G}^2)=-{\cal F}+\frac{\alpha}{60\pi}\,F_{\mu\nu}\frac{\Box}{m^2}F^{\mu\nu}+\frac{2\alpha^2}{45\,m^4}\left(4\,{\cal F}^2+7\,{\cal G}^2\right)+...
\end{equation}
Here the structure of the $\Box$-term is computed from the low-energy
limit of the standard photon vacuum polarization
function\,\cite{QFTBooks}:
\begin{equation}\label{eq:vacpolariz}
\Pi_{\mu\nu}(p)=\frac{\alpha}{15\,\pi}\,\left(p_{\mu}p_{\nu}-g_{\mu\nu}\,p^2\right)\,\frac{p^2}{m^2}+...
\end{equation}
If we would choose some particular frame and we would write the expression
(\ref{eq:EHLagrangian2}) in terms of the electromagnetic fields ${\bf E}$
and ${\bf B}$ in that frame (let us use a variable $\E$ generically
referring to both) it is easy to see that we would find an expansion in
even powers of ${\E}$, which can be roughly indicated as
\begin{equation}\label{eq:EHLagrangian3}
{\cal L}_{\rm eff}(\E)= \alpha_2 {\E}^2+ \alpha_4 {\E}^4+ {\cal O}\left({\E}^{2n}\right)...
\end{equation}
in which the coefficients $\alpha_i=A_i+B_i(p^2/m^2)$ have a constant part
($A_i$) and another that depends on some momentum scale $p$ connected with
the $\Box$-operator insertions, and satisfying $p^2/m^2\ll 1$. Similar
expansions could be done if we would treat the strong field limit, but
then the coefficients $\alpha_i$ would show the typical log behavior as a
function of $\mu=p$ derived from the renormalization group (RG) in usual
gauge theories of Particle Physics, since the Lagrangian would then
involve non-local terms of the form $F_{\mu\nu}\, \log
\Big(\Box/{m_e^2}\Big)\,F^{\mu\nu}$. However, it is pretty clear that,
again, only even powers ${\cal O}\left({\E}^{2n}\right)$ would arise in
the corresponding effective Lagrangian.

When we consider the gravitational case, the invariants $\F$ and $\G$ are
replaced by the general coordinate curvature scalar $R$ and the higher
order derivative terms, e.g. $R^2$, $R_{\mu\nu}R^{\mu\nu}$... The analog
of choosing a frame for the electromagnetic fields here is to choose the
FLRW cosmological metric. Then the curvature scalar, for example, reads
\begin{equation}\label{eq:Rset}
|R|=6\left(\,\frac{\ddot{a}}{a}+\frac{\dot{a}^2}{a^2}\,\right)
=12\,H^2+6\,\dot{H}\,.
\end{equation}
Therefore $R\sim H^2$ and also $R\sim\dot{H}$.
%That is, it is the Hubble function $H$ that plays here the role of the generic electromagnetic variable $E$ in the QED case studied above.
So when we require covariance and demand the presence of even powers of
$H$ we could as well demand powers of $\dot{H}$, this is taken for granted
in our compact language. However what we cannot  demand,  if covariance is
to be a prerequisite for our theory, is to have odd powers of $H$. A
similar consideration holds for the expansion of any higher order
invariant $R^2$, $R_{\mu\nu}R^{\mu\nu}$..., in the FLRW metric. Only
powers $\sim H^4$, $\dot{H}^2$, $H^2\,\dot{H}$ can appear (all of them to
be denoted $\sim H^4$), but no odd powers in $H$. Notice that $\ddot{H}$
and $\ddot{H}H$ would count as $\sim H^3$ and $\sim H^4$, respectively.
Therefore, the analog of (\ref{eq:EHLagrangian3}) in the case of the
cosmological effective action is expected to have the general form
\begin{equation}\label{eq:HLagrangian}
{\cal L}_{\rm eff}(H)= c_0+ c_2\,{H}^2+ c_4\,{H}^4+ {\cal O}\left({H}^{2n}\right)...
\end{equation}
where here we admit the presence of an additive term $c_0$, independent of
$H$. In the gravitational case such term is relevant, as we shall comment
later on. But we would like to warn the reader from the beginning that the
ostensibly ``simple'' formula (\ref{eq:HLagrangian}) is not as
``innocent'' and na\"if as one might think at first sight. To start with,
here we do not have a simple momentum dependence as in the QED case, and
the task of identifying the correct cosmic physical scale remains, as we
shall discuss in subsequent sections. Some more comments on the previous
expansion are given below, and then in more detail in
Sect.\,\ref{sect:DynamicalVacEner}. But let us now briefly elaborate on
the QCD infrared (IR) behavior, which can be trickier.

\newtext{At first sight}, in the QCD case the situation is entirely similar to QED.
Once more only powers of the invariants (\ref{eq:FandG}) are expected
(including the corresponding $\Box$-operator insertions for non-constant
chromoelectric and chromomagnetic fields), with the only difference that
the field strength $F_{\mu\nu}^a$ is now more complicated since the
non-Abelian gauge fields $A_{\mu}^a\,(a=1,2,3,..8)$ do couple here among
themselves. However, the IR behavior of QCD can be a bit more complicated,
and non-local physics might enter the problem and should be taken into
account, at least from the pure QCD point of view. After all QCD displays
non-trivial vacuum condensates of various sorts, e.g. $\langle 0|
F_{\mu\nu}^a\,F^{\mu\nu}_a | 0\rangle\neq 0$, whereas QED does not. It is
well-known that these condensates may actually introduce a quantum
contribution to the CC density which is of order $\sim
10^{-4}$GeV$^4$\,\newtext{\cite{Cesareo87}} and hence some $43$ orders of
magnitude larger than the observed $\rLo$ value! Obviously, even if being
12 orders of magnitude smaller than in the electroweak case (cf.
Sect.\,\ref{sect:HiggsVacuum}) the situation is still quite serious!
Whether this QCD vacuum contribution can be translated to the cosmic
vacuum is debatable, but  we cannot discard the possibility that a high IR
sensitivity in QCD  may imply that its vacuum energy depends on the
background\,\cite{Zhitnitsky12}; and hence it might, in turn, back react
on it. In this sense, the aforementioned assumptions made in the QED case
on the expansion in terms of ${\cal F}$ and ${\cal G}^2$ may not be
strictly correct in the QCD case. If so, lower dimensional operators of
the form $A^2\equiv A^a_{\mu}A_a^{\mu}$ (as a result of high IR
sensitivity) might emerge owing to the IR
renormalons\,\cite{ZhitnitskyPrivate}. In that case the kind of expansion
for the effective QCD Lagrangian involving only gluons would not
necessarily be of the simple form (\ref{eq:EHLagrangian2}), and
correspondingly this could also spoil the even power series kind of
expansion (\ref{eq:EHLagrangian3}) in the chromoelectric and
chromomagnetic fields.

The previous considerations are worth being kept in mind. However, we are
still far away from linking the tricky IR behavior of QCD (with all its
complicated, and maybe background-dependent, tunneling dynamics between
topologically distinct vacua carrying different winding numbers) with the
corresponding situation of GR. Only if the IR behavior of QCD were of the
mentioned type, and at the same time QCD could interact with the spacetime
background, would it be conceivable the presence of odd powers of $H$ in
the effective action, which would imply a violation of covariance in a
manner similar to the anomalous gauge role played by the IR-induced $A^2$
terms in the pure QCD case. In the meanwhile we have no real guarantee
that non-manifestly covariant structures are possible in the form of odd
powers of $H$.

We have already pointed out that  the even power series expansion in
$H$, Eq.\,(\ref{eq:HLagrangian}), is not as a na\"\i f expression as
it might look at first sight. If interpreted as generically
describing that part of the effective action of QFT in curved
spacetime describing the quantum corrections to the vacuum energy
density or cosmological term, we have to make clear promptly that
there is no straightforward method to prove it. The reason is that
these corrections cannot just come from local terms, not even from a
finite number of non-local terms (such as Green's function
operators)\,\cite{ShapSol09,ShapSol09b}. While a frontal
calculational approach to the renormalization of the vacuum energy
density in QFT in curved spacetime is not possible at the moment, a
roundabout way to hint at the structure of (\ref{eq:HLagrangian})
may be possible from the RG approach (cf.
Sect.\ref{sect:DynamicalVacEner} for a heuristic approach).

In the previous discussions, powers of the invariants were always
implicitly understood as integer ones. One possibility to have odd powers
of $H$ in a dynamical vacuum energy density would be, of course, to resort
to fractional powers of the invariants, say $R^{1/2}$, $R^{3/2}$... which
would provide $\sim H$, $\sim H^3$,... contributions in the FLRW metric
respectively. But, are we ready for such an eccentric possibility while
other, more amenable, options are still there to be fully exploited? For
example, the numerically successful relation (\ref{eq:geometric}) depends
quadratically on $H$ and therefore is compatible with the aforementioned
covariance. It follows that the argument that led to
Eq.\,(\ref{eq:geometric}) cannot be (without invoking contrived
assumptions) the one that we followed above through $\rL\sim
m_{\pi}^3\,H_0$ or (\ref{eq:Schutzhold}) since the latter cannot be
accepted in a natural way. It means that if there is any truth around
Eq,\,(\ref{eq:geometric}) there must exist some completely independent
pathway leading to it. We will see that there are indeed quite different
paths pointing to (\ref{eq:geometric}), or, in general, to the ``affine''
quadratic relation
\begin{equation}\label{eq:affineH2}
\rL(H)=c_0+\beta\,M_P^2\,H^2\,,
\end{equation}
with a non-vanishing $c_0$ term,  $\beta$ being here a dimensionless
coefficient. Eq.\,(\ref{eq:affineH2}) can be understood as a truncated
form of (\ref{eq:HLagrangian}) for the low energy physics of the current
universe -- see Sect.\ref{sect:EarlyUniverse}  for more details. The
presence of $c_0\neq 0$ is crucial for a realistic implementation of the
model, as it enables the transition from deceleration to acceleration in
this kind of models. The strict model (\ref{eq:geometric}), understood as
a model of the kind (\ref{eq:affineH2}) with $c_0=0$, is ruled
out, as it cannot satisfy the aforementioned
transition condition. This is also the situation with entropic force
models\,\cite{Frampton10}, for example, and many other models previously
presented on purely phenomenological grounds. In contrast, the class of
affine models (\ref{eq:affineH2}) is perfectly safe in this respect, and
in fact it has been successfully tested against the recent cosmological
data\,\cite{BPS09,GSBP11,BasPolarSola12,BasSola13a}.

\section{Zero-point energy in quantum field theory in flat spacetime}
\label{sect:ZPE2}

If we wish to go beyond the previous numerical games, things can get a bit
harder, even if we still try to keep them as simple as possible. Let us
therefore first follow a very naive formulation. Formally the ZPE of a
given quantum field, say a scalar field $\phi$, is obtained by selecting
that part of the effective potential which does not depend on the external
tails of $\phi$ (i.e. that part which is \textit{not} a function of
$\phi$). For example, the Higgs potential (cf. Section
\ref{sect:HiggsVacuum}) is in general \textit{not} a part of the ZPE
because it has a classical part (one which does \textit{not} vanish when
we set $\hbar=0$). It means that it consists of all the bubble-type
(vacuum-to-vacuum) diagrams at all orders in perturbation theory. The ZPE
is thus a pure quantum effect: it vanishes if there is no quantum theory,
$\hbar=0$. Indeed, vacuum-to-vacuum diagrams can only exist in a field
theory with vacuum fluctuations: QFT. The final result therefore can only
depend on a list of parameters $P$ (masses and coupling constants), but
not on $\phi$ itself, which can only enter virtually in the loop
propagators. If we count the loop order of perturbation theory with the
corresponding power of $\hbar$, the loopwise expansion can be presented as
a power series in $\hbar$:
\begin{equation}\label{eq:ZPE}
\ZPE(P)=\hbar\,V_P^{(1)}+\hbar^2\,V_P^{(2)}+\hbar^3\,V_P^{(3)}+....
\end{equation}
(A contribution to the 21th term of this pure vacuum-to-vacuum series
within the SM can be seen in Fig.\ref{Fig1:blobs}) From
Eq.\,(\ref{eq:ZPE}) it is obvious that even the first term depends on
$\hbar$, just linearly. This is the one-loop approximation. Since we
promised to keep things simple, let us evaluate the ZPE of a QFT for a
free scalar field at one loop only. For fermion field of spin $s_f=1/2$ an
additional factor of $4\, (={\rm Tr}\, \hat{\bf 1}_{\rm Dirac})$ and an
overall minus sign should both be inserted.

A naive calculation of the coefficient $V_P^{(1)}$ in (\ref{eq:ZPE}) is
obtained by ignoring for the moment gravity and regularizing the infinite
sum $(1/2)\sum_{\bf k} \hbar\omega_{\bf k}=(1/2)\sum_{\bf k} \hbar
\sqrt{{\bf k}^2+m^2}$ by means of a cutoff ${\CUV}$. Moving to the
continuum and trivially integrating the solid angle, it gives
\begin{eqnarray}\label{eq:ZPE1loop}
V_P^{(1)} &=& \frac{1}{2}\int \frac{{\rm d}^3k}{(2\pi)^3}\, \sqrt{{\bf
k}^2+m^2}=\frac{1}{4\pi ^2}\int _0^{\CUV} {\rm d}k\, k^2
\sqrt{k^2+m^2}\nonumber\\
& = &\frac{\CUV^4}{16
\pi^2}\left(1+\frac{m^2}{\CUV^2}-\frac14\,\frac{m^4}{\CUV^4}\,\ln\frac{\CUV^2}{m^2}
+\cdots \right)\,,
\end{eqnarray}
where in the expansion in powers of $m/\CUV$ we have explicitly kept the
important $\sim m^4\ln m^2$ term because this one does not depend on any
power of the cutoff and therefore is the term that should remain after we
attempt to remove the cutoff by some renormalization procedure. In the
1930's Pauli applied this na\"\i f approach to the electromagnetic field
($m=0$ for the photon) and choosing the inverse of the classical electron
radius for the cutoff, i.e. $\CUV=2\pi\,m_e/\alpha$ (here $\alpha$
standing for the fine structure constant). Then he plugged the result into
Einstein's universe formula (\ref{eq:EinsteinUniverse}), i.e. he replaced
$\rL$ there by the previous one-loop estimate of the ZPE, and obtained the
``radius'' of the ensuing universe:
\begin{equation}\label{eq:radiusEinsteinU}
a=\sqrt{\frac{2\pi}{G}}\,\frac{1}{\CUV^2}=\left(2\pi\right)^{-3/2}\,\left(\frac{M_P}{m_e}\right)\frac{\alpha^2}{m_e}\,.
\end{equation}
It would not even reach the Moon! (as a matter of fact the result is
appallingly small, some twenty six kilometers only!)\,\footnote{Strictly
speaking one has to include a factor of $2$ in Eq.\,(\ref{eq:ZPE1loop}) to
account for the two helicities of the photon, so the result for the
``radius'' is $1/\sqrt{2}$ smaller , i.e. some $\sim18$ kilometers, but of
course this nicety is irrelevant here.}. As we can see, the larger is the
vacuum density the smaller is the equilibrium ``radius''. This is
intuitively obvious from the fact that the negative pressure associated to
the CC value is very big even for a cutoff as small as the electron mass,
so that if gravity has to compensate such outwards vacuum pressure the
universe must be small enough to produce a gravitational (inwards) effect
of the same size. He was dismayed, but the argument is itself actually
non-rigorous. Indeed, the results
(\ref{eq:ZPE1loop})-(\ref{eq:radiusEinsteinU}) explicitly depend on the
value we can arbitrarily assign to the regulator $\CUV$; in other words,
it is merely an unphysical result obtained in the ``bare theory''.
Sometimes a particular value can better approximate the final result, but
in actual fact the physical result should be completely independent of
$\CUV$. It means that one must first renormalize the theory before jumping
to conclusions, as without renormalization the result
(\ref{eq:radiusEinsteinU}) means very little. This implies, among other
things, to get rid of the regulator before extracting meaningful
conclusions on the ZPE contribution from the electromagnetic field.
However, after doing so we are left with nothing since the photons have no
mass and hence no contribution to the ZPE could remain from them. Indeed,
upon removing the cutoff we expect that the renormalized result should be
of order
\begin{equation}\label{eq:ZPE1loopRenorm}
V_P^{(1){\rm renorm}} = -\frac{m^4}{64\pi^2}\,\ln\frac{\mu^2}{m^2}
+\cdots \,.
\end{equation}
where $\mu$ is some subtraction scale that must remain after
renormalization. From the above formula it is clear that the ZPE becomes
zero if we set $m=0$ in it. If we would instead compute the contribution,
not from the electromagnetic field, but from a scalar field of mass $m$,
the renormalized contribution should be of the order (\ref{eq:ZPE1loop}),
thus proportional to $m^4$. Unfortunately, we find that even after trying
to make some sense out of the ZPE within QFT we are back to the huge
quartic contributions to the CC which we tried to avoid in the previous
section. They are back here and this becomes of course disquieting. As for
Pauli's quantitative result, despite its roughness and severe limitations
it transmits the correct qualitative idea that any typical choice for the
cutoff scale within the particle physics domain leads to a very
unsatisfactory value for the cosmological constant as compared to the
observational measurements. We thus realize that the ZPE calculation
becomes conflictive with observations even after renormalization. Despite
we did not use a concrete renormalization scheme to guess at the
renormalized form (\ref{eq:ZPE1loopRenorm}) from the regularized
expression (\ref{eq:ZPE1loop}), it turns out that the latter is indeed
what is obtained e.g. using dimensional renormalization (see below).

Although every scheme is in principle valid, not every scheme has the
property that the renormalized quantities are in good correspondence with
the physical quantities in the particular framework of the calculation.
For example, it makes no much sense to renormalize  QCD (the gauge theory
of strong interactions) in the on-shell scheme because we never find
quarks and gluons on mass shell! One is forced to use an off-shell
renormalization scheme. This can complicate the physical interpretation,
for there can indeed be a nontrivial gap between the renormalized
parameters and the physical ones. And if all that is not enough, the CC
problem\,\cite{CCPWeinberg} is of course a problem where gravity should
play a role somewhere in the calculation of the ZPE, shouldn't it? So, at
the end of the day, we realize that the result (\ref{eq:ZPE1loop}) is in
itself far from having any physical meaning!

In an attempt to smooth out some of these problems, suppose we adopt the
MS-scheme (or Minimal Subtraction scheme) in $n$-dimensional
regularization. Then, the one-loop approximation to the ZPE renders
\begin{equation}\label{Vacfree2}
V_P^{(1)}=
\frac12\,\mu^{4-n}\, \int\frac{d^{n-1} k}{(2\pi)^{n-1}}
   \,\sqrt{{\bf k}^2+m^2}
= \frac12\,\beta_\Lambda^{(1)}\,\left(-\frac{2}{4-n}
-\ln\frac{4\pi\mu^2}{m^2}+\gamma_E-\frac32\right) \,,
\end{equation}
where $\gamma_E$ is Euler's constant, and
\begin{equation}
\label{beta4} \beta_\Lambda^{(1)}=\frac{m^4}{2\,(4\pi)^2}
\end{equation}
is the one-loop $\beta$-function for the vacuum term\,\cite{Brown92}. Notice
that $\mu$ is the characteristic 't Hooft mass unit of dimensional
regularization\,\cite{Hooft73,BolliniGiambiagi72}, and $n\rightarrow 4$ is
understood in the final result. So obviously the ZPE is UV-divergent once
more, as it could not be otherwise without renormalization.

How to get now a finite (if still not physical) value of the ZPE? We
just have to follow the renormalization program. Recall that the
Einstein-Hilbert action from which the field equations
(\ref{EEnoCC2}) are derived reads as in
Eq.\,(\ref{EH})\,\cite{CosmologyBooks}:
\begin{eqnarray} S_{\rm EH} &=& \frac{-1}{16\pi\,\Gb}\,\int d^4
x\sqrt{-g}\, \left(\,R + 2\,\BCC\,\right)=-\,\int d^4
x\sqrt{-g}\,\left(\frac{1}{16\pi\,\Gb} \,R+\rLb\right).\nonumber\\
\label{EHb}
\end{eqnarray}
Here (and hereafter) we rewrite the parameters $\CC\to\BCC$,
$\rLV\to\rLb$ and $G\to\Gb$ so as to emphasize it is the bare
action, i.e. the action before any renormalization program is
applied to account for the UV-divergences related to the quantum
matter contributions. Even though this action is not renormalizable,
we know that by including the higher order invariant terms ( i.e.
$R^2$, $R_{\mu\nu}R^{\mu\nu}$..., only relevant for the
short-distance behavior of the theory) the extended action is
renormalizable in the context of QFT in curved spacetime (where
gravity is not quantized but the matter fields yes), provided (!) we
keep the $\CC$-term in it -- see Sect.\,\ref{sect:ZPE2CURVED} for
more details. So let us maintain this term in the low-energy part of
the full renormalizable action, i.e. the long-distance EH action
(\ref{EHb}), as this part is both necessary for the renormalization
program and is accessible to observation (as we indeed know from the
observed accelerated expansion). Actually, what is accessible is the
sum of this term plus the induced quantum effects (see later on).
Only the overall quantity is physical.

The key point now, at the theoretical level, is that the CC-term in
(\ref{EHb}) is \emph{not} yet the physical quantity, it is only the
bare parameter of the bare EH action. When we include the ZPE as a
part of the full action, the overall additive term is no longer
$\rLb$ but the sum $\rLb+\ZPE^{({\rm b})}$, where $\ZPE^{({\rm b})}$
is given by (\ref{eq:ZPE}). Next we split $\rLb$ into a renormalized
part plus a counterterm, $\rLb=\rL(\mu)+\delta\rL$, where the
renormalized part depends on the arbitrary renormalization scale
$\mu$ and the counterterm depends on the regularization and
renormalization scheme. In the ${\rm MS}$ scheme one introduces, as
usual, a counterterm killing the ``bare bone'' UV-part (the pole at
$n\rightarrow 4$). In the slightly modified $\overline{\rm MS}$
scheme one collects also some additive constants. Specifically, the
counterterm reads:
\begin{equation}\label{deltaMSB}
{\delta}\rL^{\overline{\rm
MS}}=\frac{m^4\,\hbar}{4\,(4\pi)^2}\,\left(\frac{2}{4-n}+\ln
4\pi-\gamma_E\right)\,.
\end{equation}
Adopting therefore the  $\overline{\rm MS}$ scheme in dimensional
regularization we are led to use the above explicit form
(\ref{deltaMSB}) for $\delta\rL$. Finally, since the (effective
action of the) bare theory must equal (that of) the renormalized
theory (in whatever renormalization scheme), we must have
$\rLb+\ZPE^{({\rm b})}=\rL(\mu)+\ZPE(\mu)$. This condition defines
the $\overline{\rm MS}$-renormalized one-loop value of the ZPE.
Using the above equations, at one-loop it reads:
\begin{equation}\label{renormZPEoneloop}
V^{(1)}_{\rm ZPE}(\mu)=\hbar\,V_P^{(1)}+\delta\rL^{\overline{\rm MS}}=\frac{m^4\,\hbar}{4\,(4\,\pi)^2}\,\left(\ln\frac{m^2}{\mu^2}-\frac32\right)\,,
\end{equation}
which is perfectly finite. Incidentally, with this result we have checked
that the one-loop correction in dimensional regularization with minimal
subtraction indeed provides the kind of finite correction that we guessed
in Eq.\,(\ref{eq:ZPE1loopRenorm}) from the bare result
(\ref{eq:ZPE1loop}), up to additive parts related with the subtraction
procedure. This does not mean that the obtained expression is the physical
result, but at least is a renormalized and hence finite one. The price for
the finiteness of the renormalized result is its dependence on the
arbitrary mass scale $\mu$. However, while both pieces
(\ref{renormZPEoneloop}) and $\rL(\mu)$ separately depend on the scale
$\mu$, the sum
\begin{equation}\label{renormZPEoneloop2}
\rVu=\rL(\mu)+V^{(1)}_{\rm ZPE}(\mu)=\rL(\mu)+\frac{m^4\,\hbar}{4\,(4\,\pi)^2}\,\left(\ln\frac{m^2}{\mu^2}-\frac32\right)\,,
\end{equation}
does \emph{not} since by construction we started from the bare theory,
which is of course $\mu$-independent. The sum (\ref{renormZPEoneloop2})
represents the $\overline{\rm MS}$-renormalized vacuum energy of the free
field at one loop. As this quantity is the very same starting bare
expression, only rewritten now in terms of renormalized parameters, it is
overall $\mu$-independent. This is actually the main message of the
renormalization group (RG): the sum of the various $\mu$-dependencies must
cancel in the renormalized effective action, and also in the renormalized
$S$-matrix elements (when they can be defined). We can now check
explicitly that (\ref{beta4}) is indeed the one-loop $\beta$-function for
the running of the renormalized CC. Computing the logarithmic derivative
$d/d\ln\mu=\mu d/d\mu$ on both sides of (\ref{renormZPEoneloop2}) and
taking into account that this gives $d\rV^{(1)}/d\ln\mu=0$ on the
\textit{l.h.s} (for the reasons mentioned above), we immediately find the
desired result:
\begin{equation}\label{betaFunct}
\mu\frac{d\rL(\mu)}{d\mu}=\frac{\hbar\,m^4}{2\,(4\pi)^2}=\beta_\Lambda^{(1)}\,.
\end{equation}
The RG tells us something useful about the explicit $\mu$-dependence
affecting incomplete structures of the effective action. Indeed, while we
may not know the full structure of the effective action in a particular
complicated situation, our educated guess in associating $\mu$ with some
relevant dynamical variable of the system can give relevant information on
physics, similarly as we do in particle physics. This is e.g. the case
with an effective charge (or ``running coupling constant''), say the QED
or QCD renormalized gauge coupling $g=g(\mu)$, which is explicitly
$\mu$-dependent even though the full effective action or $S$-matrix
element is not.

For a more rigorous connection with the curved space-time case discussed
in the next section, it is convenient to approach Eq.\,(\ref{Vacfree2})
from a more formal point of view, namely from the notion of effective
action\,\cite{QFTBooks}. The desired form for the effective action at
one-loop reads:
\begin{equation}\label{EAoneloop}
\Gamma_{\rm eff}[\pc]=S[\pc]+\frac{i\,\hbar}{2}\,Tr\,\ln {\cal K}(x,x')\,.
\end{equation}
where $S[\pc]$ is the classical action
\begin{eqnarray}
S[\pc]=\int
d^4x\,{\cal L}=\int
d^4x\,\,\left[\frac12\,\gmnu\,\partial_{\mu}\pc\,\partial_{\nu}\pc-V_c(\pc)\right]\,,
\label{Sclassical}
\end{eqnarray}
and
\begin{equation}\label{eq:KGmatrix}
\K(x,x')=\left[\Box_x+V''(\pc)\right]\,\delta(x-x')\,,
\end{equation}
is essentially the inverse propagator in the presence of the background
matter field $\pc$.  Here $V_c$ is the tree-level or classical part of the
potential for that field, typically of the form (\ref{Poten}). For a free
field ($\lambda=0$) it contains only the mass term: $V(\phi)=(1/2)\,m^2\,
\phi^2$ and Eq.\,(\ref{eq:KGmatrix}) simplifies to
\begin{equation}\label{eq:KGmatrixFree}
\K(x,x')=\left[\Box_x+m^2\right]\,\delta(x-x') \ \ \ \ \ \  ({\rm
free\ QFT})\,.
\end{equation}
The result (\ref{EAoneloop}) was expected, namely the effective action at
this order is the sum of the classical action (\ref{Sclassical}) and the
one-loop term (\ref{eq:KGmatrix}) -- notice the presence of $\hbar$ in
front of it in Eq.\,(\ref{EAoneloop}). The quantum correction term, call
it $\Gamma^{(1)}$,  is generated exclusively by the vacuum diagrams, and
therefore represents the ZPE. We can understand that indeed $\Gamma^{(1)}$
is associated to vacuum-to-vacuum diagrams (i.e. closed loop diagrams
without external tails of quantum matter) by the fact that we have to
compute a trace over all indices, including the spacetime ones:
\begin{equation}\label{EAoneloopFlat}
\Gamma^{(1)}=\frac{i\,\hbar}{2}\,Tr\,\ln {\cal K}(x,x')=\frac{i\,\hbar}{2}\,\int d^4 x \lim_{x\to x'}\,\ln [{\cal K}(x,x')]\,.
\end{equation}
Diagrammatically we can interpret we are following a closed line starting
at one point and ending at the same point, i.e. a vacuum-to-vacuum diagram
or ``blob''. Thereafter we have to sum these blobs over all of the
spacetime points, as indicated by the above integral. Setting $\pc=$const.
the classical action (\ref{Sclassical}) boils down to (minus) the
classical potential times the spacetime volume $\Omega$, i.e.
$S[\pc]=-\int d^4x V(\pc)=-\Omega V(\pc)$. The \textit{l.h.s.} of
(\ref{EAoneloop}) can then be written as $-V_{\rm eff}\,\Omega$, where
$V_{\rm eff}$ is the so-called effective potential. Clearly $V_{\rm
eff}=V_c+\hbar V^{(1)}$, where the one-loop correction reads
\begin{equation}\label{eq:V1}
V^{(1)}=-\,\frac{i}{2}\,\Omega^{-1}\,Tr\ln{\cal K}\,.
\end{equation}
Next the integral can be conveniently worked out in momentum space. For
the free field case (\ref{eq:KGmatrixFree}) it immediately leads to
\begin{eqnarray}\label{eq:TrV1}
Tr\,\ln{\cal K}=\Omega\,\int \frac{d^n
k}{(2\pi)^n}\,\ln\left(-k^2+m^2\right)\,,
\end{eqnarray}
where we moved to $n$ dimensions to regularize the result. The spacetime
volume cancels in (\ref{eq:V1}) and after a simple calculation one finally
retrieves the formula (\ref{Vacfree2}) and the corresponding
$\overline{\rm MS}$-renormalized final result(\ref{renormZPEoneloop2}).

\section{ZPE and the full effective Higgs potential for QFT in flat
spacetime} \label{sect:effHiggsFLAT}

The ZPE calculation for free fields considered in the previous section is
the simplest kind of quantum vacuum effect one can deal with. However, the
general renormalized effective potential extending to the quantum domain
the classical potential, $V_c\rightarrow \EP$, takes on the form
\begin{equation}\label{Veff}
 \EP=V_c+\hbar\,V_1+ \hbar^2\,V_2+ \hbar^3\,V_3+...
\end{equation}
The quantum effects to all orders of perturbation theory arrange
themselves in the form of a loopwise expansion where the number of loops
is tracked by the powers of $\hbar$. Thus, at one loop we have only one
power of $\hbar$, at two loops we have two powers of $\hbar$ etc. For
$\hbar=0$, however, there are no loops and the effective potential reduces
to the classical potential, $V_c$, given by equation (\ref{Poten}) in the
electroweak standard model. On the other hand, each of the loop terms in
(\ref{Veff}) can be split into two independent contributions, one
consisting of loops with no external legs ({vacuum-to-vacuum} parts
$V_P^{(i)}$, i.e. the ZPE contribution at ith-order) and the other
involving loops with external legs of the Higgs field $\phi$ (i.e.\ the
ith-loop correction $V_{\rm scal}^{(i)}(\phi)$ to the classical Higgs
potential):
\begin{equation}\label{splitloop}
V_1=V_P^{(1)}+V_{\rm scal}^{(1)}(\phi)\,,\ \ \ \
V_2=V_P^{(2)}+V_{\rm scal}^{(2)}(\phi)\,,\ \ \ \
V_3=V_P^{(3)}+V_{\rm scal}^{(3)}(\phi)...\,.
\end{equation}
As a result, the effective potential (\ref{Veff}) at the quantum level
splits naturally into two parts, one which is $\phi$-independent and
another that is $\phi$-dependent:
\begin{equation}\label{EPZPE}
\EP(\phi)=\ZPE+\tEP(\phi)\,,
\end{equation}
where
\begin{equation}\label{ZPE}
\ZPE=\hbar\,V_P^{(1)}+\hbar^2\,V_P^{(2)}+\hbar^3\,V_P^{(3)}+....
\end{equation}
is the full \textit{zero-point energy} (ZPE) contribution. It is a number,
it only depends on the set of parameters $P=m,\lambda,...$ of the
classical potential, not at all on the fields. As mentioned, the latter
consists in {the sum} of all the vacuum-to-vacuum parts of the effective
potential. The ZPE part is sourced exclusively from closed loops of matter
fields (i.e.\ vacuum loops without external $\phi$-legs). In the previous
section we have computed the one-loop contribution to the ZPE and for free
fields, i.e. just the term $V_P^{(1)}$. The ZPE receives {in general}
contributions to all orders of perturbation theory, except at zero loop
level since $\ZPE$  is a pure quantum effect that vanishes for $\hbar=0$.
Now, besides the ZPE there is the scalar field dependent part of the
effective potential:
\begin{equation}\label{loopwisepot}
V_{\rm scal}(\phi) = V_c(\phi)+\hbar\,V_{\rm scal}^{(1)}(\phi)+
\hbar^2\,V_{\rm scal}^{(2)}(\phi)+ \hbar^3\,V_{\rm
scal}^{(3)}(\phi)+...
\end{equation}
This one is not purely quantum (i.e.\ it does not vanish for $\hbar=0$) as
the first term (the classical potential) is, of course, {\emph not}
proportional to $\hbar$. The above $\phi$-dependent part of $V_{\rm eff}$
receives {in general} also contributions to all orders of perturbation
theory, and vanishes for $\phi=0$ since in this case all the loops have
external $\phi$ legs, {including} the tree-level part. Thus, in the
absence of SSB the effective potential boils down to just the ZPE,
\begin{equation}\label{eq:EPzerophi}
\EP(\phi=0)=\ZPE
\end{equation}
which is a number entirely constructed from a power series of $\hbar$. On
the other hand, the expression (\ref{loopwisepot}) is the effective
potential excluding that ZPE number. The full effective potential
(\ref{EPZPE}) contains both contributions.

In the above formulae, all the field theoretical ingredients ($m$,
$\lambda$, $\phi$ and $V_{\rm eff}$) are in fact bare quantities ($m_0$,
$\lambda_0$, $\phi_0$ and $V_{\rm eff\, 0}$) that require renormalization,
as the loopwise expansion is UV-divergent order by order. Renormalization
means that we replace all the bare quantities with renormalized ones (in
some given renormalization scheme with a specific set of renormalization
conditions) plus counterterms (which are also scheme dependent and are
partially fixed by the condition of canceling the UV-divergences):
$m_0=m+\delta m$, $\lambda_0=\lambda+\delta \lambda$,
$\phi_0=Z_{\phi}^{1/2}\,\phi=(1+\delta Z_{\phi}/2)\,\phi$... Of course, a
similar splitting occurs with the vacuum term, which was originally a bare
term $\rLb$. We must also split it into a renormalized piece plus a
counterterm: $\rLb=\rL(\mu)+\delta\rL$. The full set of counterterms is
essential to enable the loop expansion to be finite order by order in
perturbation theory. For instance, if we would renormalize the theory in
the $\overline{\rm MS}$ scheme in dimensional regularization, the suitable
counterterm for the vacuum parameter was given in Sect.\,\ref{sect:ZPE2},
see Eq.\,(\ref{deltaMSB}).

\subsection{Effective potential and renormalization group invariance} \label{sect:EffPotentRGinvariance}

For a practical calculation of the first (one-loop) quantum correction to
the Higgs potential (\ref{Poten}), we have to compute the one-loop term
(\ref{EAoneloopFlat}) in the presence of the $\lambda\neq0$ term in the
potential. In this case,
\begin{equation}\label{eq:KGmatrixNonFree}
\K(x,x')=\left[\Box_x+V''(\pc)\right]\,\delta(x-x')=\left[\Box_x+m^2+\frac12\,\lambda\,\pc^2\right]\,\delta(x-x')\,.
\end{equation}
For $\pc=$const. this will lead us to determine the explicit form for
$V_{\rm scal}^{(1)}(\phi)$ in the above language. This introduces some
complications but the calculation can still be carried out without much
problems. In the constant mean field limit we may equate the bare and
renormalized effective action and we obtain
\begin{equation}\label{RGpostulate}
\rLb+\EP(\phi_0, m_0, \lambda_0)=\rL(\mu)+\EP(\phi(\mu), m(\mu),
\lambda(\mu); \mu)\,.
\end{equation}
As always the renormalized result depends on an arbitrary mass scale $\mu$
The overall $\mu$-dependence, however, must eventually cancel. In fact,
the renormalized parameters are finite quantities which are also functions
of $\mu${:} $\phi=\phi(\mu)$, $m=m(\mu)$, $\lambda=\lambda(\mu)$,
$\rL=\rL(\mu)$, and since the vacuum energy cannot depend on the arbitrary
scale $\mu$, the sum of the renormalized vacuum term and the renormalized
potential must be globally scale-independent (i.e.\ $\mu$-independent).
This is obviously so because the bare vacuum term and bare effective
potential were scale-independent to start with. Thus, from
(\ref{RGpostulate}) we have
\begin{equation}\label{RGequation}
\mu\frac{d}{d\mu}\left[\rL(m(\mu), \lambda(\mu);
\mu))+\EP(\phi(\mu); m(\mu), \lambda(\mu); \mu)\right]=0\,.
\end{equation}
This relation implies that the full effective potential is actually
\textit{not} renormalization group (RG) invariant (contrary to some
inaccurate statements in the literature), but it becomes so only after we
add up to it the renormalized CC vacuum part $\rL$. In reality, the
structure of the effective potential (\ref{EPZPE}) is such that the
previous relation splits into two independent RG equations:
\begin{equation}\label{RGequation1}
\mu\frac{d}{d\mu}\left[\rL(m(\mu), \lambda(\mu); \mu))+\ZPE(m(\mu),
\lambda(\mu); \mu)\right]=0
\end{equation}
and
\begin{equation}\label{RGequation2}
\mu\frac{d}{d\mu}\, V_{\rm scal}(\phi(\mu); m(\mu), \lambda(\mu);
\mu)=0\,.
\end{equation}
Equation (\ref{RGequation1}) shows that it is only the strict
vacuum-to-vacuum part (i.e.\ the ZPE) the one that needs the renormalized
vacuum term $\rL$ to form a finite and RG-invariant expression, whereas
the renormalized $\phi$-dependent part of the potential (i.e.\ the
tree-level plus the loop expansion with external $\phi$-tails) is finite
and RG-invariant by itself. This is of course the essential message from
the renormalization group. Explicitly, equation (\ref{RGequation2}) reads
\begin{equation}
\left\{ \mu\frac{\partial }{\partial \mu}+\beta _{P}
\,\frac{\partial }{\partial P} - \gamma_{\phi} \,\phi\,
\frac{\partial }{\partial \phi}\right\} \, V_{\rm scal}
\left[P(\mu),\phi(\mu);\mu \right] =0\,,\label{RGVeff}
\end{equation}
where as usual $\beta _{P} = \mu{\partial P}/{\partial\mu}$
($P=m,\lambda,...$) and $\gamma_{\phi}=\mu{\partial \ln
Z_{\phi}^{1/2}}/{\partial\mu}$. Similarly, equation (\ref{RGequation1})
can be put in the form (\ref{RGVeff}), except that the $\phi$ term is
absent.

Plugging equation (\ref{renormZPEoneloop}) in the general RG equation
(\ref{RGequation1}), we find immediately that the renormalized vacuum term
$\rL(\mu)$ obeys the one-loop RG-equation which we have found previously,
i.e. Eq.\,(\ref{betaFunct}).

The next step is the one-loop renormalization of the effective potential.
This is standard\,\cite{QFTBooks}, of course, although the usual
discussions on this subject rarely pay much attention to disentangle the
ZPE part from it. Let us do it. Once more we have to compute the one-loop
expression (\ref{EAoneloopFlat}), but in this case for the operator
(\ref{eq:KGmatrixNonFree}), so we have
\begin{eqnarray}\label{eq:TrV1Pot}
Tr\,\ln{\cal K}=\Omega\,\int \frac{d^n
k}{(2\pi)^4}\,\ln\left(-k^2+V''(\pc)\right)\,.
\end{eqnarray}
Substituting this expression in (\ref{eq:V1}), we may split the result in
the suggestive form
\begin{equation}\label{V3}
V_1(\pc)=-\,\frac{i}{2}\,\int\frac{d^n
k}{(2\pi)^n}\,\ln\left[-k^2+m^2\right]-\,\frac{i}{2}\,\int\frac{d^n
k}{(2\pi)^n}\,\ln\left(\frac{k^2-V''(\pc)}{k^2-m^2}\right)\,.
\end{equation}
Notice that the first term is independent of $\phi$, i.e. it only depends
on the parameters of the potential. In the one-loop case, it depends only
on $m$, but at higher loops it would also depend on $\lambda$ (in the
interactive theory). Diagrammatically, it corresponds to a closed one-loop
diagram without external $\phi$-tails, i.e. to a vacuum-to-vacuum one-loop
diagram. As could be expected, this term can be identified with the
one-loop contribution for the $\ZPE$ in Eq.\,(\ref{ZPE}), previously
addressed in Sect.\,\ref{sect:ZPE2}. Similarly, the second integral on the
\textit{r.h.s.} of Eq.\,(\ref{V3}) gives the one-loop correction to the
$\phi$-dependent part of the potential, i.e. the second term on the
\textit{r.h.s.} of Eq.\,(\ref{loopwisepot}). Therefore Eq.\,(\ref{V3})
contains the full one-loop correction, which includes the ZPE part and the
$\phi$-dependent contribution to the potential. To it we still have to add
the classical potential, cf.\,Eq.\,(\ref{loopwisepot}).

Sticking to the $\overline{MS}$ scheme in dimensional regularization to
fix the counterterms, one finds the final result for the renormalized full
effective potential (\ref{EPZPE}) up to one-loop:
\begin{eqnarray}\label{VOVR2}
\EPR(\phi)=
\frac12\,m^2(\mu)\,\phi^2+\frac{1}{4!}\,\lambda(\mu)\,\phi^4
+\hbar\,\frac{\,\left(m^2+\frac12\,\lambda\,\phi^2\right)^2}{4(4\pi)^2}\left(\ln\frac{m^2+\frac12\,\lambda\,\phi^2}{\mu^2}-\frac32\right)\,,
\end{eqnarray}
Notice the implicit $\mu$-dependence of the masses and couplings. Together
with the explicit $\mu$-dependent parts of the effective potential, this
insures the full RG-invariance of  the effective action in the constant
mean field limit, i.e. the fulfilment of Eq.\,(\ref{RGequation}).

As shown in Eq.\,(\ref{V3}), the two kind of one-loop effects are built in
the calculation. Therefore, the expression (\ref{VOVR2}) must boil down to
the renormalized ZPE for $\phi=0$, and indeed we verify that in this limit
we recover the one-loop term on the \textit{r.h.s.} of
Eq.\,(\ref{renormZPEoneloop2}):
\begin{eqnarray}\label{VeffRen2}
\EPR(\phi=0)=\frac{\hbar\,m^4}{4(4\pi)^2}\left(\ln\frac{m^2}{\mu^2}-\frac32\right)\,.
\end{eqnarray}
This is accordance with the expectation in Eq.\,(\ref{eq:EPzerophi}). We
are now ready for addressing the CC fine-tuning problem in the context of
a well-defined, renormalized, and RG-invariant vacuum energy density in
flat space.

Once the full effective potential has been renormalized, the two loopwise
expansions (\ref{ZPE}) and (\ref{loopwisepot}) become finite to all orders
of perturbation theory. Furthermore, the basic equation (\ref{rLphclass})
remains formally the same in the quantum theory, i.e.\ the physical energy
density associated to the CC is the sum of the vacuum part plus the
induced part. The only difference is that the induced part now contains
all the quantum effects, i.e.\ it reads $\rLI=\langle V_{\rm eff}^{\rm
ren}(\phi)\rangle$, {where} $V_{\rm eff}^{\rm ren}(\phi)\equiv
\EP(\phi(\mu); m(\mu), \lambda(\mu); \mu)$ is the renormalized effective
potential. {Notice that} the latter includes the (renormalized) ZPE part,
which was absent in the classical theory. Thus, the physical CC emerging
from the renormalization program (in any given subtraction scheme) reads
\begin{equation}\label{rLphquant}
\rV=\rL^{\rm ren}+\langle V_{\rm eff}^{\rm
ren}(\phi)\rangle=\rL^{\rm ren}+\ZPE^{\rm ren}+\langle V_{\rm
scal}^{\rm ren}(\phi)\rangle\,.
\end{equation}
The formal structure of this renormalized result is valid to all orders of
perturbation theory. For simplicity we have obviated the $\mu$-dependence,
which appears implicitly in all couplings and fields, and explicitly in
the structure of the terms beyond the tree-level. RG-invariance of
physical quantities, initially formulated in the bare theory, tells us
that such $\mu$-dependence must finally cancel among all terms in the
renormalized theory. Thus $d\rV/d\ln\mu=0$. This cancelation, however,
does not mean that the role played by the scale $\mu$ is necessarily
irrelevant. In fact, since the entire structure of the renormalized $\rV$
is parameterized through $\mu$, this fact can help us to unveil the form
of the quantum effects in $\rV$ and their dependence on relevant physical
scales of the problem. In the cosmological case this is only possible if
the theory is formulated in curved spacetime and when we consider the
renormalization effects on the full effective action, not just the
effective potential. We will come back to this point later on.

\subsection{The ``Mother'' of all the fine tuning problems} \label{sect:FineTuningMother}

We are now ready to formulate the extreme severity of the fine-tuning
problem, which we address here only in QFT in flat spacetime. Indeed,
being the expression (\ref{rLphquant}) the precise QFT prediction of the
physical value of the vacuum energy density to all orders of perturbation
theory, it must be equal to the observational measured
value\,\cite{PLANCK2013}, i.e.\ $\rV=\rLo\simeq 2.5\times 10^{-47}$
{GeV}$^4$. We have already seen in section \ref{sect:HiggsVacuum} that the
lowest order contribution from the Higgs potential is $55$ orders of
magnitude larger than $\rLo$, and that this enforces us to choose the
vacuum term $\rL$ with a precision of $55$ decimal places such that the
sum $\rL+\rLI$ gives a number of order $10^{-47}$ {GeV}$^4$. The problem
is that the fine-tuning game, ugly enough  already at the classical level,
becomes devastating at the quantum level. Indeed, recall that we have the
all order expansions (\ref{ZPE}) and (\ref{loopwisepot}). Therefore, the
quantity that must be equated to $\rLo$ is not simply (\ref{eq:Higgstree})
but the full \textit{r.h.s.}\ of (\ref{rLphquant}), which as we said is a
well-defined (finite and RG-invariant) expression. It means that, instead
of the ``simple'' equation (\ref{finetuningclassical}), we must now
fulfill the much more gruesome one:
\begin{eqnarray}\label{eq:finetuning}
10^{-47}\,{\rm GeV}^4=\rL-10^8\,{\rm GeV}^4&+&\hbar\,V_P^{(1)}+\hbar^2\,V_P^{(2)}+\hbar^3\,V_P^{(3)}....\nonumber\\
&+&\hbar\,V_{\rm scal}^{(1)}(\phi)+\hbar^2\,V_{\rm
scal}^{(2)}(\phi)+\hbar^3V_{\rm scal}^{(3)}(\phi)...
\end{eqnarray}

As compared to Eq.\,(\ref{finetuningclassical}), on the \textit{r.h.s.} of
the equation we now have, in addition, two independent perturbatively
renormalized series contributing to the observed value of the vacuum
energy density to one-loop, two-loops, three-loops etc... up to some
nth-loop order (both for the ZPE series and the series associated to the
quantum corrected Higgs potential). It follows that the numerical value
for the ``renormalized vacuum counterterm'' $\rL$ must be changed
accordingly order by order in perturbation theory. Specifically, the
number $\rL$ on the \textit{r.h.s.}\ of equation (\ref{eq:finetuning})
must be re-tuned with $55$ digits of precision as many times as the number
of diagrams (typically thousands) that contribute to the highest nth-loop
still providing a contribution to the CC that is at least of the order of
the experimental number placed on the \textit{l.h.s.}\ of that equation.
For example, let us roughly assume that each electroweak loop beyond the
first one (where no couplings are present for the ZPE) contributes on
average a factor $g^2/(16\pi^2)$ times the fourth power of the electroweak
scale $v\equiv\langle\phi\rangle\sim 250$ GeV (see section
\ref{sect:HiggsVacuum}), where $g$ is either the $SU(2)$ gauge coupling
constant or the Higgs self-coupling, or a combination of both. It follows
that the order, $n$, of the highest loop diagram that may contribute to
the measured value of the vacuum energy density, and that therefore could
still be subject to fine-tuning, can approximately be derived from
\begin{equation}\label{nloops}
\left(\frac{g^2}{16\,\pi^2}\right)^n\, v^4=10^{-47}\,GeV^4\,.
\end{equation}
%
%%%%%%%%%%%%%%%%%%%%%%%%%%%%%%%%%%%%%%%%%%%%%%%%%%%%%%%%%%%%%%%%%%%
\FIGURE[t]{
  %\begin{center}
    %\begin{tabular}{cc}
      \resizebox{0.8\textwidth}{!}{\includegraphics{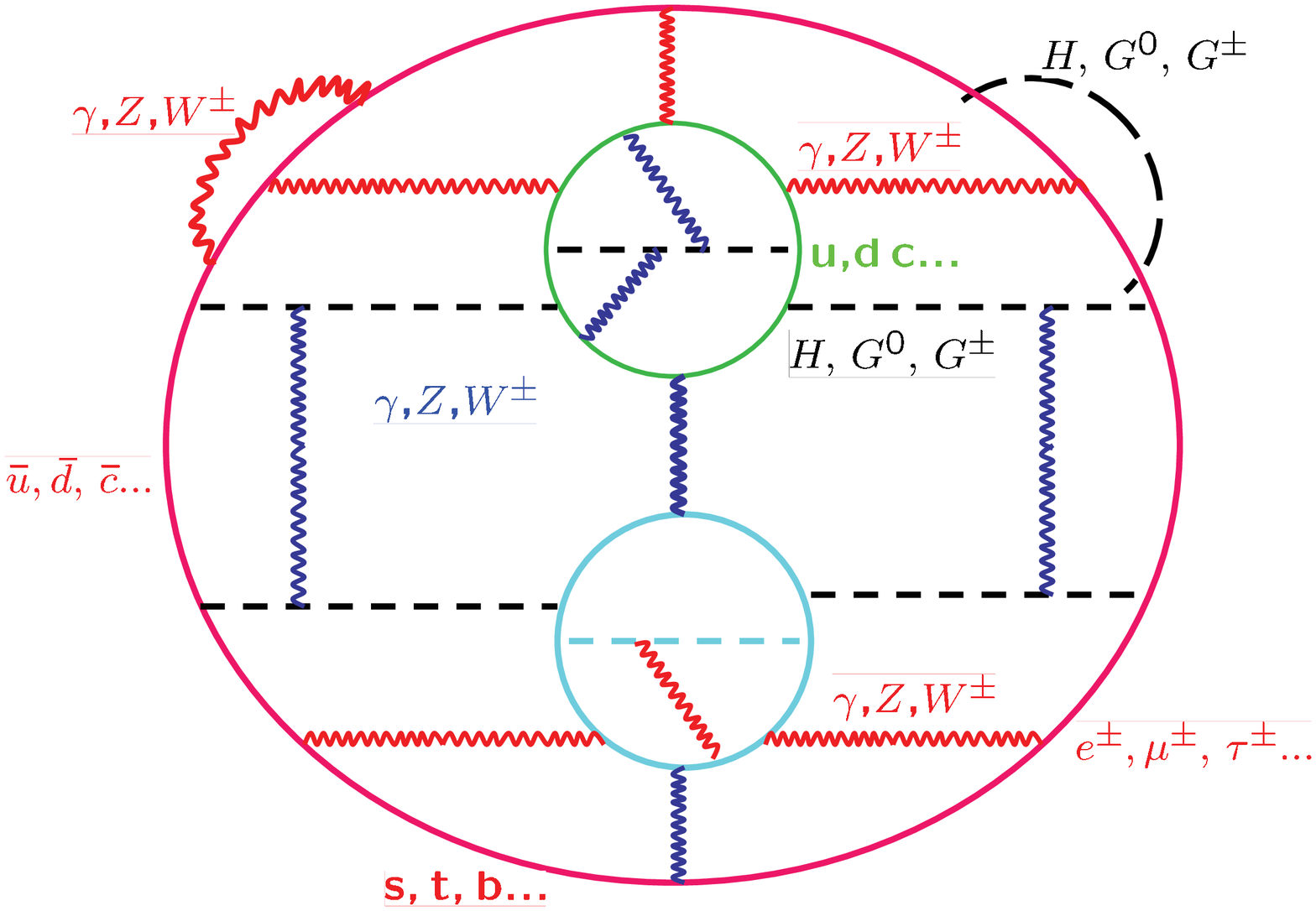}}
      %&
      %\hspace{0.3cm}
      %\resizebox{0.45\textwidth}{!}{\includegraphics{fig2.eps}} \\
      %(a) & (b)
    %\end{tabular}
    \caption{One of the many thousand 21th-loop vacuum-to-vacuum diagram still contributing to the value of the cosmological constant
    in the Standard Model of Electroweak interactions. }
%\end{center}
  \label{Fig1:blobs}}
%\end{figure}
%%%%%%%%%%%%%%%%%%%%%%%%%%%%%%%%%%%%%%%%%%%%%%%%%%%%%%%%%%%%%%%%%%%%
%
Take e.g.\ $g$ equal to the $SU(2)$ gauge coupling constant of the
electroweak SM, which satisfies $g^2/(16\pi^2)=\alpha_{\rm
em}/(4\pi\sin^2\theta_W)\simeq 2.5\times 10^{-3}$. This is actually a
conservative assumption because in practice there are larger contributions
in the SM associated to the big top quark Yukawa coupling $Y_t$. But it
will suffice to illustrate the situation. Since
$v\equiv\langle\phi\rangle\sim 250$ GeV, we find $n\simeq 21$. Therefore
all of the (many thousand) loop diagrams pertaining to the 21th
electroweak order (see Fig.\ref{Fig1:blobs} for a typical example) are
still contributing sizeably to the value of the CC, and must therefore be
readjusted by an appropriate choice of the renormalized value of the
vacuum term $\rL$. This looks preposterous and completely unacceptable. It
goes without saying that this situation worsens even more for higher
energy extensions of the SM, such as in GUT's. Actually, replacing the
cosmological term by a cosmic scalar field with some peculiar potential
only iterates the same kind of fine tuning problem, let alone that it does
not explain why e.g. the electroweak vacuum energy can be hidden under the
rug with no relation to the CC value.

Special symmetries such as Supersymmetry (SUSY)\cite{SUSYTheories}, for
example, are of little help to solve the CC problem, despite some early
hopes\,\cite{WessZumino74}, since SUSY is necessarily broken, and hence
all the above problems replicate very similarly to the SM
case\,\cite{CCPWeinberg}. Only dynamical mechanisms could really help
here, namely mechanisms capable of automatically adjusting the CC to the
present tiny value even starting from an arbitrarily large one in the
early universe. As we know, they were first attempted using scalar fields,
but these suffer from a ``no-go theorem'' \,\cite{CCPWeinberg}. However,
these mechanisms are technically possible within generalized dynamical
vacuum models with modified gravity and may lead to a new intriguing kind
of proposal called the ``Relaxed Universe'' -- see
Ref.\,\,\cite{RelaxedUniverse}. The idea goes far beyond the usual
frameworks of modified gravity\,\cite{ModifiedGravity} in which the DE is
a late time effect only. {Relaxation is a mechanism that offers indeed an
entirely different perspective} of the old CC problem; in particular, it
provides the extremely appealing possibility to solve the CC problem
dynamically, hence without the horrendous fine tuning problem mentioned
above. Whatever it be the large value of the initial vacuum energy,
$\rLi$, the value we measure is not $\rLi$ but the effective one $\rL^{\rm
eff}=\rLi+\rL(H)$. The latter remains always very small, namely $|\rL^{\rm
eff}|/\rLi\ll 1$. Here $\rL(H)$ is the dynamical compensating term, that
is to say, a function of $H$ which remains (automatically) very close to
$\rLi$ and opposite in sign to it \textit{at any epoch} of the cosmic
history, in particular in our time. The construction of $\rL(H)$ is based
on a new class of modified
gravities\,\,\cite{RelaxedUniverse,RelaxOld,RelaxFlorian}). As a result,
the effective CC that we measure at present can perfectly be of the order
of the current one, i.e. $\rL^{\rm eff}\simeq \rLo$, no matter what is the
initial value $\rLi$. In point of fact, in these relaxation models the
current value of the expansion rate, i.e. $H_0$, becomes determined by an
approximate relation of the form $H_0\sim 1/\left(\rLi\right)^n\,(n> 0)$.
Quite remarkably, this relation explains why $H_0$ is small, precisely
because the initial vacuum energy $\rLi$ was very large! Interestingly,
this kind of relaxation mechanism can also be extended to the
astrophysical domain, say to our Solar System, and one can show how to
make the (arbitrarily large) initial vacuum energy completely innocuous
for the Celestial Mechanics of planets and stars\,\cite{RelaxAstro}.

\newtext{There are other interesting ideas along the dynamical approach.}
Let me mention a particularly appealing one put forward  by Barrow and
Shaw\,\cite{Barrow2011ab}, based on extending the Einstein-Hilbert
variational principle for GR by promoting $\CC$ from being a parameter to
a field, and then only include causal variations which are on and inside
our past light cone\,\cite{BarrowPrivate}. This produces a natural
estimate of the expected value of $\CC$ observed at time $t$ to be of
order $1/t^2$ in Planck units, which is consistent with the observed
classical history of the universe. In GR, as we know, $\CC$ is a true
constant and is not seen to evolve, although the constant value it takes
depends on the cosmic time of observation according to the aforesaid
relation. Hence, in that modified variational approach the resulting
history is indistinguishable from GR with a constant value of $\CC$  put
in by hand at the right value $\sim 1/t^2$ at each observational
time\,\cite{Barrow2011c}.

Unfortunately, we still don't know which fundamental theory could
naturally accommodate any of the above options, but they surely offer an
encouraging starting point for further considerations. On the other hand,
we should also consider if the excruciating fine tuning problem just
described above can be reformulated in more amenable terms within the
standard framework of renormalization theory. We shall have occasion to
say some words on it, but above all we have to address first the situation
in curved spacetime, if only in a summarized way.

\section{ZPE in curved space-time} \label{sect:ZPE2CURVED}

So far so good, but where does the curvature of spacetime enter the
previous discussion on the ZPE in Sect.\,\ref{sect:ZPE2} or the Higgs
potential in Sect.\,\ref{sect:effHiggsFLAT}? Nowhere! However, it should
be quite natural to discuss the vacuum energy in a curved background,
chiefly if we aim at elucidating its possible connection to the value of
the cosmological term, shouldn't it? In this section, we are going to
summarize the effects of curvature for the calculation of the ZPE. The
discussion of the Higgs potential part in curved spacetime does not change
in any fundamental way the kind of problems that we will meet for the
discussion of the ZPE in a curved background, so we shall limit here to
briefly consider this last aspect of the vacuum problem and leave the
(somewhat bulky) details for a more comprehensive exposition
elsewhere\,\cite{SolaReview2013b}.

The first and most important observation is that in the presence of vacuum
energy (and hence with a nonvanishing value of the CC in Einstein's
equations) the constant Minkowski metric $\eta_{\mu\nu}=(+,-,-,-)$
obviously is not a solution of the field equations (\ref{EEnoCC2}). It
means that, with $\CC\neq 0$, the metric $g_{\mu\nu}$ solving them is a
nontrivial one and therefore the spacetime is intrinsically curved.
However, this is not what we have assumed in the calculation leading to
the result (\ref{renormZPEoneloop2}). Therefore, that calculation is, in
principle, not appropriate to reflect any dynamical aspect of the
expanding spacetime. In particular, the arbitrary mass scale $\mu$ has no
immediate connection with any physical quantity of relevant interest, in
contrast to the situation of a cross-section calculation in, say, QED or
QCD, where one tries to make an association with some characteristic
energy scale of the colliding or decaying process in order to estimate the
quantum corrections with the help of the RG.

This is certainly a relevant issue for QFT in curved
spacetime\,\cite{QFTBook1,QFTBook2,QFTBook3}, i.e. the theory of quantized
fields on a curved classical background, where the gravitational field
itself does not participate of the quantization process. In the functional
formulation this means that the generating functional of the theory (from
which we can derive the Green's functions from the functional derivatives
of it with respect to a classical source $J(x)$) can be written as a path
integral over only the quantum matter fields (collectively represented
here by $\phi$):
\begin{equation}\label{eq:ZJ}
    Z[J]=\int [D\phi]\,\exp\left\{\frac{i}{\hbar}\left[S\left[\phi,g_{\mu\nu}\right]+ \int d^4x\sqrt{-g(x)}\,J(x)\,\phi(x)\right]\right\}\,.
\end{equation}
Once more $S[\pc]$ in that formula is the classical action for the scalar
field, but adapted to the curved spacetime:
\begin{eqnarray}
S[\phi,g_{\mu\nu}]=\int
d^4x\,\sqrt{-g(x)}\,\left[\frac12\,\gmnu\,\partial_{\mu}\phi\,\partial_{\nu}\phi+\frac12\,\xi\,\phi^2\,R-V_c(\phi)\right]\,.
\label{StotalCurved}
\end{eqnarray}
A first complication is that this matter action generalizes the original
form (\ref{Sclassical}) in that we have included in it the generally
invariant integration measure $d^4 x \sqrt{-g(x)}$; and also a possible
non-minimal coupling term $\xi\,\pc^2\,R$, where $\xi$ is a dimensionless
coefficient, which is necessary for renormalizability (see below).

In principle, the problem under discussion should also be a focus of
attention for Quantum Gravity\,\cite{DeWitt07}, the theory where the
gravity field is also quantized together with the matter fields.
Unfortunately, QG is essentially non-perturbative and
nonunitary\,\cite{Smolin03Giddings11}, although some interesting
perturbative results have been discussed in the
literature\,\cite{Woodard11and12}. Despite countless efforts, QG
does not exist as a consistent theory yet. String theory seems to be
able to address some of the fundamental problems posed by QG, but it
is not obvious that it can solve all of them, not even a significant
part. In any case we do not wish to discuss them here. Our objective
right now is much more modest, and yet not trivial. We deal with the
problems of QFT in curved spacetime. In contrast to QG, the former
exists as a renormalizable theory in the perturbative sense. It can
be considered a fairly successful theory of the external
gravitational field in interaction with quantum matter. That
beautiful theory has already a long time-honored tradition,
collected in several books and
reviews\,\cite{QFTBook1,QFTBook2,QFTBook3,ReviewsQFTCurved,QFTBook4,Shapiro08}.
However, not even at present such semiclassical approach to gravity
is completely understood. In fact, it is far from being
so\,\,\cite{ShapSol09,ShapSol09b,Asorey12}. A serious hint is the
fact that the ZPE contribution to the vacuum energy density in QFT
in curved spacetime turns out to be exactly the same (at least if
interpreted in the traditional way) as the one we have obtained for
flat space, i.e. Eq.\,(\ref{renormZPEoneloop2}), up to the
renormalization of the higher order geometric operators $R^2$,
$R_{\mu\nu}R^{\mu\nu}...$ in the vacuum action, including of course
the low energy part represented by the EH term with cosmological
constant (\ref{EHb}).

\subsection{The extended vacuum action and the geometric one-loop effective action} \label{sect:ExtendedVacuumAction}

First of all, let us recall that in order that the QFT theory
becomes renormalizable in curved spacetime, the classical vacuum
action must also contain the standard higher derivative (HD) part:
\begin{equation}\label{HigherDeriv}
S_{\rm HD}\, =\, \int d^4x\sqrt{-g}\, \left\{ \alpha_1^{\rm (b)} C^2
+  \alpha_2^{\rm (b)}\,R^2 +\alpha_3^{\rm (b)} E + \alpha_4^{\rm
(b)} {\nabla^2} R \right\}\,,
\end{equation}
where, $\alpha^{\rm (b)}_{1,2,3,4}$ are bare parameters,
$C^2=R_{\mu\nu\rho\sigma}R^{\mu\nu\rho\sigma}-2R_{\mu\nu}R^{\mu\nu}+(1/3)\,R^2$
is the square of the Weyl tensor and
$E=R_{\mu\nu\rho\sigma}R^{\mu\nu\rho\sigma}-4R_{\mu\nu}R^{\mu\nu}+R^2$ is
the topological Euler's density.  The coefficients $\alpha_i$ of these
terms are then renormalized by the quantum effects, which have the same
structure, and in this way we can absorb the new infinities. The theory is
thus one-loop renormalizable in curved spacetime. The HD terms are short
distance effects which have no impact at low energies, i.e. they do not
lead to significant corrections to the Einstein equations (\ref{EEnoCC2}),
valid at long distances (the situation of the present, low-energy,
universe). In the FLRW metric all these terms are of order $R^2\sim H^4$
(including $\Box R\sim H^4$), which are negligible for the entire
post-inflationary history of the universe. The terms we are interested in
are those which are at least of order $R\sim H^2$.

To achieve a finite renormalized theory we proceed as follows. As usual we
split each bare coefficient into a renormalized term plus a counterterm:
\begin{equation}\label{eq:split1}
\alpha_1^{\rm (b)}=\alpha_1(\mu)+\delta\alpha_1\,,\ \ \ \ \ \ \ \ \
\ \alpha_2^{\rm (b)}=\alpha_2(\mu)+\delta\alpha_2\
\end{equation}
and similarly,
\begin{equation}\label{eq:split2}
\frac{1}{\Gb}=\frac{1}{G(\mu)}+\delta\left(\frac{1}{G}\right)\,,\ \
\ \ \ \ \ \ \ \ \rLb=\rL(\mu)+\delta\rL\,,
\end{equation}
We disregard here the details on the topological term and the total
derivative part of the HD action. In doing these splittings the
renormalized quantities are $\mu$-dependent because they are supposed to
be defined at a given renormalization point $\mu$ in RG-space, where they
should hypothetically make contact with some experimental input. The point
$\mu$ is arbitrary, and in fact the wisdom of the RG is that the physics
should not depend on its election. However, $\mu$ can have a more or less
physical meaning depending on the renormalization scheme that is used.
Finally, the counterterms can be chosen to cancel the UV divergences.
These have been dimensionally regularized in our approach, \newtext{i.e.}
we have to understand once more that $d^4x$ is replaced by $d^nx$ in the
action, then we perform the computation in $n$ dimensions and the UV
divergences appear as poles at $n=4$, as we did before in the flat
spacetime case.
\newtext{In this way} the structure of the above counterterms
can be easily identified. We simply require that $\delta\alpha_{1,2}$, as
well as $\delta(1/G)$ and $\delta\rL$, are chosen to cancel those poles at
$n=4$ so as to make finite the values of the final one-loop parameters.

The use of dimensional regularization,\cite{Hooft73,BolliniGiambiagi72} is quite standard and
relatively simple, but on the other hand it is a source of ambiguity
when we face the physical interpretation of the results. We know about these problems
since long ago in the usual practice in e.g. Particle Physics and the renormalization of the
SM in various subtraction schemes\,\cite{Hollik95}. The
ambiguity reaches a climax when we next use minimal subtraction to
renormalize the theory. Again, minimal subtraction (whether in the
$\overline{\rm MS}$ form or pole subtraction followed with whatever
finite part we like) is simple enough, but the renormalized
quantities (which become functions of the arbitrary mass scale
$\mu$) suffer from a lack of direct physical interpretation. Let us
note that we can do all this without still mentioning what is the
physical system we have behind, we only know that we are
renormalizing the classical vacuum action. The system can be the
universe or some other gravitational framework in which we may be
interested to study the quantum matter effects.

It is obvious that at some point we have to put our system in context, and
then assess what could be the physical quantity with which we could dream
establishing a possible association with $\mu$. For example, $\mu$ in the
expression (\ref{renormZPEoneloop2}) for the vacuum energy density in flat
spacetime has no physical meaning at all. How could it possibly have, if
the spacetime is just Minkowskian and there is no exchange of information
between the matter loops and the constant metric? In that case $\mu$ is
just a formal parameter. To let it gain some physical meaning (not by
itself, of course, but as a device helping to parameterize the vacuum
energy in terms of physical quantities) we need some nontrivial spacetime
dynamics, therefore a curved background.

A contextual interpretation of $\mu$ in MS-like subtraction schemes would be unnecessary if $\mu$ had been
a physical subtraction point from the very beginning (say some momentum subtraction point
within a more physical effective field theory
approach\,\cite{Manohar96,Pich97,Gorbar03}), although this has a
calculational price. But if we want to still stay with the
mathematical simplicity of dimensional regularization with ${\rm
MS}$-like subtraction, we are forced to appeal to a contextual
physical interpretation in the last stage of the calculation.
Unfortunately, not even solving this problem should mean the end of
our troubles; there are still some other headaches in the list,
maybe the most severe ones.

Let us however move on and see. Following the aforementioned procedure,
the total effective action (classical plus one-loop corrections) can be
conveniently organized as follows:
\begin{eqnarray}\label{eq:classplusquant}
\Gamma= S[\pc]+ S_{\rm HD}&+& S_{\rm EH}+\Gamma^{(1)}_{\rm eff}
=S[\pc]+S_{\rm HD}^{(1)}+ S^{(1)}_{\rm EH}\,,
\end{eqnarray}
where\,\footnote{These are the quantum effects from the matter field
$\phi$ encoded in the explicit computation of the one-loop part of the
effective action, Eq.\,(\ref{EAoneloopFlat}). In the curved space-time
case the replacement $d^4x\to\sqrt{-g}\,d^4x$ must be done in that
formula, and the inverse propagator $\K(x,x')$ now is also more
complicated than in flat spacetime. This is of course the hardest part of
the calculation, that I am fully sparing to the reader here -- see e.g.
\cite{QFTBook1,QFTBook2}; confer also \cite{SolaReview2013b} for the
details along the present lines.}
\begin{eqnarray}\label{eq:Gammar1comp}
\Gamma^{(1)}_{\rm eff}=\frac{\hbar}{2(4\pi)^2}\int {\rm d}^4 x \sqrt{-g}
&&\left(\frac{2}{4-n}+\ln\frac{4\pi\,\mu^2}{m^2}-\gamma_E\right)\,\nonumber\\
&&\times\left(\frac12\,m^4-m^2\,a_1(x)+a_2(x)+\cdots \right)\,.\nonumber\\
\end{eqnarray}
Here $a_{1,2}$ are the so-called Schwinger-DeWitt coefficients
(coming from the adiabatic expansion of the matter field propagator
in the curved background). They have a pure geometric form:
\begin{eqnarray}
&&a_1(x)= -\left(\frac16-\xi\right)\,R\,, \nonumber\\
&&a_2(x)=\frac12\,\left(\frac16-\xi\right)^2\,R^2+
\frac16\left(\frac15-\xi\right)\Box R+\frac{1}{180}\left(R^{\alpha\beta\gamma\delta}R_{\alpha\beta\gamma\delta}-R^{\alpha\beta}R_{\alpha\beta}\right)\nonumber\,.\nonumber\\ \label{eq:SDWcoeff2}
\end{eqnarray}
Furthermore, in the second step of Eq.\,(\ref{eq:classplusquant}) we have
reorganized the complicated expressions entering (\ref{eq:Gammar1comp}) in
a suitable way. Part of the result can be absorbed in the higher
derivative vacuum part
\begin{eqnarray}\label{eq:SHSoneloop}
S^{(1)}_{\rm HD}=
\int d^4x\sqrt{-g}\, \left( \alpha_1^{\rm (1)} C^2 +  \alpha_2^{\rm (2)}\,R^2 +...\right)\nonumber\\
\end{eqnarray}
and the remaining is absorbed in the EH action:
\begin{eqnarray}\label{EHextended}
S_{\rm EH}^{(1)} =-\int d^4x \sqrt{-g}\left(\frac{1}{16\pi
\Gbu}R+\rVu\right)\,.
\end{eqnarray}
To be more precise, in the above equations we have absorbed the bare terms and the quantum matter effects  in the one-loop
coefficients $\alpha_{1}^{(1)}$ and $\alpha_{2}^{(1)}$ of the HD action and the one-loop Newton's constant and CC term,
$\Gbu$ and $\rVu$, of the low energy EH action. After some calculations, the final result can be cast
as\,\cite{SolaReview2013b}
\begin{eqnarray}
\alpha_{1}^{(1)}&=&\alpha_1(\mu)-\frac{\hbar}{2(4\pi)^2}\,\frac{1}{120}\,\left(\ln\frac{m^2}{\mu^2}+{\rm finite\ const.}\right)\label{eq:Renormalpha1}\\
\alpha_{2}^{(1)}&=&\alpha_2(\mu)-\frac{\hbar}{4(4\pi)^2}\,\left(\frac16-\xi\right)^2\,\,\left(\ln\frac{m^2}{\mu^2}+{\rm finite\ const.}\right)\label{eq:Renormalpha2}
\end{eqnarray}
and
\begin{eqnarray}
\frac{1}{16\pi\Gbu}&=&\frac{1}{16\pi G(\mu)}+\frac{\hbar\,m^2}{2(4\pi)^2}\left(\frac16-\xi\right)\,\left(\ln\frac{m^2}{\mu^2}+{\rm finite\ const.}\right)\label{eq:RenormG}\\
\rVu&=&\rL(\mu)+\frac{m^4\,\hbar}{4\,(4\,\pi)^2}\,\left(\ln\frac{m^2}{\mu^2}+{\rm
finite\ const.}\right)\label{eq:RenormVacuum}\,.
\end{eqnarray}

The parameters  $P_i^{(1)}=\alpha_{1}^{(1)},\ \alpha_{2}^{(1)},\
{1}/G^{(1)},\ \rVu$  on the \textit{l.h.s.} of these expressions are the
final ``one-loop parameters''; they are finite at one loop because we have
used the counterterms to cancel the divergences coming from the one-loop
contributions to them. The very fact that such redefinition of the
original coefficients of (\ref{HigherDeriv}) and (\ref{EHb}) can be made
shows the practical possibility to renormalize the theory. It does not
mean, however, that we can get an immediate physical interpretation of the
renormalized expressions. The arbitrariness of the finite constant terms
in the expressions (\ref{eq:RenormVacuum}) is only a small hint of the
ambiguity and lack of direct interpretation of these formulas.

If we would continue the calculation at $2$-loops, the parameters
$P_i^{(1)}$ would play the role of bare parameters, in a similar way
as the bare parameters (\ref{eq:split1})-(\ref{eq:split2}) of the
classical action, and therefore they would be UV-divergent
quantities that split once more into a renormalized parameter and a
corresponding counterterm. These counterterms would then be used to
cancel the $2$-loop divergences etc. As parameters of the bare
action, all of the $P_i^{(1)}$ are of course independent of the
arbitrary renormalization point $\mu$. The explicit $\mu$-dependence
of the various terms in each $P_i^{(1)}$ must cancel in the overall
expression. One source of $\mu$-dependence comes from the
corresponding ``renormalized parameters'', $P_i(\mu)$, which
therefore ``run'' with the scale $\mu$. However, as already warned
before, at this point such ``running'' has no obvious relation with
the variation of any physical quantity. On the other hand, whether
such connection is possible cannot be unambiguously decided in the
absence of a concrete physical framework. In the next section,
however, we place these formulae in the cosmological context and
only then some chance exists for a possible physical interpretation.

\subsection{Renormalization group equations} \label{sect:RGequations}

Mathematically, the $\mu$-running is determined by the corresponding
renormalization group equation (RGE) for each of the parameters, and
follows from setting the total derivatives of the one-loop parameters
$P_i^{(1)}$ with respect to $\mu$ to zero. For convenience we compute the
logarithmic derivatives of each one of them, i.e. $
{dP_i^{(1)}}/{d\ln\mu}=0$. The explicit RGE's are immediately obtained
from  equations (\ref{eq:Renormalpha1})-(\ref{eq:RenormVacuum}):
\begin{eqnarray}
\frac{d\alpha_1(\mu)}{d\ln\mu}&=&-\frac{\hbar}{120(4\pi)^2}\equiv\beta_1^{(1)}\label{eq:RGElpha1}\\
\frac{d\alpha_2(\mu)}{d\ln\mu}&=&-\frac{\hbar}{2(4\pi)^2}\,\left(\frac16-\xi\right)^2\equiv\beta_2^{(1)}\label{eq:RGEalpha2}
\end{eqnarray}
and
\begin{eqnarray}
\frac{d}{d\ln\mu}\left(\frac{1}{16\pi G(\mu)}\right)&=&\frac{\hbar\,m^2}{(4\pi)^2}\left(\frac16-\xi\right)\equiv\beta_{G^{-1}}^{(1)}\label{eq:RGEG}\\
\frac{d\rL(\mu)}{d\ln\mu}&=&\frac{\hbar\,
m^4}{2\,(4\pi)^2}\equiv\beta_\Lambda^{(1)}\label{eq:RGEVacuum}\,.
\end{eqnarray}
Notice that the obtained RGE's do not depend at all on the
unspecified finite terms indicated in equations
(\ref{eq:Renormalpha1})-(\ref{eq:RenormVacuum}).  The
\textit{r.h.s}'s of the above expressions define the corresponding
one-loop $\beta$-functions that control the running of these
parameters in the renormalization scheme we have used, an ${\rm
MS}$-based one. Note, in particular, the RGE for $\rL$,
Eq.\,(\ref{eq:RGEVacuum}). It coincides exactly with Eq.\,
(\ref{betaFunct}), the flat spacetime result. Integration of
(\ref{eq:RGEVacuum}) immediately furnishes the one-loop finite
result
\begin{equation}\label{renormZPEoneloopCurved}
\rVu=\rL(\mu)+\frac{m^4\,\hbar}{4\,(4\,\pi)^2}\,\left(\ln\frac{m^2}{\mu^2}+{\rm finite\ const.}\ \right)\,.
\end{equation}
The finite additive constant in this formula is not very important, as it
depends on the kind of ${\rm MS}$-based subtraction scheme we use. For
exact $\overline{\rm MS}$, the constant must of course be the same one as
in Eq.\,(\ref{renormZPEoneloop2}), but this is quite irrelevant as it does
not change at all the worrisome aspect of the result, namely the fact that
we get a quantum ``correction'' to the vacuum energy density growing as
$\sim m^4$, for a particle of mass $m$. This kind of result is completely general,
the one-loop $\beta_\Lambda$-function for a particle of spin $J$ and mass $M_J$ is found to be\,\cite{ShapSol00}
\begin{equation}\label{eq:BetaFunctionSpinJ}
\beta_\Lambda^{(1)}=(-1)^{2J}\,\left(J+1/2\right)\,n_c\,n_J\,\frac{\hbar\,M_J^4}{(4\pi)^2}
\end{equation}
with $(n_c, n_{1/2}) = (3, 2)$ for quarks, $(1, 2)$ for leptons and $(n_c, n_{0,1}) = (1, 1)$ for scalar and vector
fields. The corresponding energy density is proportional to $M_J^4$.

So we are back to the numerological disaster first struck by Zeldovich
(cf. Sect.\ref{sect:ZPEold}), which forced him to change his strategy to
estimate the vacuum energy in particle physics. Somehow we have not
advanced a single step since then, despite much QFT in curved spacetime!
This is the result we mentioned before, and it comes as a kind of an
``unexpected'' surprise. It certainly does not help shedding any light on
our poor understanding of the vacuum energy density in QFT in curved
spacetime\,\cite{QFTBook1,QFTBook2} no matter how we arrange our
renormalization conditions around flat space $g_{\mu\nu}=\eta_{\mu\nu}$,
despite the practical efficiency of some renormalization
techniques\,\cite{MarkkanenTranberg2013}. The underlying problem, however,
does remain. We will come back to this central issue in the next section.

In view of the situation, it would be unwise to rush into conclusions at
this point.  For example, it would make little sense to evaluate the found
formula for the vacuum energy density by using some value of $\mu$ and,
say, the full collection of particle masses of the SM of strong and
electroweak interactions, or whatever extension of it. First, because on
the face of the obtained result we feel that we are somehow in deep water
and we don't know what is the next surprise we'll come across in this
story; and, second, because it is obvious that the contribution from a
single mass, being proportional to the quartic power of it, is completely
out of range (cf. our ``numerology discussion'' in
Sect.\,\ref{sect:ZPEold}). The only possible exception is perhaps a very
light neutrino mass, see Eq.\,(\ref{eq:neutrino}), or the existence of new
degrees of freedom at a similar mass scale that would determine the
behavior of the cosmological term \,\cite{ShapSol99}.

We already mentioned quantum gravity as a dreamed theory in this
field, and string theory as a promising approach to it. But
unfortunately neither QG nor string theory have been able to solve
the vacuum energy problem either. As a matter of fact no theoretical
framework at present is capable to provide a fully consistent and
realistic account for the vacuum energy density in the cosmological
context. Although some of the technical problems can be glimpsed in
the more modest curved QFT arena\,\cite{ShapSol09,ShapSol09b}, the
situation is far from being completely (not even substantially)
understood. We feel that if the problem cannot be minimally dealt
with at the level of unquantized gravity, the role of QG may not be
such decisive to settle this issue, especially if we limit the scope
of the CC problem to the scale of the SM of strong and electroweak
interactions (which is extremely far away from the Planck scale).

We should judiciously expect that QFT in curved spacetime is competent
enough in this energy domain so as to provide a first reliable hint of the
possible solution, i.e. something that goes beyond the ``cul de sac''
situation we have now. \newtext{One could argue} that such solution
already exists in QFT\,\cite{Rovelli10} in the sense that the $\CC$-term,
together with the higher derivative part of the vacuum action, is all we
need to renormalize the complete action (\ref{eq:classplusquant}) and
``solve'' the old CC-problem upon standard renormalization procedures.
Something of it is probably true, and is actually one of the leitmotifs of
this work, as we shall see in the next sections. However, the scientific
community did make a big hype about the old CC
problem\,\cite{CCPWeinberg,CCproblem2,PeeblesRatra03}, see also
\cite{CCPWitten}, and we should not go too fast and jump to conclusions.
So, even if the value of the vacuum energy can be fixed by renormalization
at a particular time or energy (thus dispatching the old CC problem in a
rather unceremonious manner, actually in the very same manner we would
perform the computation of the electron's mass by ``renormalizing away''
the huge radiative corrections and adjusting the mass counterterm
appropriately so as to reproduce the measured value of the electron's
mass), there is still a huge enterprise ahead us after we fix the value of
the vacuum energy density at some point of the cosmic history; namely, we
should be able to understand its dynamical variation through the cosmic
expansion.

This dynamics is something expected within  QFT in curved
spacetime\,\cite{SolaReview2011,ShapSol09,ShapSol09b,ShapSol00,Fossil07},
but the whole issue remains unsolved yet, neither in the positive nor in
the negative sense. It is one of the hot points of this work to try to
convince the reader that the answer is positive, even if a rigorous proof
is not available yet. In the next sections I will try to elaborate on it
in the most simple and pedagogical way I can think of. I will try to show
by semi-heuristic arguments that a cosmic time evolution of the vacuum
energy density is to be considered as the most natural prediction within
QFT in curved spacetime. In this sense, the so-called dark energy (DE)
should be nothing more, but nothing less, than vacuum energy in cosmic
expansion.

\section{Dynamical vacuum energy in an expanding universe}
\label{sect:DynamicalVacEner}

Let us focus here on the vacuum part of the effective cosmological term,
i.e. we are now mainly concerned with the ZPE in curved spacetime. With
the traditional approach the metric expansion is usually made around flat
space, i.e. $g_{\mu\nu}=\eta_{\mu\nu}+h_{\mu\nu}$. If one traces the
details of the calculation\,\cite{SolaReview2013b}, one can easily see
that this is the reason why the result (\ref{renormZPEoneloopCurved}) is
the same as in flat space, Eq.\,(\ref{renormZPEoneloop}). In fact, the
result is entirely dependent on the free propagator in flat spacetime,
which appears as the first term of the adiabatic expansion of the Green's
function in the curved case. This can be troublesome since we are dealing
with the renormalization of the cosmological term, namely a term  which
does \textit{not} exist in Minkowski space. Therefore, in those situations
when the spacetime background is unavoidably curved, sticking to the
expansion $g_{\mu\nu}=\eta_{\mu\nu}+h_{\mu\nu}$ could miss the very
dynamical correction to the vacuum energy density that we are looking for.
In other words, we cannot rely on a perturbative expansion around
a background field configuration which is not a solution of the equations
of motion!

Although the root of the problem has been
identified\,\,\cite{ShapSol09,ShapSol09b}, the remedy for it is not
available yet. One needs to compute not only the divergences, but also the
first finite correction using a physical renormalization scheme (beyond
$\overline{\rm MS}$ or the like); and, above all, one has to learn how to
cope with the technical limitations of the current QFT methods for
computing the renormalization corrections around a non-trivial background.
For de Sitter background, for example, one expects $\rV\sim H^4$ by pure
dimensional analysis, and indeed particular calculations reach that kind
of conclusion\,\cite{Woodard08,Serreau2011,AliKaya13}. But no realistic
attempt has ever been made for a FLRW background, where we should expect
both $M_i^2H^2$ and $H^4$ terms, where $M_i$ are the particle masses. Such
calculation should be a rather nontrivial one, perhaps unreachable by the
presently known methods. However we expect that it should lead to a RGE of
the form\,\footnote{Notice that integration of Eq.\,(\ref{seriesRLH})
immediately leads to a vacuum effective Lagrangian of the form
(\ref{eq:HLagrangian}) discussed in Sect.\,\ref{sect:oddH} from an analogy
with QED and QCD in which only even powers of the electromagnetic fields
could be involved.}:
\begin{eqnarray}\label{seriesRLH}
\frac{d\rL}{d\ln
H^2}=\frac{1}{(4\pi)^2}\sum_{i}\left[\,a_{i}M_{i}^{2}\,H^{2}
+\,b_{i}\,H^{4}+{\cal O}\left(\frac{H^{6}}{M_{i}^{2}}\right)\right]\,,
\end{eqnarray}
in which $\mu=H$ should sensibly act as the natural running scale in the
cosmological context. Obviously the ${\cal O}({H^{6}}/{M_{i}^{2}})$ terms
represent the decoupling contributions that should appear in a physical
renormalization scheme. Notice that the scheme used in Sect.
\ref{sect:ZPE2CURVED} was off-shell and it could not trace the decoupling
effects. The above equation describes the ``change in the value of the
vacuum energy density'' (triggered by the quantum matter effects)
associated to a ``change in the curvature of the spacetime'', this
curvature being of order $R\sim H^2$ in the FLRW metric. It is well known
that renormalization theory can only be predictive if we first input the
value of the relevant parameters at a given renormalization point. After
that we can predict the value at another point. For instance, the e.m.
fine structure constant at the scale of the electron mass, $m_e\simeq 0.5$
MeV $=5\times 10^{-4}$ GeV, is $\alpha(m_e)\simeq 1/137$, whereas its
value at the scale of the $Z$-boson mass, $M_Z\simeq 90$ GeV, is  $\sim
7\%$ larger: $\alpha(M_Z)\simeq 1/128$. The RG approach cannot pretend to
``compute'' $\alpha(m_e)$ or $\alpha(M_Z)$, only the change from one value
to the other when moving from $\mu=m_e$ to $\mu=M_Z$, where here we make a
direct association of $\mu$ with the physical scale of the particle
masses. Similarly, renormalization theory cannot aim at computing ``the
value'' itself of the vacuum energy density and the CC, but only the
evolution or running of this value after we have measured it at some
cosmic energy scale. Such RG approach to
cosmology\,\cite{Nelson1982,ParkerToms1984,Buchbinder1986} should suffice
for an effective study of the running vacuum energy density within QFT in
curved spacetime, quite different from the more ambitious attempts at
predicting its current value (the old CC problem\,\cite{CCPWeinberg}).
These attempts are probably doomed to fail until we can first cope with
the prediction, or a more fundamental understanding, of the fundamental
parameters of the SM, if that is possible at all.

In the following we consider some aspects of the RG-running
framework to cosmology from the viewpoint of QFT in curved
space-time by employing the standard perturbative RG-techniques of
Particle
Physics\cite{ShapSol09,ShapSol09b,ShapSol00,Fossil07,ShapSol03,ShapSol03b,SS05,BabicET02,Guberina2003},
see e.g. \cite{SolaReview2011,SolaReview05} for a short review. The
ensuing RG-based dynamical vacuum energy models emphasize the
evolution of the vacuum energy as a particularly well motivated
function of the Hubble rate, i.e. $\rL=\rL(H)$. In the next section
we shall elaborate on motivating its structure.

\subsection{Running gravitational coupling and vacuum energy}
\label{sect:VacEnerCurvature}

While it is not possible to derive the general RGE for $\rL$ in QFT in
curved spacetime of the form given in (\ref{seriesRLH}) yet, we may be
able to hint at the expected dynamical effects induced by the quantum
corrections in some indirect way.

If we look at the vacuum diagrams of Fig.\ref{Fig1:blobsHair} we notice
the following. The simplest diagram there is the classic ``blob'' with
just a closed loop line of the scalar field. This one is already present
in the flat spacetime case. However, the presence of the external
gravitational field introduces ``hair'' (i.e. $h_{\mu\nu}$-tails of the
classical field departing from the flat space structure). This modifies
the originally ``bald'' quantum vacuum blob, and for this reason we have
many other blob diagrams in Fig.\ref{Fig1:blobsHair}, which we call the
``haired'' ones. The infinitely many tails are induced on all possible
diagrams that are preexisting in flat space, and are generated by the
non-polynomial expansion of the factor $\sqrt{-g}=1+\frac12\,h+{\cal
O}\left(h^2\right)+{\cal O}\left(h_{\mu\nu}\,h^{\mu\nu}\right)+...$ in the
action (\ref{StotalCurved}) of the scalar field, which we take as a free
field here, i.e. $V(\phi)=(1/2)\,m^2\,\phi^2$. For the particular case of
the vacuum diagrams under consideration, the expansion of $\sqrt{-g}$ is
performed in the one-loop correction term of the effective action in flat
space, Eq.\,(\ref{EAoneloopFlat}), after we replace $d^4x\to
\sqrt{-g}\,\,d^4x$ in it so as to account for the curvature effects. The
previously computed renormalization effects from the matter field $\phi$
on the parameters of the purely geometric vacuum action $S_{EH}+S_{HD}$
(conf. Sect.\,\ref{sect:ZPE2CURVED}) can now be viewed diagrammatically in
Fig.\ref{Fig1:blobsHair}, which leads more intuitively to a possible
physical interpretation after we  put our external gravity system in
context -- the FLRW cosmological one in this case.

%%%%%%%%%%%%%%%%%%%%%%%%%%%%%%%%%%%%%%%%%%%%%%%%%%%%%%%%%%%%%%%%%%
%\begin{figure}[]
\FIGURE[t]{
  %\begin{center}
    %\begin{tabular}{cc}
      \resizebox{0.7\textwidth}{!}{\includegraphics{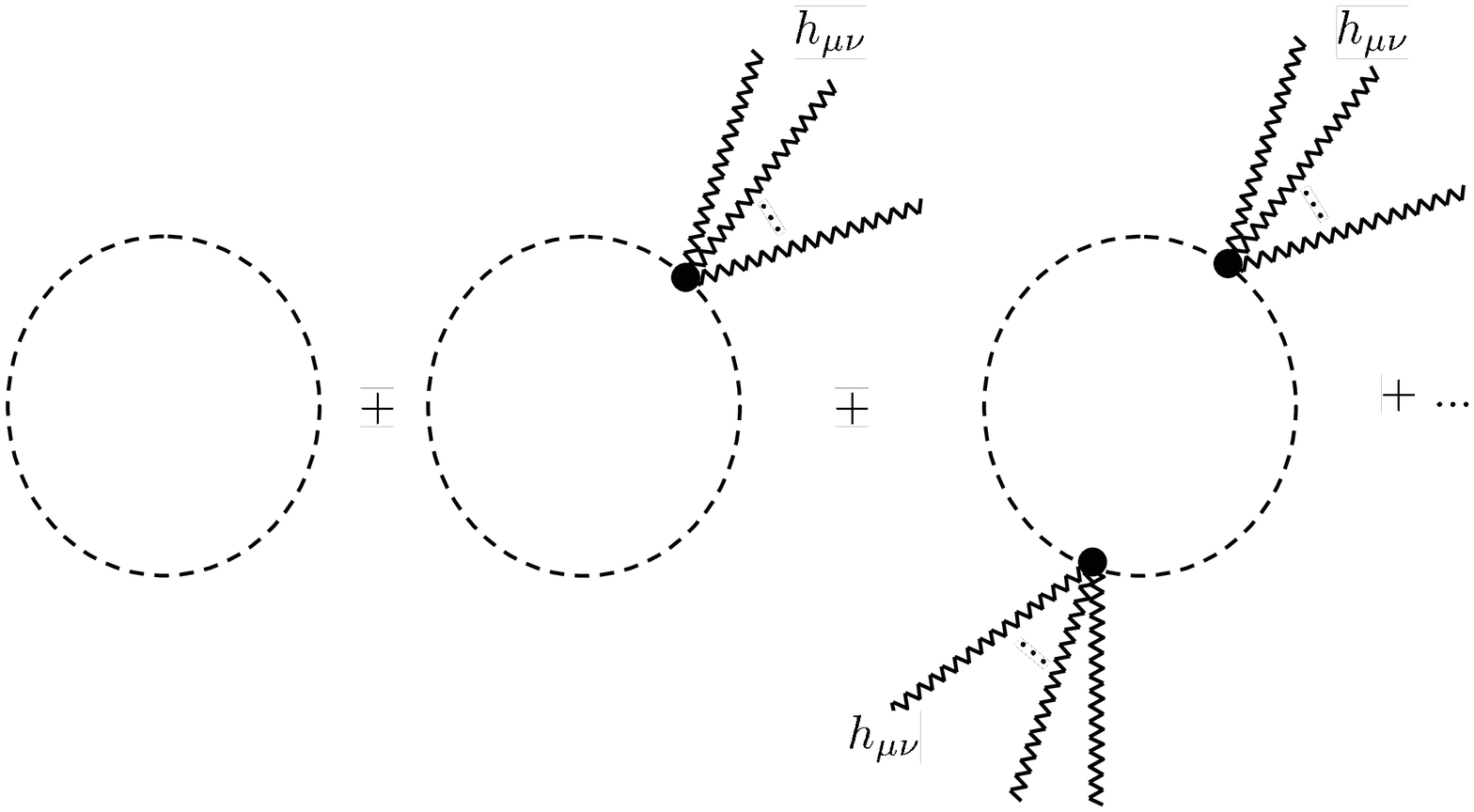}}
      %&
      %\hspace{0.3cm}
      %\resizebox{0.45\textwidth}{!}{\includegraphics{fig2.eps}} \\
      %(a) & (b)
    %\end{tabular}
    \caption{The one-loop vacuum-to-vacuum diagrams of the scalar matter field in the presence of an external gravitational field.
    The first type of contribution is the ``bald blob''
    with no external tails, which is quartically divergent.  The other contribution are the  ``haired blobs'' with one or more external
    field insertion, where an arbitrary number of $h_{\mu\nu}$-tails of the background field  are attached to one or more points.
    They appear from the expansion of the
    metric in the form $g_{\mu\nu}=\eta_{\mu\nu}+h_{\mu\nu}$, and the corresponding determinant in the action:
    $\sqrt{-g}=1+\frac12\,h+{\cal O}\left(h^2\right)+{\cal O}\left(h_{\mu\nu}\,h^{\mu\nu}\right)+...$
    (with $h\equiv g_{\mu\nu}h^{\mu\nu})$.
    The first haired vacuum diagram with one insertion is quadratically divergent, as there is one propagator of the scalar field.
    The bunch of tails organize themselves in a covariant
    way to generate an action term of the form $\sqrt{-g}\,m^2\,R$, and hence renormalize the inverse
    gravitational Newton's coupling $1/G$ in the EH action (\ref{EHextended}). The third type of diagram
    is the ``doubly haired blob'' and contains two insertion points.
    With two propagator lines, it is only logarithmically
    divergent; it renormalizes the coefficients $\alpha_{1,2}$ of the HD-action (\ref{eq:SHSoneloop}).
    The diagrams with three or more insertions
    of the external field are finite. In the flat spacetime case only the ``bald blob'' contributes,
    so the dynamics of the vacuum energy in an expanding
    universe must come from the ``haired blobs'', which may pump in energy from the expanding background into the vacuum matter loops,
    thereby informing them that there is a non-trivial geometry around them.}
%\end{center}
  \label{Fig1:blobsHair}}
%\end{figure}
%%%%%%%%%%%%%%%%%%%%%%%%%%%%%%%%%%%%%%%%%%%%%%%%%%%%%%%%%%%%%%%%%%%%

The situation is similar to the Euler-Heisenberg-Weisskopf effective
Lagrangian of QED at low energies, discussed in Sect.\,\ref{sect:oddH} --
see Eq.\,(\ref{eq:EHLagrangian}). We can also interpret this Lagrangian
diagrammatically. At one-loop we can imagine a series of blobs with
``hair'' (in this case represented by the tails of the external
electromagnetic field). The number of tails is arbitrary, but must be even
(owing to Furry's Theorem\,\cite{QFTBooks}, which is well-known to be a
consequence of charge conjugation symmetry of the electromagnetic
interaction). There is however an important difference, namely at each
point we attach a single tail of the external electromagnetic field, as
there is no equivalence here for the non-polynomial expansion of the
metric determinant. What we have now is a low energy theory of soft
photons interacting -- in any (even) number -- through loops of ``heavy''
electrons, i.e. an effective theory of light-by-light scattering. Mind
also that in contrast to the gravitational case, all QED loops have
external tails, starting form the loop having $2$ tails, then another that
has $4$, another with $6$, etc.

But, in contradistinction  to the effective QED theory of photons, in the
gravitational case the first loop in Fig.\ref{Fig1:blobsHair} can indeed
have zero tails (``bald'' blob). Such blob does not interact with the
curved background at all, as it is only through the tails that there can
be energy-momentum transfer from the background field into the loops,
thereby informing the quantum matter fields that there is a non-trivial
background around them.  It follows that any result that depends solely on
that ``bald'' diagram contains the same information as in flat space. This
is what happens with the unsuccessful Eq.\,(\ref{renormZPEoneloopCurved}).
That equation is {not} wrong, but it is \textit{not} the vacuum energy
density. As a matter of fact, it has no direct physical meaning at all. We
will see in Sect.\,\ref{sect:Casimir} what is its eventual fate.

In contrast, the `haired blobs'' do interact with the background geometry.
Let us concentrate on the first ``haired blob''. It contains one insertion
point where infinitely many tails of the external gravitational field
$h_{\mu\nu}$ can be attached. That blob, as all the other `haired'' ones,
is pumping external energy-momentum from the FLRW background. The typical
magnitude of the momentum should be of the order of $H$, which has
dimension of energy in natural units. In this sense the association
$\mu=H$ looks quite reasonable and physically intuitive, but only after we
have placed the system in the cosmological context. This is similar to the
correspondence of $\mu$ with a momentum variable $q$ in particle physics
processes, although admittedly in the cosmological case the interpretation
is less straightforward, specially in the low energy domain.

Fig.\ref{Fig1:blobsHair} shows indeed that the Feynman diagrams
potentially contributing to the renormalization of the vacuum energy
density involve an infinite series of closed loops with arbitrary amount
of ``hair'', i.e. with an unrestricted number of external legs of the
classical gravitational field associated to the FLRW metric, with a
corresponding infinite number of Green's function insertions of the matter
fields. The subtle resummation properties of this series determine the
physical renormalization properties of the vacuum energy. A frontal
approach to accomplish the task of this resummation in a physical
renormalization scheme is not possible at present, if the metric expansion
is not performed with respect to the flat spacetime background
($g_{\mu\nu}=\eta_{\mu\nu}+h_{\mu\nu}$). Such expansion is relatively
simple, but fully inappropriate in the presence of a CC term, as it does
not satisfy the field equations -- thus missing the entire income from the
relevant contributions! In the meanwhile we can hint at the possible
generic structure of the final result, Eq.\,(\ref{seriesRLH}), from the RG
properties of particular terms in the effective action, such as the EH
term, in combination with the Bianchi identity (\ref{BianchiGeneral}),
serving the latter as a consistency requirement to preserve the general
covariance of the theory.

The bunch of $h_{\mu\nu}$-tails in the first ``haired'' diagram organize
themselves in a covariant way to generate an action term of the form
$\sqrt{-g}\,m^2\,R$, and hence renormalize the inverse gravitational
Newton's coupling $1/G$ in the EH action (\ref{EHextended}). The precise
RGE for this coupling was given before in Eq.\,(\ref{eq:RGEG}). At this
point this does not yet lead to a renormalization of the vacuum energy,
but it can be related to it in an indirect way, as we shall see. Whereas
in the absence of a physical context the $\mu$-dependence would not be of
much help, if we now follow the aforementioned ansatz $\mu=H$ within the
cosmological context, we can immediately integrate (\ref{eq:RGEG}) to find
(in natural units):
\begin{equation}\label{eq:IntegrationRGforG1}
\frac{1}{G(H)}=\frac{1}{G_0}+\frac{m^2}{2\pi}\left(\frac16-\xi\right)\,\ln\frac{H^2}{H_0^2}\,,
\end{equation}
where $G_0\equiv G(H_0)=1/M_P^2$ is the current value of the gravitational
coupling. Equivalently,
\begin{equation}\label{eq:IntegrationRGforG2}
G(H)=\frac{G_0}{1+\nu\,\ln{\left(H^2/H_0^2\right)}}\,,
\end{equation}
where $\nu=(1/2\pi)\,\left(\frac16-\xi\right) m^2/M_P^2$ is a
dimensionless coefficient which acts as the reduced $\beta$-function for
the running of the gravitational coupling with the physical scale $H$.
Since the coefficient $\xi$ in the previous equation is not determined and
the number of participating matter fields is arbitrary, we can generalize
$\nu$ in the form
\begin{equation}\label{eq:nuloopcoeff1}
\nu=\frac{1}{2\pi}\, \sum_{i}
\left(\frac16-\xi_i\right)\frac{m_i^2}{M_P^2}\,.
\end{equation}
It is important to realize that if the running gravitational coupling as a
function of $\mu=H$ is given by Eq.\,(\ref{eq:IntegrationRGforG2}), we
cannot continue with Eq.\,(\ref{renormZPEoneloopCurved}) as a valid RGE
for the running vacuum energy density, since there is a link between the
running of $G(H)$ and the running of $\rL(H)$ which must be preserved.
That link is enforced by the Bianchi identity satisfied by the Einstein
tensor on the \textit{l.h.s.} of Eq.\, (\ref{EEnoCC1}), namely
$\nabla^{\mu}G_{\mu\nu}=0$. There are also the contributions provided by
the HD terms in the action (\ref{HigherDeriv}), which modify of course the
complete field equations. These higher order effects are represented in
diagrammatic form by the ``double haired'' diagrams in
Fig.\ref{Fig1:blobsHair}. But at low energies we can disregard them since
they entail ${\cal O}(H^4)$ corrections which are negligible. In this way
we obtain the following relation for the source term on the
\textit{r.h.s.} of Einstein's equations:
\begin{equation}\label{GBI}
\nabla^{\mu}\left(G\,\tilde{T}_{\mu\nu}\right)=\nabla^{\mu}\,\left[G\,(T_{\mu\nu}+g_{\mu\nu}\,\rL)\right]=0\,.
\end{equation}
Using the FLRW metric (\ref{eq:FLRWmetric}) and the standard
energy-momentum tensor for matter in the form of a perfect fluid,
Eq.\,(\ref{eq:Tmunu}), a straightforward calculation from (\ref{GBI})
provides the following ``mixed'' local conservation law:
\begin{equation}\label{BianchiGeneral}
\frac{d}{dt}\,\left[G(\rmr+\rL)\right]+3\,G\,H\,(\rmr+\pmr)=0\,.
\end{equation}
Therefore, if $G(t)=G(H(t))$ is evolving with the expansion rate in some
particular way, the previous identity enforces $\rL=\rL(H(t))$ to evolve
accordingly. Let us assume that $G(t)=G(H(t))$ runs with $H$ as in
Eq.\,(\ref{eq:IntegrationRGforG2}). Let us also assume at this point that
matter is covariantly conserved, i.e.
\begin{equation}\label{standardconserv}
\dot{\rho}_m+3\,H\,(\rmr+\pmr)=0\ \ \Rightarrow\ \
\rmr(a)=\rmr^0\,a^{3(1+\wm)}\,,
\end{equation}
where $\pmr=\wm\rmr$ is the equation of state of matter. It is then easy
to show that (\ref{BianchiGeneral}) boils down to
\begin{equation}\label{Bianchi1}
(\rmr+\rL)\dot{G}+G\dot{\rL}=\left[(\rmr+\rL)\frac{dG}{dH}+G\frac{d\rL}{dH}\right]\dot{H}=0\,.
\end{equation}
Obviously $\dot{H}\neq 0$, so we can equate to zero the expression in the
parenthesis.

In the following we stick to flat space geometry, i.e. we take $K=0$ in
Eq.\,(\ref{eq:FriedmannK}), and hence the metric of spacetime becomes
$ds^{2}=dt^{2}-a^{2}(t)d\vec{x}^{2}$. After all, this seems to be the most
plausible possibility in view of the present observational
data\,\cite{PLANCK2013} and the natural expectation from the inflationary
universe. Friedmann's equation for flat space simply reads
\begin{equation}\label{Friedmann}
H^2=\frac{8\pi G}{3}(\rmr+\rL)\,.
\end{equation}
Combining this equation with the Bianchi identity (\ref{Bianchi1}) a
simple differential equation for $\rL$ as a function of the Hubble rate
emerges. Let us note that $G$ in (\ref{Friedmann}) is not constant in the
present instance, but given by (\ref{eq:IntegrationRGforG2}). Solving for
$\rL=\rL(H)$ we find a simple ``affine'' quadratical law:
\begin{equation}\label{eq:SolrL}
\rL(H)=c_0+\frac{3\nu}{8\pi}\,M_P^2\,H^2\,,
\end{equation}
$\nu$ being here, of course, the same coefficient defined in
(\ref{eq:nuloopcoeff1}). This is a nice equation. With it we have found an
explicit realization of the kind of dreamed law for the vacuum energy
density suggested in Eq.\,(\ref{eq:affineH2}), with $\beta=3\nu/8\pi$.

Interestingly enough the obtained equation (\ref{eq:SolrL}) is of the same
form as the one that would follow from solving the previously mentioned
general RGE (\ref{seriesRLH}) provided we restrict the latter to the
current universe, namely when the ${\cal O}(H^4)$ terms can be neglected,
which is consistent with the approximation we used to derive
(\ref{eq:SolrL}). Integrating (\ref{seriesRLH}), and comparing with
(\ref{eq:SolrL}) it follows that $\nu$ must also be given by
\begin{equation}\label{eq:nuloopcoeff2}
\nu=\frac{1}{6\pi}\, \sum_{i=f,b} a_i\frac{M_i^2}{M_P^2}\,,
\end{equation}
in which only the coefficients $a_i$ of the first term on the
\textit{r.h.s} of (\ref{seriesRLH}) are involved.  We shall adopt this
equation as it is more general and assumes independent contributions from
bosons and fermions with different multiplicities.

It is interesting to compare equation (\ref{eq:SolrL}) with the
Heisenberg-Euler-Weisskopf Lagrangian of QED at low energies, which we
outlined in Sect.\,\ref{sect:oddH} -- confer Eq.\,(\ref{eq:EHLagrangian}).
The $\F^2$ and $\G^2$ contributions in that effective QED Lagrangian are
the equivalent to the $H^2$  terms in (\ref{eq:SolrL}). In both cases
there is suppressing coupling ($\alpha^2$ in one case, and $\nu$ in the
other). But there is an important difference. While in QED we consider a
correction to the Maxwell Lagrangian $\F$, and therefore the correction
involves terms of order $\F^2\sim \E^4$ (using the notation of
Sect.\,\ref{sect:oddH}) suppressed by the ``heavy mass'' power $1/m^4$
(perfectly consistent with the dimensional analysis), in the gravity case
the situation is different. Here we consider the vacuum energy density, so
we address a correction to a constant term $c_0$ (of mass dimension $4$ in
natural units), and hence the first correction is of order $H^2$. As a
result the coefficient of this term must have dimension $2$, and that is
why the correction must involve a square power of the mass, i.e. of the
form $\sim M_i^2\,H^2$. Summing over fields and their multiplicities  as
in (\ref{eq:nuloopcoeff2}) this leads to an expression of the form
(\ref{eq:SolrL}). The remarkable fact is that while in the QED case we
have a correction subdued by the traditional Appelquist-Carazzone
decoupling theorem\,\cite{AppCarazz74}, i.e. a correction that dies out
with the mass of the heavy degrees of freedom, in the vacuum energy case
we instead have a ``soft-decoupling'' correction which actually increases
with the masses. This is the reason why the heaviest masses $M_i$
(presumably from some GUT near the Planck scale) dominate the sum
(\ref{eq:nuloopcoeff2}) and the reduced $\beta$-function coefficient $\nu$
is small, but not necessarily too small. Theoretically we expect $|\nu|\ll
1$, typically $\sim 10^{-3}$, and indeed this is also what is presently
tolerated when we confront the model with
observations\,\cite{BPS09,BasPolarSola12} -- see
Sect.\ref{sect:SolvingVacEnerCurvature}.

Normalizing the obtained result (\ref{eq:SolrL}) with respect to the
current CC value $\rL(H_0)=\rLo$, we can rewrite it as
\begin{equation}\label{RGlaw2}
 \rL(H)=\rLo+ \frac{3\nu}{8\pi}\,M_P^2\,(H^{2}-H_0^2)\,.
\end{equation}
where  $c_0$ is related with the current value of the vacuum energy as
$\rLo=c_0+\left(3\nu/8\pi\right)\,M_P^2\,H_0^2$. The present value of the
Hubble rate is $H_0\equiv 100$\,\,h$\, Km/s/Mpc=1.0227\,h\times 10^{-10}$
yr$^{-1}$. The observations give $h\simeq 0.70$ (e.g. $h=0.673\pm 0.012$
from PLANCK\,\cite{PLANCK2013}).

As explained around Eq.\,(\ref{eq:affineH2}) in Sect.\,\ref{sect:oddH},
the presence of the affine term $c_0\neq 0$ is crucial for a realistic
implementation of the model. A vacuum energy evolving only as $\sim H^2$
(with $c_0=0$) would be incompatible with the transition from deceleration
to acceleration\,\cite{BasPolarSola12}.

Let us emphasize that the affine quadratic law (\ref{RGlaw2}) insures a
mild evolution of the vacuum energy density, specially if the parameter
$|\nu|<1$. We shall consider observational limits on this parameter in
Sect.\,\ref{sect:SolvingVacEnerCurvature}, but it is obvious from its
definition (\ref{eq:nuloopcoeff1}) or (\ref{eq:nuloopcoeff2}) that the
natural theoretical range expected for it is $|\nu|\ll1$. This expectation
will be confirmed by the phenomenological analysis. What is utmost
important to note is that no term of the form $\sim M_i^4$ drives now the
evolution of the vacuum energy. Indeed, this kind of terms are
incompatible with the mild running of the gravitational coupling, as
determined by Eq.\, (\ref{eq:IntegrationRGforG2}). This equation, together
with the covariant constraint imposed by the Bianchi identity
(\ref{Bianchi1}), imply that the cosmological term must evolve in the form
(\ref{RGlaw2}).

Furthermore, from the renormalization group point of view the $\sim M_i^4$
terms are not expected on the \textit{r.h.s.} of the general
RGE\,(\ref{seriesRLH}) either, as otherwise this would mean that a
particle with mass $M_i$ is an active degree of freedom for the running of
$\rL$. Since, however, the running is parameterized by the scale $\mu=H$,
this would entail $H>M_i$, which is an impossible condition to satisfy for
any known particle mass at any time in the matter dominated epoch (recall
that $H_0\sim 10^{-42}$ GeV). Let us note that even if we go to the
radiation dominated epoch, at temperature $T$, we find from Friedman's
equation (with $\rho_m\sim T^4$) that to satisfy the condition $H>M_i$
roughly means $T^2/M_P>M_i$, or equivalently
\begin{equation}\label{eq:activeMi}
\frac{M_i^4}{T^4}<\frac{M_i^2}{M_P^2}\ll1\,,
\end{equation}
for any particle of mass $M_i$. Hence, at the time when the $\sim M_i^4$
vacuum contributions start being active they are negligible as compared to
the radiation contribution $\sim T^4$. We conclude that, in the RG
formulation, the terms $\sim M_i^4$ remain innocuous throughout the entire
cosmic history: for, at low energies, these terms are not allowed by the
condition $H>M_i$, which can never be fulfilled, whilst the effect becomes
completely irrelevant at high energies (when that condition is possible).

To summarize, while a direct calculation of the dynamical renormalization
effects on the cosmological term is not possible at the moment without
expanding around a non-flat background in the presence of massive fields
and within a physical renormalization scheme, at least an indirect hint of
the result should be glimpsed by requiring the consistency of the
renormalization effects on the different terms of the effective action. As
these terms are linked by the Bianchi identity, the possible quantum
effects are tied to the general covariance of the theory. This requirement
could give us an indirect clue, which can be read off on more physical
grounds from the diagrams of Fig.\ref{Fig1:blobsHair}. We suggested that
as long as for flat spacetime only the ``bald blob'' contributes, the
dynamics of the vacuum energy in an expanding universe should emerge from
the ``haired blobs'', which may pump in energy from the expanding
background into the vacuum matter loops. At low energies we found that
this procedure indicates that the renormalization effects impinged upon
the cosmological term of the vacuum action consists of the ${\cal O}(H^2)$
terms reflected in Eq.\,(\ref{eq:SolrL}). These are of course the same
kind of quantum effects that should be found from a direct computational
approach, when it will be technically feasible.

\subsection{A comment on, and an analogy with, the Casimir effect}
\label{sect:Casimir}

In the meanwhile our roundabout path gives us a handle on the type of
effects we are looking for. It is encouraging to see that the obtained
result by the indirect procedure is free from the unwanted $\sim m^4$
contributions that plague the traditional approach. Somehow the ${\cal
O}(H^2)$ terms are the real effects that remain in the expanding spacetime
after we subtract the flat spacetime result --- quite in the same manner
as when in the calculation of the Casimir effect (see e.g.
\,\cite{Bordag01,Elizalde95} for a review) one removes the divergence of
the result upon subtracting the vacuum energy density when there are no
plates. Recall that in the Casimir effect the (attractive) force between a
pair of neutral, parallel conducting planes is due to the disturbance of
the quantum electrodynamics (QED) vacuum caused by the presence of the
boundaries. Since at zero temperature there are no real photons in between
the plates, it should be the vacuum alone, i.e., the ground state of QED
which causes the plates to attract each other\,\footnote{For this reason
the Casimir effect is usually advocated as experimental evidence for the
ZPE. This has been disputed in\,\cite{Jaffe2005} though. It is known since
long that Casimir forces can be derived as macroscopic manifestations of
van der Waals interatomic forces\,\cite{Bordag01}. While these need not
use the ZPE concept, the simplest way to derive them makes use of it, as
shown in London's theory\,\cite{London1930}. The situation with the
Casimir effect is entirely analogous\,\cite{Milonni96}; and thanks to it
one can replace a complicated many-body problem with a boundary value
problem. It thus seems likely that the Casimir effect is ultimately a
physical manifestation of a ``differential ZPE effect''. Furthermore, the
reality of the vacuum structure can manifest in subtler Casimir forces of
topological nature, which cannot be removed by subtraction or redefintion
of observables\,\cite{Zhitnitsky2013}.}. The pressure, or force per unita
area ($F/A$), on the plates is a pure quantum effect (proportional to
$\hbar$) that goes as the inverse of the quartic power of the distance $a$
between the plates, i.e.
\begin{equation}\label{eq:CasimirForce}
F/A=-\frac{1}{A}\frac{d E}{da}=-\hbar\,c\,\frac{\pi^2}{240\,a^4}\,.
\end{equation}
The Casimir effect is caused by the difference between the vibrational
modes of the QED vacuum in between the plates as compared to the region
outside. While the ZPE itself may not be measurable, ``changes in the
ZPE'' are detectable. The effect would, of course, be dynamical if the
distance between the plates would change in time, $a=a(t)$. Similarly, we
may view the evolution of the vacuum energy in an expanding background
with (dynamical) curvature $R\sim H^2(t)$, as the change that remains of
the disturbed vacuum energy density after we remove the flat spacetime
result -- which is also contained in the curved spacetime calculation,
Eq.\, (\ref{renormZPEoneloopCurved}). The mass term $m^4$ there is
replaced here by $1/a^4$ (note that no mass contribution is possible for
the Casimir effect in QED since photons are massless). Now, while the
$m^4$ term is removed when the flat spacetime result is subtracted, the
term $1/a^4$ remains in the Casimir effect because the plates, of course,
stay. In the Casimir effect one regularizes the infinite mode sum
$\sum_{\bf k} \frac12\,\hbar\omega_{\bf k}$ by subtracting the infinite
value of the ZPE when the plates are infinitely apart. This infinite
quantity cancels against the infinite ZPE value corresponding to the two
plates being at finite distance $a$, and in this way the final, and
finite, result (\ref{eq:CasimirForce}) emerges -- reflecting the
``differential vibrational effect'' caused by the inner modes only. But
notice that if letting $a\to 0$, we would strike another infinite value.
This is a short distance effect or UV-divergence. The new infinity ought
again to be subtracted because it corresponds to a situation where the
space between the plates disappears, so no ``distinctive'' standing waves
can form in that limit.

Such situation can be thought of as the equivalent of subtracting the unphysical
$\sim m^4$ term in the vacuum energy density of free fields, as that term
appears in taking care of the UV divergences through the
MS-renormalization procedure both in Minkowski and curved spacetime
cases\,\cite{SolaReview2013b}. Recall that in our discussion of the
renormalization of the ZPE in flat spacetime (cf. Sect.\,\ref{sect:ZPE2})
we introduced the counterterm  (\ref{deltaMSB}) for the vacuum energy
density, in such a way so as to kill the ``bare bone'' UV-part (the pole
at $n\rightarrow 4$), or, more generally, that part plus some (arbitrary)
finite remainder. In point of fact, we could have been maximally
consistent and arrange our renormalization condition such as to cancel any arbitrary
finite remnant, i.e. the entire value of the flat space-time result,
thereby leaving no $\sim m^4$ terms at all! In this way the physically
renormalized ZPE of flat spacetime would not be Eq.\,
(\ref{renormZPEoneloop2}) but just zero, and Einstein's equations in vacuo
can then be satisfied for the Minkowski metric!

This should not be viewed as a radical renormalization setup, it is
actually a completely standard one. Let us recall once more that the
renormalized parameters are not necessarily physical parameters. In the
on-shell scheme, the renormalized masses coincide with the physical
masses, but this is not so in e.g. the $\overline{\rm MS}$-scheme. A
similar situation occurs here, the renormalized result
(\ref{renormZPEoneloop2}) cannot be directly interpreted as the vacuum
energy density because it is incompatible with Einstein's equations in
Minkowskian spacetime. Therefore, we are forced to choose our
``renormalization condition'' to completely kill such contribution and
leave not only a finite, but an exactly vanishing physical vacuum energy
density in flat spacetime. In the notation of Sect.\,\ref{sect:ZPE2}, this
means
\begin{equation}\label{eq:zeroVacEnergFlatSpacetime}
\rV=\rLb+\ZPE^{({\rm b})}=\rL(\mu)+\delta\rL+\ZPE^{({\rm
b})}(\mu)=0\,.
\end{equation}
At one loop this reads
\begin{equation}\label{deltaMSBPhysical}
\rL(\mu)+\delta\rL+\frac{m^4\,\hbar}{4\,(4\pi)^2}\,\,\left(-\frac{2}{4-n}
-\ln\frac{4\pi\mu^2}{m^2}+\gamma_E-\frac32\right)=0\,.
\end{equation}
The presence of the counterterm insures that the expression on the
\textit{l.h.s.} can be rendered finite. After that there is no reason to
leave any finite part in Minkowski space.

The above procedure is not so different to renormalizing, say, the electron mass and charge. We know with accuracy the
physical value of $m_e$ and $\alpha_{\rm em}$ in the low energy regime (Thomson's limit). This is the limit where we fix the
renormalization conditions in this case. If we use on-shell renormalization, we have to arrange the mass and charge
counterterms to exactly cancel the ``infinite tower'' of possible radiative corrections to any of these parameters at all
orders in perturbation theory, both the UV-divergent parts and all the finite parts as well, such that the corresponding
renormalized parameters exactly coincide with the measured ones. If we use another (off-shell) renormalization scheme
($\overline{\rm MS}$, for instance) the renormalized parameters will not be the physical ones. Can we still find them? Sure,
but we've got to do some homework! We have to compute the dressed electron propagator, the dressed photon propagator and the
dressed QED vertex $\gamma e^{-}e^{-}$ in the $\overline{\rm MS}$-scheme. Then go to the Thomson limit and identify the
poles of the propagators with the physical masses (in this case $m_e$ and $0$, respectively), and finally identify the
electromagnetic coupling of the dressed vertex with the classical electron charge. It goes without saying that these
operations imply an adjustment of the parameters of the theory comparable to the on-shell renormalization. The arrangement
made at low energy does not, of course, preclude that at high energies we still have the finite contributions to the running
mass and charge. Despite one might view with suspicion these adjustments as fine tuning, they are in fact called
renormalization.

What we have done  with the CC parameter in Minkowski space is to renormalize it to zero. So, again, this is not fine
tuning, it is renormalization. The ``only'' difference is that in the cosmological case the radiative ``corrections'' are
seen as huge effects when compared to the physical result (the value of $\rLo$), but as we shall comment below this is a
``mirage'' that appears because we make the comparison now, i.e. in our very low energy universe under the assumption that
the CC is strictly constant. For a dynamical vacuum term (only possible in the curved spacetime case) the situation can
change dramatically.

In fact, in curved spacetime we can have a non-trivial result sourced by the ``distinctive vibrational modes'' of the vacuum
energy density with respect to the flat spacetime case (the only measurable quantity). To insure, however, that the physical
flat spacetime result in vacuo is retrieved (viz. exactly zero vacuum energy density), we still have to remove the original
flat spacetime ZPE result (\ref{renormZPEoneloopCurved}). For this we must use once more the renormalization condition
(\ref{eq:zeroVacEnergFlatSpacetime}), but this does not mean that the vacuum energy density in the curved case is again
zero. Being the spacetime curved, a nonvanishing value is now perfectly consistent with the Einstein's equations in vacuo.
The expression that we claim here is, of course, Eq.\,(\ref{eq:affineH2}) for the low energy theory, which we have motivated
indirectly and is pending from a rigorous direct calculation in a suitable setup. What is important at this point is that no
trace is to be left of the $\sim m^4$ terms. And this fact cannot depend on anything else but on the renormalization
condition established in the first step by which the Minkowskian vacuum was properly defined. In other words, we should not
expect that the absence of the unwanted terms has anything to do with the details of the curved spacetime calculation. The
latter should only account for the dynamical effects $\sim H^2$ of the vacuum energy density in an expanding spacetime.

At the end of the day only the distinctive ${\cal O}(H^2)$ disturbance,
along with the additive $c_0$ term in Eq.\,(\ref{eq:SolrL}), remain at low
energy as the final measurable output for the curved spacetime case. This
is ultimately the meaning of the analogy with the Casimir effect after removing the
contribution in the short distance limit $a\to 0$. Let us emphasize once
more that both physical terms are correctly motivated in the RG
formulation, and in particular the additive term $c_0$ is absolutely
essential for a correct phenomenological description of the cosmic
transition from deceleration to acceleration. This term appears in a
natural way in the present framework from the integration of the general
RGE (\ref{seriesRLH}) for the vacuum energy in an expanding spacetime. After the ugly $\sim m^4$
terms are definitely banished from the horizon, to call the final result $\rL(H)$ still vacuum energy density
(or something else) is not very important. What matters is that it is a part of the quantum effects
on the effective action which results from the dynamics of the expanding universe.

\subsection{Extension to the early universe} \label{sect:EarlyUniverse}

Interestingly, a kind of equation such as (\ref{RGlaw2}) was
suggested in \cite{ShapSol09,ShapSol09b}, and previously in
\cite{ShapSol00,ShapSol03,ShapSol03b,SS05} on more phenomenological
grounds. But the first hint that an expression of that sort could be
heuristically derived from a specific QFT framework was provided in
Ref.\,\cite{Fossil07} in a conformal field theory context. More
recently there have appeared alternative QFT frameworks suggesting a
similar kind of evolution of the vacuum energy leading once more to
dynamical terms $\sim H^2$\,\cite{Maggiore,Bilic1,Bilic2,Hack13}.
See also \cite{Paddy05} for other interesting considerations along
these lines in a different context.
%
%%%%%%%%%%%%%%%%%%%%%%%%%%%%%%%%%%%%%%%%%%%%%%%%%%%%%%%%%%%%%%%%%%%
\FIGURE[t]{
  %\begin{center}
    %\begin{tabular}{cc}
      \resizebox{1.0\textwidth}{!}{\includegraphics{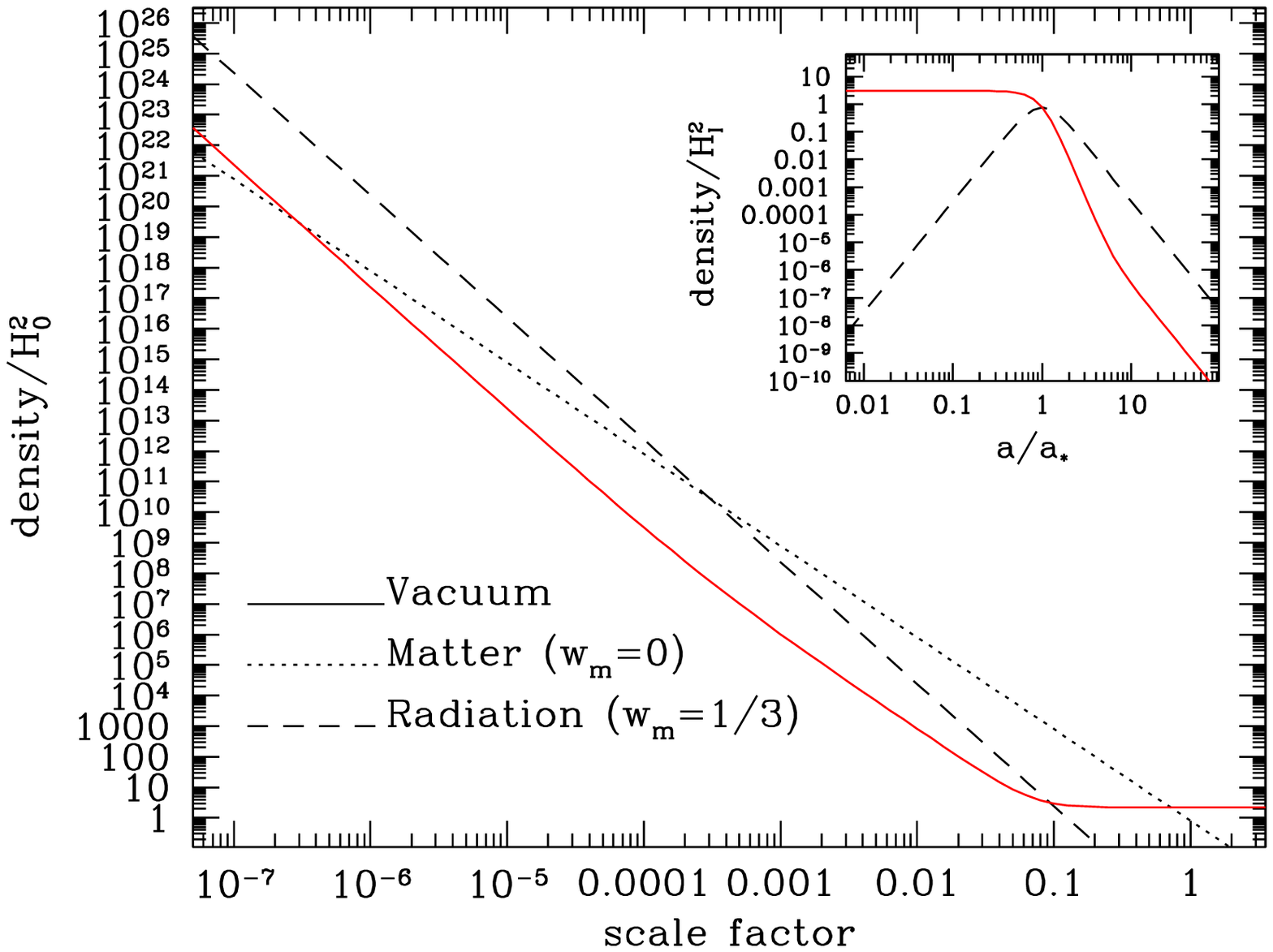}}
      %&
      %\hspace{0.3cm}
      %\resizebox{0.45\textwidth}{!}{\includegraphics{fig2.eps}} \\
      %(a) & (b)
    %\end{tabular}
    \vspace{0.2cm}
    \caption{{\bf Outer plot}: The evolution of the radiation,
non-relativistic matter and vacuum energy densities, for the unified
vacuum model (\ref{eq:UnifiedRunning}) with  $n=2$ in units of
$H^{2}_{0}$\,\,\cite{LimaBasSola13a,HMention2013}. Note that similar
behavior is found also $\forall n$\,\cite{Perico2013a}. The curves shown
are: radiation (dashed line), non-relativistic matter (dotted line) and
vacuum (solid line, in red). To produce the lines the following inputs
have been used: $\nu=10^{-3}$, $\Omega^{0}_{m}=0.27$,
$\Omega^{0}_{R}=(1+0.227N_{\nu})\,\Omega^{0}_{\gamma}$,
$\Omega^{0}_{\Lambda}=1-\Omega^{0}_{m}-\Omega^{0}_{R}$,
$(N_{v},\Omega^{0}_{\gamma},h)\simeq (3.04,2.47\times 10^{-5}h^{-2},0.71)$
(cf. Komatsu et al. \cite{Kom11}). {\bf Inner Plot:} the primeval vacuum
epoch (inflationary period) into the FLRW radiation epoch. Same notation
for curves as before, although the densities are now normalized with
respect to $H_I^2$ and the scale factor with respect to $a_{*}$, which is
is the scale factor at the transition time when the inflationary period
ceases. We have set $\alpha=1$ because the value of the expansion rate at
the inflationary epoch, $H_I$, corresponds to an arbitrarily high energy
near the Planck scale. For convenience we used $8\pi G=1$ units in the
plots. }
%\end{center}
  \label{Fig:CosmicHistory}}
%\end{figure}
%%%%%%%%%%%%%%%%%%%%%%%%%%%%%%%%%%%%%%%%%%%%%%%%%%%%%%%%%%%%%%%%%%%%

We expect that the low energy running vacuum model (\ref{eq:SolrL}) can be
generalized in the form (\ref{seriesRLH}) so as to include the important
effects from the higher powers of $H$ at the early inflationary times.
\newtext{It is interesting to note} that the next-to-leading higher order power,
i.e. $H^4$, can be motivated on fundamental QFT grounds, as shown in
\cite{Fossil07} within the framework of the modified anomaly-induced
inflation
scenarios\,\cite{ShapSol01,DESYTomsk,GracefulExitRuss02,PeliShTak03}.
These are a generalization of Starobinsky's model type of
inflation\,\cite{Starobinsky80}, in which the vacuum effective
action for massive quantum fields can be computed using the
conformal representation of the fields action\,\cite{PSW87}. A
unified model of this kind, where the dark energy emerges at late
times as a ``fossil'' of the early inflationary universe was
presented in \cite{Fossil07}. These models of the early universe,
combined with the low energy theory, suggest that the effective
expression of the vacuum energy density should be a combination of
even powers of $H$, essentially $H^2$ and $H^4$.  Recently, in
Ref.\,\cite{LimaBasSola13a,HMention2013,Perico2013a} a detailed
analysis of the complete cosmic history of the universe has been
presented by considering the class of models of the form
\begin{equation}\label{eq:UnifiedRunning}
\rL(H) = c_0 + \frac{3\nu}{8\pi}\,M_P^2 H^{2} + \frac{3\alpha}{8\pi}\,M_P^2
\frac{H^{2n}}{H_{I}^{(2n-2)}} \,,
\end{equation}
in which $n\geqslant 2$. Clearly, this expression is an extension of
Eq.\,(\ref{eq:SolrL}) along the lines of the general RGE
(\ref{seriesRLH}). The higher power of $H$ should obviously be operative
only for large values of $H$ near the inflationary scale $H_I$ (presumably
a GUT scale not very far from the Planck scale). Typically we expect
$n=2$, i.e. a high energy behavior $\sim H^4$\,\,\,\cite{LimaBasSola13a}.
At present, the $H^4$ term is of course negligible and we are effectively
left with the low energy theory (\ref{eq:SolrL}). Notice that the
dimensionless coefficient $\alpha$ (enabling the running of the vacuum
energy near the GUT scale) can be related to the coefficients of the
general RGE (\ref{seriesRLH}) as follows:
\begin{equation}\label{eq:alphaloopcoeff}
\alpha=\frac{1}{12\pi}\, \frac{H_I^2}{M_P^2}\sum_{i=f,b} b_i\,.
\end{equation}
We point out that while we use only even powers of the Hubble rate, odd
powers have also been considered phenomenologically in the old literature
for the treatment of the inflationary stages\,\cite{LM94,LT96}. The case
with $c_0=0$ and an arbitrary high power of $H$ was  phenomenologically
investigated in \cite{MaiaLima2000}. More recently one can also find
models where odd powers are used for describing the CC evolution for the
present universe, or in combination with even powers for the early
universe\,\cite{Schutzhold02,OddHpowers,Carneiro09,Basilakos09}. Finally,
let us mention that powers of $H$, including linear ones, have also been
used to describe the late time evolution in terms of the so-called bulk
viscosity parameters, but again this is a purely phenomenological
treatment of the purported DE fluid\,\cite{RenMeng06}. Let us insist that,
barring some particular situations mentioned in Sect\,\ref{sect:oddH}, the
general form that we propose in (\ref{eq:UnifiedRunning}) involves only
even powers of the expansion rate, at all stages of the cosmological
evolution, as a most {\textit{fundamental requirement} from the general
covariance of the effective action of QFT in curved
spacetime\,\cite{ShapSol09,ShapSol09b}. Recall once more our analogy with
QED and QCD sketched in Sect.\,\ref{sect:oddH} where a similar property
holds.

The class of universes whose vacuum energy density evolves as in
Eq.\,(\ref{eq:UnifiedRunning}) is very appealing. It turns out that the
differential equation for $H$ has a constant solution in the early
stages\,\cite{LimaBasSola13a},
therefore an inflationary solution. Furthermore, there is a transition
point $a=a_{*}$ from the inflationary epoch towards a standard (FLRW)
radiation dominated stage, as shown in the inner plot of
Fig.\,\ref{Fig:CosmicHistory} (for $n=2$). In between these two eras, we
can have either huge relativistic particle production $\rho_r\propto a^4$
in the deflation period (namely around the time when the inflationary
phase progressively weakens), followed by standard dilution $\rho_r\propto
a^{-4}$ well in the radiation era. The outer plot of
Fig.\,\ref{Fig:CosmicHistory} depicts the lengthy cosmic history after the
inflationary regime up to the final de Sitter epoch, which started near
our time and extends into the future.

We shall not dwell here on the interesting phenomenological features of the class of unified vacuum
models (\ref{eq:UnifiedRunning}), see the detailed analyses presented in
Ref. \,\cite{LimaBasSola13a,HMention2013,Perico2013a}\,\footnote{See also
Ref.\,\cite{Pavon2013} for thermodynamical considerations on this kind of
$\CC(H)$ dynamical vacuum models.}. It suffices to say that all these
models start from an inflationary phase and, for all $n>2$, they
automatically reach the situation of ``graceful exit'' from the
inflationary phase into a standard FLRW radiation dominated epoch.
\newtext{This is} already a significant achievement! In general they
provide an effective framework containing most of the basic ingredients
that should probably be desirable for a future fundamental theory of the
cosmic evolution, namely a theory capable of tackling efficiently the
important cosmological problems which are still pending. And, most
important, the late time cosmic expansion history is very close to the
standard $\Lambda$CDM model, with small departures that should be
accessible to the future  observations.

\subsection{Different scenarios for running cosmological parameters at
low energies} \label{sect:DifferentScenarios}

We discuss now some possible scenarios for running cosmological
parameters. We shall focus here on the implications for the low
energy regime of the cosmological evolution and therefore we do not
consider the highest powers of $H$ introduced in
Eq.\,(\ref{eq:UnifiedRunning}), it will be enough to consider the
$H^2$ dynamical effects. Once more we adopt the spatially flat FLRW
metric, i.e. $K=0$ in Eq.\,(\ref{eq:FLRWmetric}), for all models to
be discussed henceforth. In such conditions, the relevant
Friedmann's equation providing the Hubble rate reads as in
Eq.\,(\ref{Friedmann}). As stated, we assume that $\rL=\rL(t)$ and
$G=G(t)$ can be functions of the cosmic time $t$. In addition, the
dynamical equation for the acceleration of the universe is given by
the expression (\ref{eq:acceleration}), or equivalently
\begin{equation}\label{eq:acceleration2}
\frac{\ddot{a}}{a}=-\frac{4\pi\,G}{3}\,(\rmr+3\pmr-2\rL)=-\frac{4\pi\,G}{3}\,(1+3\wm)\,\rmr+\frac{8\pi\,G}{3}\,\rL\,.
\end{equation}
In the late universe ($\rmr\to 0$) the vacuum energy density $\rL$
dominates. It accelerates the cosmos for $\rL>0$. This may occur either,
because $\rL$ is constant, and for a sufficiently old universe one finally
has $\rmr(t)<2\,\rL$, or because $\rL(t)$ evolves with time, and the
situation $\rL(t)>\rmr(t)/2$ is eventually reached sooner or later than
expected.

Let us come back to the general covariant conservation law,
Eq.\,(\ref{BianchiGeneral}), and consider various
possibilities\,\footnote{The local covariant conservation equation
(\ref{BianchiGeneral}) is not independent from equations
(\ref{eq:FriedmannK}) and (\ref{eq:acceleration}), as it is actually a
first integral of the system formed by the last two. For example, it is
easy to check from equations (\ref{eq:FriedmannK}) and
(\ref{BianchiGeneral}) that the acceleration equation for the scale factor
takes the usual form (\ref{eq:acceleration}) -- or equivalently
(\ref{eq:acceleration2}) -- even for time evolving $G$ and $\rL$ (and for
nonvanishing spatial curvature $K$).}:
\begin{itemize}

\item \textbf{Model I}:  $G=$const. {and} $\rL=$const.:

Under these conditions and in the absence of other components in the
cosmic fluid, apart from matter and a strictly constant $\CC$-term,
the local covariant conservation law of matter-radiation is strictly
fulfilled, i.e. Eq.\,(\ref{standardconserv}). If, on top of this, we
have zero spatial curvature, $K=0$, then Model I becomes the almost
thirty years old flat $\CC$CDM, or ``concordance'' cosmological
model\,\cite{Peebles84}, viz. the currently reigning standard
cosmological model.
\item  \textbf{Model II}: $G=$const {and} $\dot{\rho}_{\CC}\neq 0$:

Here Eq.(\ref{BianchiGeneral}) leads to the mixed conservation law:
\begin{equation}\label{mixed conslaw}
\dot{\rho}_{\CC}+\dot{\rho}_m+3\,H\,(\rmr+\pmr)=0\,.
\end{equation}
An exchange of energy between matter and vacuum takes place. The model
can be solved only if more information is provided on e.g. the cosmic
evolution of $\rL$, see the next section.

\item \textbf{Model III}: $\dot{G}\neq 0$ {and} $\rL=$const.:
\begin{equation}\label{dGneqo}
\dot{G}(\rmr+\rL)+G[\dot{\rho}_m+3H(\rmr+\pmr)]=0\,.
\end{equation}
Since $G$ does not stay constant here, this equation implies matter
non-conservation. It could be solved e.g. for $G$, if the (anomalous)
cosmic evolution of $\rmr$ would be given by some particular ansatz.

\item \textbf{Model IV}: $\dot{G}\neq 0$ {and} $\dot{\rho}_{\CC}\neq
    0$:

Although several possibilities are available here, the simplest one is
    of course the framework that has motivated our analysis in the
    previous section, where matter is covariantly conserved and the
    dynamical interplay occurs between $G$ and $\rL$ through
Eq.\,(\ref{Bianchi1}). The cosmological equations for this model were
solved in Sect.\,\ref{sect:VacEnerCurvature}. In the next section we
solve also the ``running'' Models II and III and compare with the
present one.

\end{itemize}

\subsection{Solving the cosmological equations for Models II and III}
\label{sect:SolvingVacEnerCurvature}

The class of Models II and III is quite general and we cannot solve them
unless we provide some more information, similarly to the situation with
Model IV. Let us first concentrate on solving the class of scenarios
denoted as Model II. Let $\rMo$ be the total matter density of the present
universe, which is essentially non-relativistic ($\wm\simeq 0$). The
corresponding normalized density is $\OMo=\rMo/\rco\simeq 0.3$ , where
$\rco$ is the current critical density. Similarly, $\OLo=\rLo/\rco\simeq
0.7$ is the current normalized vacuum energy density, for flat space. If
$\rL$ evolves with the Hubble rate in the form indicated in
Eq.\,(\ref{RGlaw2}), the non-relativistic matter density and vacuum energy
density evolve with the redshift as follows\,\cite{BPS09,ShapSol03}:
\begin{equation}\label{mRG2}
\rM(z;\nu) =\rMo\,(1+z)^{3(1-\nu)}\,,
\end{equation}
and
\begin{equation}\label{CRG2}
\rL(z)=\rLo+\frac{\nu\,\rM^0}{1-\nu}\,\left[(1+z)^{3(1-\nu)}-1\right]\,.
\end{equation}
The corresponding Hubble function reads
\begin{equation}
{H^2(z)}=\frac{8\pi\,G}{3\,(1-\nu)} \left[{\rLo-\nu\,\rco}+\rmo\,
(1+z)^{3(1-\nu)}\right] \;. \label{nomalHflow}
\end{equation}
The crucial parameter is $\nu$, which we have introduced in
sect.\,\ref{sect:VacEnerCurvature}. It is responsible for the time
evolution of the vacuum energy. From Eq.\,(\ref{mRG2}) we confirm, that it
accounts also for the non-conservation of matter. For $\nu=0$ it leads to
the exact local covariant conservation, which for non-relativistic matter
reads
\begin{equation}\label{mRG2b}
\rM(z) =\rMo\,(1+z)^{3}\,.
\end{equation}
Next we note that $\delta\rho_M\equiv \rM(z;\nu)-\rM(z)$ is the net amount
of non-conservation of matter per unit volume at a given redshift. This
expression must be proportional to $\nu$, since we subtract the conserved
part. At this order we have $\delta\rho_M=-3\,\nu\, \rMo (1+z)^3\ln(1+z)$.
{We differentiate it} with respect to time and expand in $\nu$, and divide
the final result by $\rM$. {This provides} the relative time variation:
\begin{equation}\label{eq:deltadotrho}
\frac{\delta\dot{\rho}_M}{\rM}=3\nu\,\left(1+3\ln(1+z)\right)\,H+{\cal O}(\nu^2)\,.
\end{equation}
Here we have used $\dot{z}=(dz/da)\dot{a}=(dz/da)aH=-(1+z)H$. Assuming
relatively small values of the redshift, we may neglect the log term and
are left with:
\begin{equation}\label{eq:deltadotrho2}
 \frac{\delta\dot{\rho}_M}{\rM}\simeq 3\nu\,\,H\,.
\end{equation}
From (\ref{CRG2}) we find:
\begin{equation}\label{eq:deltaLambda}
\frac{\dot{\rho}_{\CC}}{\rL}\simeq -3\nu\,\frac{\OMo}{\OLo}\,(1+z)^3\,H+{\cal O}(\nu^2)\,.
\end{equation}
It is of the same order of magnitude as (\ref{eq:deltadotrho2}) and has
the opposite sign. From a detailed analysis of the combined data on type
Ia supernovae, the Cosmic Microwave Background (CMB), the Baryonic
Acoustic Oscillations (BAO) and the structure formation data a direct
cosmological bound on $\nu$ has been obtained in the
literature\,\,\cite{BPS09}:

\begin{equation}\label{eq:numodeliicosmology}
|\nu|^{\rm cosm.}\lesssim {\cal O}(10^{-3})\,,\ \ \ \ \ ({\rm Model\ II})\,.
\end{equation}
It is consistent with the theoretical expectations\,\cite{Fossil07}.

Let us now analyze Model III, which can also accommodate matter
non-conservation in the form (\ref{mRG2}), but at the expense of a time
varying $G$. We compare it with a similar model where $G$ is also running,
Model IV, but where matter is conserved.

Within the class of scenarios indicated as Model III  the parameter $\rL$
remains constant ($\rL=\rLo$) and $G$ is variable. This is possible due to
the presence of the \textit{non} self-conserved matter density
(\ref{mRG2}). Trading the time variable by the scale factor, we can
rewrite Eq.\,(\ref{dGneqo}) as follows:
\begin{equation}\label{dGneqo2}
G'(a)\left[\rM(a)+\rLo\right]+G(a)\left[{\rho}'_M(a)+\frac{3}{a}\,\rM(a)\right]=0\,.
\end{equation}
The primes indicate differentiation with respect to the scale factor. We
insert equation (\ref{mRG2}) in (\ref{dGneqo2}), integrate the resulting
differential equation for $G(a)$ and express the final result in terms of
the redshift:
\begin{equation}\label{GzfixedL}
G(z)=G_0\,\left[\OMo\left(1+z\right)^{3(1-\nu)}+\OLo\right]^{\nu/(1-\nu)}\,.
\end{equation}
Here $G_0=1/M_P^2$ is the current value of the gravitational coupling. The
previous equation is correctly normalized: $G(z=0)=G_0$, due to the cosmic
sum rule in flat space: $\OMo+\OLo=1$. For $\nu=0$ the gravitational
coupling $G$ remains constant: $G=G_0$. {Since $\rL$ is constant in the
current scenario}, the small variation of $G$ is entirely due to the
non-vanishing value of the $\nu$-parameter in the matter non-conservation
law (\ref{mRG2}). This leads to the dynamical feedback of $G$ with
matter\,\footnote{The matter non-conservation law (\ref{mRG2}) was first
suggested and analyzed in \cite{ShapSol03,ShapSol03b}, and later on in
\cite{WangMeng04} and in \cite{Guberina06}.}. For the present model
Friedmann's equation (\ref{Friedmann}) becomes:
\begin{equation}\label{FriedmannModeliii}
H^2(z)=\frac{8\pi
G(z)}{3}\left[\rMo\,(1+z)^{3(1-\nu)}+\rLo\right]=H_0^2\frac{G(z)}{G_0}\,\left[\OMo\,(1+z)^{3(1-\nu)}+\OLo\right].
\end{equation}
Combining (\ref{GzfixedL}) and (\ref{FriedmannModeliii}), we find the
Hubble function of this model in terms of $z$:
\begin{equation}\label{HzModeliii}
H^2(z)=H_0^2\,\left[\OMo\left(1+z\right)^{3(1-\nu)}+\OLo\right]^{1/(1-\nu)}\,,
\end{equation}
and we obtain:
\begin{equation}\label{GrelatedH}
\frac{G(z)}{G_0}=\left[\frac{H^2(z)}{H_0^2}\right]^{\nu}\,.
\end{equation}
Since $\nu$ is presumably  small in absolute value (as in the previous
section), we can expand (\ref{GrelatedH}) in this parameter:
\begin{equation}\label{GzfixedLsmallnu}
G(H)\simeq G_0\,\left(1+\nu\,\ln\frac{H^2}{H_0^2}+{\cal
O}(\nu^2)\right)\,.
\end{equation}
At leading order in  $\nu$ this expression for the variation of $G$ is
identical to the one found for Model IV, see
Eq.\,(\ref{eq:IntegrationRGforG2}), except for the sign of $\nu$. The
equation(\ref{GzfixedLsmallnu}) allows us to estimate the value of the
parameter $\nu$ by confronting the model with the experimental data on the
time variation of $G$. Differentiating (\ref{GzfixedLsmallnu}) with
respect to the cosmic time, we find in leading order in $\nu$:
\begin{equation}\label{GdotoverGo}
\frac{\dot{G}}{G}= 2\nu\,\frac{\dot{H}}{H}=-2\,(1+q)\,\nu\,H\,,
\end{equation}
where we have used the relation $\dot{H}=-(1+q)H^2$, in which
$q=-\ddot{a}/(aH^2)$ is the deceleration parameter.  From the known data
on the relative time variation of $G$ the bounds indicate that
$|\dot{G}/G|\lesssim 10^{-12}\,{\rm yr}^{-1}\,$\,\cite{FundConst}. If we
take the present value of the deceleration parameter, we have
$q_0=3\OMo/2-1=-0.595\simeq -0.6$ for a flat universe with $\OMo=0.27$. It
follows:
\begin{equation}\label{GdotoverGo2}
\left|\frac{\dot{G}}{G_0}\right|\lesssim \,0.8 |\nu|\,H\,.
\end{equation}
Taking the current value of the Hubble parameter: $H_0\simeq 7\times
10^{-11}\,{\rm yr}^{-1}\,$ (for $h\simeq 0.70$), {we obtain $|\nu|\lesssim
10^{-2}$. The real value of $|\nu|$ can be smaller, but to compare the
upper bound that we have obtained with observations makes sense in view of
the usual interpretation of $\nu$ in sect.\,\ref{sect:VacEnerCurvature}
and the theoretical estimates indicated there. The constraints from Big
Bang nucleosynthesis (BBN) for the time variation of $G$ are more
stringent and lead to the improved bound. Since Models III and IV share a
similar kind of running law for the gravitational coupling (except for the
sign of $\nu$)  we can extract the same bound for $|\nu|$ in the two
models following the method of Ref.\,\cite{GSFS10} and references therein,
particularly\,\cite{{Iocco:2008va}}. The final result is
\begin{equation}\label{eq:nuboundGvariable}
|\nu|^{\rm BBN}\lesssim 10^{-3}\ \ \ \ \ ({\rm Models\ III\ and\ IV})\,.
\end{equation}
The cosmological data from different sources furnish about the same upper
bound on $|\nu|$ for the two  running models where matter is
non-conserved, i.e. Models II and III. In both cases the upper bound on
$|\nu|$ is $\sim 10^{-3}$, as shown by equations
(\ref{eq:numodeliicosmology}) and (\ref{eq:nuboundGvariable}).

Although the order of magnitude of the bounds on $|\nu|$ are sometimes
coincident for different models, the interpretation can be quite
different. For example, Model IV cannot -- in contrast to Models II and
III -- be used to explain the possible time variation of the fundamental
constants of the strong interactions and the particle masses (see next
section). It can only be used to explain the time variation of the
cosmological parameters $\rL$ and $G$ in a way which is independent, in
principle,  from the microphysical phenomena in particle physics and
nuclear physics.

\section{Dynamical vacuum energy and the time variation of the
fundamental constants}\label{sect:DynamicalVacFundConst}

As we have seen in Sect.\,\ref{sect:VacEnerCurvature}, in an
expanding universe the vacuum energy density $\rL$ is expected to be
a dynamical quantity, and should exhibit a slow evolution determined
by the expansion rate of the universe $H$. While a formal and
completely rigorous proof of this contention is at the moment not
feasible, the hints that it must be so are sufficiently encouraging
to further spur these investigations at least from the
phenomenological side\,\cite{SolaReview2011,SolaReview05}, let alone
the theoretical studies pointing to that possibility in particular
QFT frameworks\,\cite{Fossil07,GrandSola13}. If so, the dynamical
vacuum in QFT in curved spacetime could provide an alternative
scenario (beyond the usual quintessence kind of approaches) for
implementing dynamical dark energy (DE) as a general paradigm for
curing or alleviating the old CC problem, as well as the so-called
cosmic coincidence problem. Interestingly enough, this possibility
has been phenomenologically tested and profusely confronted with the
latest accurate data on type Ia supernovae (SNIa), the Baryonic
Acoustic Oscillations (BAOs), and the anisotropies of the Cosmic
Microwave Background (CMB), see particularly
\cite{BPS09,GSBP11,BasPolarSola12,BasSola13a,GPS08,GPS09,GOPS07,FSS06,GSS06}
for the most recent and comprehensive studies (which have also been
applied to alternative DE models\,\cite{BLP10}), see also
\cite{PelOpher1,PelOpher2}. Furthermore, a unified vacuum framework
capable of describing the complete cosmic history as of the early
inflationary times to the present dark energy epoch has also
recently been proposed within this same kind of
approach\,\cite{LimaBasSola13a,HMention2013,Perico2013a}.

If that is not enough, quite intriguingly the idea could also be tested
from an entirely different vein, namely one capable of providing a new and
completely independent insight into the whole subject. This novel and
recent approach was first suggested in\,\cite{FritzschSola2012}, and it is
related with the precise measurements of the so-called Fundamental
Constants of Nature and their possible time variation, see
e.g.\,\cite{FundConst} (and the long list of references therein). I would
like to devote some time in this review to this particular
phenomenological approach to dynamical vacuum energy.

\subsection{Cosmic running of the fundamental ``constants''}\label{sect:CosmicRunningLQCD}

Recent measurements on the time variation of the fine structure constant
and of the proton-electron mass ratio suggest that basic quantities of the
Standard Model, such as the QCD scale parameter $\LQCD$, may not be
conserved in the course of the cosmological evolution\,\cite{Fritzsch11}.
The masses of the nucleons $m_N$ and of the atomic nuclei would also be
affected. Matter is not conserved in such a universe. In the framework of
ideas that we have developed in Sect.\,\ref{sect:DifferentScenarios},
these measurements could be interpreted as a leakage of matter into vacuum
or vice versa. In the following we wish to show that the amount of leakage
necessary to explain the measured value of the time variation
$\dot{m}_N/m_N$ could be of the same order of magnitude as the
observationally allowed value of $\dot\rho_{\CC}/\rL$, with a possible
contribution from the dark matter particles\,\cite{FritzschSola2012}.
Therefore, the dark energy in our universe could be the dynamical vacuum
energy in interaction with ordinary baryonic matter as well as with dark
matter.

The QCD scale parameter is related to the strong coupling constant
$\alpha_s=g_s^2/(4\pi)$. To lowest (1-loop) order one finds:
\begin{equation}\label{alphasLQCD}
\alpha_s(\mu_R)=\frac{1}{\beta_0\,\ln{\left(\LQCD^2/\mu_R^2\right)}}=\frac{4\pi}{\left(11-2\,n_f/3\right)\,\ln{\left(\mu_R^2/\LQCD^2\right)}}\,,
\end{equation}
where $\mu_R$ is the renormalization point and $\beta_0\equiv-b_0=
-(33-2\,n_f)/(12\,\pi)$ ($n_f$ being the number of quark flavors) is the
lowest order coefficient of the $\beta$-function. If we include all the
know flavors, we obviously have $\beta_0<0$ and therefore when the
renormalization scale $\mu_R$ increases the strong coupling constant
diminishes (this is the famous asymptotic freedom property of QCD).
However, for lower scales, namely for $\mu_R$ near $\LQCD\sim {\cal
O}(100)$ MeV, the strong coupling constant increases and in practice it
becomes arbitrarily large. This is precisely the regime which we cannot
explore with perturbation theory, and as we see the two regimes are
roughly separated by the $\LQCD$ scale. Not surprisingly this scale is the
one that rules the calculation of the strongly interacting hadronic bound
states like the proton, and in fact it dominates the calculation of its
mass, see Eq.\,(\ref{eq:ProtonMass}) below. The QCD scale parameter
$\LQCD$ has actually been experimentally measured with $\sim 10\%$
precision: $\LQCD = 217\pm 25$ MeV.

The next observation is crucial for our
discussion\,\cite{FritzschSola2012}. When we embed QCD in the FLRW
expanding background, the value of $\LQCD$ need not remain rigid anymore.
The value of $\LQCD$ could change with $H$, and this would mean a change
in the cosmic time. If $\LQCD=\LQCD(H)$ is a function of $H$, the coupling
constant $\alpha_s=\alpha_s(\mu_R;H)$ is also a function of $H$ (apart
from a function of $\mu_R$). The relative cosmic variations of the two QCD
quantities are related (at one-loop) by:
\begin{equation}\label{eq:timealphaLQC}
\frac{1}{\alpha_s}\frac{d\alpha_s(\mu_R;H)}{dH}=\frac{1}{\ln{\left(\mu_R/\LQCD\right)}}\,\left[\frac{1}{\LQCD}\,\frac{d{\Lambda}_{\rm
QCD}(H)}{dH}\right]\,.
\end{equation}
The potential significance of this relation is out of discussion: if the
QCD coupling constant $\alpha_s$ or the QCD scale parameter $\LQCD$
undergo a small cosmological time shift, the nucleon mass and the masses
of the atomic nuclei of the universe would also change in proportion to
$\LQCD$.

The cosmic dependence of the strong coupling $\alpha_s(\mu_R;H)$ can be
generalized to the other couplings
$\alpha_i=\alpha_i(\mu_R;H)$\,\cite{{CalmetFritzsch06}}, in particular the
electromagnetic fine structure constant $\alpha_{\rm em}$. In a grand
unified theory (GUT) these couplings converge at the unification point.
Let $d\alpha_i$ be the cosmic variation of $\alpha_i$ with $H$. Each of
the $\alpha_i$ is a function of $\mu_R$, but the expression
$\alpha_i^{-1}\left(d\alpha_i/\alpha_i\right)$ is easily seen to be
independent of $\mu_R$. As a result we can see, using
Eq.\,(\ref{eq:timealphaLQC}), that the running of $\alpha_{\rm em}$ is
related to the corresponding cosmic running of $\LQCD$ as follows:
\begin{equation}\label{eq:timealpha}
\frac{1}{\alpha_{\rm em}}\frac{d\alpha_{\rm
em}(\mu_R;H)}{\,dH}=\frac83\,\frac{\alpha_{\rm
em}(\mu_R;H)/\alpha_s(\mu_R;H)}{\ln{\left(\mu_R/\LQCD\right)}}\,\left[\frac{1}{\LQCD}\,\frac{d{\Lambda}_{\rm
QCD}(H)}{dH}\right]\,.
\end{equation}
At the renormalization point $\mu_R=M_Z$, where both $\alpha_{\rm em}$ and
$\alpha_s$ are well-known, one finds:

\begin{equation}\label{eq:timealpha2}
\frac{1}{\alpha_{\rm em}}\frac{d\alpha_{\rm
em}(\mu_R;H)}{\,dH}\simeq
0.03\left[\frac{1}{\LQCD}\,\frac{d{\Lambda}_{\rm
QCD}(H)}{dH}\right]\,.
\end{equation}
Thus the QCD scale $\LQCD$ runs more than $30$ times faster with the
cosmic expansion than the electromagnetic fine structure constant.
Searching for a cosmic evolution of $\LQCD$ is thereby much easier than
searching for the time variation of $\alpha_{\rm em}$. It is important to
note that the cosmic running properties of the QCD scale parameter could
have nontrivial implications on the physics of the fundamental constants
of Nature. This is sooner said than done: how do we search for a possible
cosmic running of the QCD scale? The answer is: searching for a possible
running of the proton mass, a parameter which has been (and is being)
monitored by many astrophysical and laboratory experiments (or more
specifically, what is traced is the ratio of the proton mass to the
electron mass, where the latter is of course independent of
QCD)\,\cite{FundConst}. Let us take for instance the proton mass
$m_p\simeq 938$ MeV, supposed to be a fundamental constant. It can be
computed from $\LQCD$, the quarks masses and the electromagnetic
contribution. The precise formula reads
\begin{eqnarray}\label{eq:ProtonMass}
m_p &=&c_{\rm QCD}\LQCD+c_u\,m_u+c_d\,m_d+c_s\,m_s+c_{\rm em}\LQCD\nonumber\\
&=&\left(860+21+19+36+2\right)\,{\rm MeV}\,,
\end{eqnarray}
from which we learn that it is largely dominated by the QCD scale, namely
the bulk ($\sim 92\%$) of the proton mass is given by $m_p \simeq c_{\rm
QCD}\LQCD$. It follows that there is a close relation between the possible
cosmic time variation of these two quantities. We will take advantage of
this fact in what follows.

\subsection{Non-conservation of matter in the
universe}\label{sect:Nonconservation of matter}

Let us focus on the impact of the cosmological Models II and III of
Sect.\,\ref{sect:DifferentScenarios} on the non-conservation of matter in
the universe. Recall that we have considered bounds on the ``leakage
parameter'' $\nu$ within the class of these models based on the
non-conservation matter density law (\ref{mRG2}). We must be careful in
interpreting such a non-conservation law. For example, if we take the
baryonic density in the universe, which is essentially the mass density of
protons, we can write $\rM^B=n_p\,m_p$, where $n_p$ is the number density
of protons and $m_p^0=938.272 013(23)$ MeV is the current proton mass. If
this mass density is non-conserved, either $n_p$ does not exactly follow
the normal dilution law with the expansion, i.e. $n_p\sim a^{-3}=(1+z)^3$,
{but} the anomalous law:
\begin{equation}\label{eq:nonconservednumber}
n_p(z)=n_p^0\,(1+z)^{3(1-\nu)}\ \ \ \ ({\rm at\ fixed\ proton\ mass}\ m_p=m_p^0)\,,
\end{equation}
and/or the proton mass $m_p$ does not stay constant with time and
redshifts with the cosmic evolution:
\begin{equation}\label{eq:nonconservedmp}
m_p(z)=m_p^0\,(1+z)^{-3\nu}\ \ \ \ ({\rm with\ normal\ dilution}\ n_p(z)=n_p^0\,(1+z)^3)\,.
\end{equation}
In all cases it is assumed that the vacuum absorbs the difference (i.e.
$\rL=\rL(z)$ ``runs with the expansion''). The first possibility implies
that during the expansion a certain number of particles (protons in this
case) are lost into the vacuum (if $\nu<0$; or ejected from it, if
$\nu>0$), whereas in the second case the number of particles is strictly
conserved. The number density follows the normal dilution law with the
expansion, but the mass of each particle slightly changes (decreases for
$\nu<0$, or increases for $\nu>0$) with the cosmic evolution.

For the following considerations we adopt the second point of view, {i.e.}
Eq.\,(\ref{eq:nonconservedmp}). Being the matter content of the universe
dominated by the dark matter (DM), we cannot exclude that the particle
masses of which is made also vary with cosmic time. Let us denote the mass
of the average DM particle $m_X$, and let $\rX$ and $n_X$ be its mass
density and number density, respectively. The overall matter density of
the universe can be written as follows:
\begin{eqnarray}\label{eq:MdensityUniv}
\rM&=&\rB+\rLep+\rR+\rX=\left(n_p\,m_p+n_n\,m_n\right)+ n_e\,m_e+\rR+n_X\,m_X\nonumber\\
&\simeq& n_p\,m_p+n_n\,m_n+n_X\,m_X\,.
\end{eqnarray}
Here $n_p, n_n, n_e, n_X\, (m_p,m_n,m_e,m_X)$ are the number densities
(and masses) of protons, neutrons, electrons and DM particles. The
baryonic and leptonic parts are $\rB=n_p\,m_p+n_n\,m_n$ and
$\rLep=n_e\,m_e$ respectively. The small ratio $m_e/m_p\simeq 5\times
10^{-4}$ implies that the leptonic contribution to the total mass density
is negligible: $\rLep\ll \rB$. We have also neglected the relativistic
component $\rR$ (photons and neutrinos).

If we assume that the mass change through the cosmic evolution is due to
the time change of $m_p$, $m_n$ and $m_X$, we can compute the mass density
anomaly per unit time, i.e. the deficit or surplus with respect to the
conservation law,  by differentiating (\ref{eq:MdensityUniv}) with respect
to time and subtracting the ordinary (i.e. fixed mass) time dilution of
the number densities. The result is:
\begin{equation}\label{eq:timeMdensityUniv}
\delta\dot{\rho}_M=n_p\,\dot{m}_p+n_n\,\dot{m}_n+n_X\,\dot{m}_X\,.
\end{equation}
The relative time variation of the mass density anomaly can be estimated
as follows:
\begin{equation}\label{eq:reltimeMdensityUniv1}
\frac{\delta\dot{\rho}_M}{\rM}=\frac{n_p\,\dot{m}_p+n_n\,\dot{m}_n+n_X\,\dot{m}_X}{n_p\,m_p+n_p\,m_p+n_X\,m_X}\simeq
\frac{n_p\,\dot{m}_p+n_n\,\dot{m}_n+n_X\,\dot{m}_X}{n_X\,m_X}\,\left(1-\frac{n_p\,m_p+n_n\,m_n}{n_X\,m_X}\right)\,.
\end{equation}
The current normalized DM density $\ODMo=\rX/\rc\simeq 0.23$ is
significantly larger than the corresponding normalized baryon density
$\OMBo=\rB/\rc\simeq 0.04$. Therefore $n_X\,m_X$ is larger than
$n_p\,m_p+n_n\,m_n$  by the same amount. If we assume
$\dot{m}_n=\dot{m}_p$, we find approximately:
\begin{equation}\label{eq:reltimeMdensityUniv2}
\frac{\delta\dot{\rho}_M}{\rM}=\frac{n_p\,\dot{m}_p}{n_X\,m_X}\,\left(1+\frac{n_n}{n_p}-\frac{\OMB}{\ODM}\right)+
\frac{\dot{m}_X}{m_X}\left(1-\frac{\OMB}{\ODM}\right)\,.
\end{equation}
In the approximation $m_n=m_p\,$ we can rewrite the prefactor on the
\textit{r.h.s} of Eq.\,(\ref{eq:reltimeMdensityUniv2}) as follows:
\begin{equation}\label{eq:prefactor}
\frac{n_p\,\dot{m}_p}{n_X\,m_X}=\frac{\OMB}{\ODM}\,\frac{\dot{m}_p}{m_p}\left(1-\frac{n_n/n_p}{1+n_n/n_p}\right)\simeq
\frac{\OMB}{\ODM}\,\frac{\dot{m}_p}{m_p}\left(1-\frac{n_n}{n_p}\right)\,.
\end{equation}
The ratio $n_n/n_p$ is of order $10\%$ after the primordial
nucleosynthesis. Since $\OMB/\ODM$ is also of order $10\%$, we can neglect
the product of this term with $n_n/n_p$\,. When we insert the previous
equation into (\ref{eq:reltimeMdensityUniv2}), the two $n_n/n_p$
contributions cancel each other. The expression ${1-\OMB}/{\ODM}$
factorizes in the two terms on the \textit{r.h.s} of
Eq.\,(\ref{eq:reltimeMdensityUniv2}). The final result is:
\begin{equation}\label{eq:reltimeMdensityUniv}
\left(1-\frac{\OMB}{\ODM}\right)^{-1}\frac{\delta\dot{\rho}_M}{\rM}=\frac{\OMB}{\ODM}\,\frac{\dot{m}_p}{m_p}+\frac{\dot{m}_X}{m_X}
=\frac{\OMB}{\ODM}\,\frac{\dot{\Lambda}_{\rm QCD}}{\LQCD}+\frac{\dot{m}_X}{m_X}\,.
\end{equation}
As mentioned, we have approximated $m_p\simeq c_{\rm QCD}\,\LQCD$.

The expression ${\delta\dot{\rho}_M}/{\rM}$ in
Eq.\,(\ref{eq:reltimeMdensityUniv}) must be the same as the one we have
computed in (\ref{eq:deltadotrho}), if we consider the models based on the
generic matter non-conservation law (\ref{mRG2}). Therefore the two
expressions should be equal, and we obtain approximately:
\begin{equation}\label{eq:reltimeMdensityUniv3}
3\nueff\,H=\frac{\OMB}{\ODM}\,\frac{\dot{\Lambda}_{\rm QCD}}{\LQCD}+\frac{\dot{m}_X}{m_X}\,,
\end{equation}
where we have defined
\begin{equation}\label{eq:nueff}
\nueff=\frac{\nu}{1-\OMB/\ODM}\,,
\end{equation}
and numerically $\nueff\simeq 1.2\,\nu$. The differential equation
(\ref{eq:reltimeMdensityUniv3}) describes approximately the relationship
between the matter non-conservation law (\ref{mRG2}), the evolution of the
vacuum energy density $\rL$ (and/or $G$) and the time variation of the
nuclear and particle physics quantities.

\subsection{Cosmic running of the QCD scale and of the baryonic matter}\label{sect:CosmicRunningLQCDBaryons}

Various scenarios are possible. Suppose that the dark matter particles do
not vary with time, i.e. $\dot{m}_X=0$, and only the cosmic evolution of
$\LQCD$ accounts for the non-conservation of matter. Trading the cosmic
time for the scale factor through $\dot{\Lambda}_{\rm
QCD}=\left(d{\Lambda}_{\rm QCD}/da\right)\,a\,H$ and integrating the
resulting equation, we can express the final result in terms of the
redshift:

\begin{equation}\label{eq:LQCDz}
\LQCD(z)=\LQCD^0\,\left(1+z\right)^{-3\,(\ODMo/\OMBo)\,\nueff}\,.
\end{equation}
Since the contribution of the quark masses $m_u$,$m_d$ and $m_s$ to the
proton mass is small -- cf. Eq.\,(\ref{eq:ProtonMass}) -- we can
approximate the proton mass by $m_p\simeq c_{\rm QCD}\,\LQCD$. Therefore,
for the protons we have
\begin{equation}\label{eq:mN}
m_p(z)=m_p^0\,\left(1+z\right)^{-3\,(\ODMo/\OMBo)\,\nueff}\,.
\end{equation}
Here $\LQCD^0$ and $m_p^0$ are the QCD scale and proton mass at present
($z=0$); $\ODMo$ and $\OMBo$ being the current values of these
cosmological parameters.

The presence of the factor ${\OMBo}/{\ODMo}$ in the power law makes eq.
(\ref{eq:mN}) more realistic than eq. (\ref{eq:nonconservedmp}). In the
case $\nu=0$ the QCD scale and the proton mass would not vary with the
expansion of the universe, but for non-vanishing $\nu$ it describes the
cosmic running of $\LQCD=\LQCD(z)$ and $m_p=m_p(z)$. For $\nu>0$ ($\nu<0$)
the QCD scale and proton mass decrease (increase) with the redshift. This
is consistent, since for $\nu>0$ ($\nu<0$) the vacuum energy density is
increasing (decreasing) with the redshift -- cf. Eq.\,(\ref{CRG2}) --, and
it is smaller (larger) now than in the past.

We can write down the variation of the QCD scale in terms of the Hubble
rate $H$. With the help of Eq.\,(\ref{nomalHflow}) it is easy to see that
Eq.\,(\ref{eq:LQCDz}) can be turned into an expression for $\LQCD$ given
explicitly in terms of the primary cosmic variable $H$:
\begin{equation}\label{eq:LQCDH}
\LQCD(H)=\LQCD^0\,\left[\frac{1-\nu}{\OMo}\,\frac{H^2}{H_0^2}-\frac{\OLo-\nu}{\OMo}\right]^{-(\ODMo/\OMBo)\,\nueff/(1-\nu)}\,,
\end{equation}
with $\OMo=\OMBo+\ODMo$. $\nu$ and $\nueff$ are involved in
(\ref{eq:LQCDH}), since they come from different sources. This equation
satisfies the normalization condition $\LQCD(H_0)=\LQCD^0$ due to the
cosmic sum rule for flat space: $\OMo+\OLo=1$.

{Obviously the cosmic time variation of the $\LQCD$ scale is very small in
our framework. This can be more easily assessed if we use
Eqs.\,(\ref{eq:LQCDz}) and (\ref{eq:LQCDH}) to compute the relative time
variation of the QCD scale with respect to the present value. Since $\nu$
is small it it easy to show that}
\begin{equation}\label{eq:deltaLQCDH}
\frac{\LQCD(z)-\LQCD^0}{\LQCD^0}=-\frac{\ODMo}{\OMBo}\,\frac{\nueff}{1-\nu}
\ln\left[\frac{1-\nu}{\OMo}\,\frac{H^2(z)}{H_0^2}-\frac{\OLo-\nu}{\OMo}\right]\,.
\end{equation}
{As a concrete example, let us consider the studies made in
Ref.\,\cite{Reinhold06} on comparing the $H_2$ spectral Lyman and Werner
lines observed in the Q 0347-383 and Q 0405-443 quasar absorption systems.
The comparison with the corresponding spectral lines at present may be
sensitive to a possible evolution of these lines in the last twelve
billion years and involves redshifts in the range $z\simeq 2.6-3.0$. A
positive result could be interpreted as a small variation of the proton to
electron mass ratio  between two widely separated epochs of the
cosmological evolution\,\cite{Reinhold06} . Assuming that $|\nu|={\cal
O}(10^{-3})$, as suggested by Eq.\,(\ref{eq:numodeliicosmology}), it
follows from the previous formulae that the relative variation of $\LQCD$
in this lengthy time interval is only at the few percent level with
respect to its present day value. From Eq.\,(\ref{CRG2}) we can then
easily check that the corresponding variation of $\rL(z)$ with respect to
the current value $\rLo$ is also of a few percent. As expected, the two
scales undergo tiny variations over very long periods of time, in fact
cosmological periods, and therefore the large hierarchy between them at
present -- namely $\LQCD={\cal O}(100)$ MeV$={\cal O}(10^8)$ eV and
$\rL^{1/4}={\cal O}(10^{-3})$ eV -- is essentially preserved over the
cosmological evolution. However, even this small crosstalk between these
two widely separated scales could be sufficient for being eventually
detected by the aforementioned high precision experiments aiming at
measuring very tiny variation of the proton to electron mass ratio.

Indeed, this is suggested by the fact that the expected range of values of
$\nu$ is within the scope of the precision of these experiments. }
Consider the state of the art in the current laboratory tests, using
atomic clocks. According to our estimate (\ref{eq:timealpha2}), the
largest effect is expected to be a cosmological {redshift} ({hence time
variation}) of the nucleon mass, which can be observed by monitoring
molecular frequencies. These are precise experiments in quantum optics,
e.g. obtained by comparing a cesium clock with 1S-2S hydrogen transitions.
In a cesium clock the time is measured by using a hyperfine
transition\,\footnote{Recall that the cesium hyperfine clock provides the
modern definition of time. In SI units, the second is defined to be the
duration of $9.192631770\times 10^{9}$ periods of the transition between
the two hyperfine levels of the ground state of the $^{133}$Cs atom}.
Since the frequency of the clock depends on the magnetic moment of the
cesium nucleus, a possible variation of the latter is proportional to a
possible variation of $\LQCD$. A hyperfine splitting is a function of
$Z\,\alpha_{\rm em}$ ($Z$ being the atomic number) and is proportional to
$Z\,\alpha_{\rm em}^2(\mu_N/\mu_B)(m_e/m_p)\,R_{\infty}$, where
$R_{\infty}$ is the Rydberg constant, $\mu_N$ is the nuclear magnetic
moment and $\mu_B=e\hbar/2m_pc$ is the nuclear magneton. We have
$\dot{\mu}_N/\mu_N\propto -\dot{\Lambda}_{\rm QCD}/\LQCD$. The hydrogen
transitions are only dependent on the electron mass, which we assume to be
constant. The comparison over a period of time between the cesium clock
with hydrogen transitions provides an atomic laboratory measurement of the
ratio $\mu_{\rm pe}\equiv m_p/m_e$. The most sophisticated atomic clock
experiments aim soon to reach a sensitivity limit of
%
%\begin{equation}\label{eq:timeLambda}
$\left|{\dot{\Lambda}_{\rm QCD}}/{\LQCD}\right|<10^{-14}\,{\rm
yr}^{-1}$\,\,\cite{FritzschSola2012}.
%\end{equation}
%
Since the proton mass is given essentially by $\LQCD$, as indicated by
Eq.\,(\ref{eq:ProtonMass}), we have $\dot{m}_p\simeq
c_{\LQCD}\,\dot{\Lambda}_{\rm QCD}$ and the corresponding time variation
of the ratio of the proton mass would be:
\begin{equation}\label{eq:timeLambda2}
\left|\frac{\dot{\mu}_{pe}}{\mupe}\right|=\left|\frac{\dot{m}_p}{m_p}\right|\simeq
\left|\frac{\dot{\Lambda}_{\rm QCD}}{\LQCD}\right|<
10^{-14}\,{\rm yr}^{-1}\,.
\end{equation}
The atomic clock result (\ref{eq:timeLambda2}) would indicate a time
variation of the ratio $\mupe$, which is consistent (in absolute value)
with the astrophysical measurements\,\cite{Reinhold06}.

Using the above equations and Eq.\,(\ref{alphasLQCD}), we can obtain the
corresponding evolution of the strong coupling constant $\alpha_s$ with
the redshift:
\begin{equation}\label{eq:alphasz}
\frac{1}{\alpha_s(\mu_R;z)}=\frac{1}{\alpha_s(\mu_R;0)}+6\,b_0\,
\frac{\ODMo}{\OMBo}\,\nueff\,\ln{(1+z)}\,,
\end{equation}
where $\alpha_s(\mu_R;0)$ is the value of $\alpha_s(\mu_R;z)$ today
($z=0$). Since $b_0>0$, we observe that for $\nu>0$ ($\nu<0$) the strong
interaction $\alpha_s(\mu_R;z)$ decreases (increases) with $z$, i.e. with
the cosmic evolution. The corresponding variation of the strong coupling
with the Hubble rate can also be determined \,\footnote{We point out that
a similar running of the strong coupling constant with the cosmic
expansion was pointed out in a different context by J.D. Bjorken in
\cite{Bjorken2002}.}:
\begin{equation}\label{eq:alphasH}
\frac{1}{\alpha_s(\mu_R;H)}=\frac{1}{\alpha_s(\mu_R;H_0)}+2\,b_0\,
\frac{\ODMo}{\OMBo}\,\frac{\nueff}{1-\nu}\,\ln{\left[\frac{1-\nu}{\OMo}\,\frac{H^2}{H_0^2}-\frac{\OLo-\nu}{\OMo}\right]}\,.
\end{equation}
Here $\alpha_s(\mu_R;H_0)$ is the current value of $\alpha_s(\mu_R;H)$.

Remarkably, the above expression displays the logarithmic running of the
strong coupling as a function of \textit{two} energy scales: one is the
ordinary QCD running scale $\mu_R$, the other is the cosmic scale defined
by the Hubble rate $H$, which has dimension of energy in natural units.
The second term on the \textit{r.h.s.} depends on the product of the two
$\beta$-function coefficients, the one for the ordinary QCD running
($b_0$) and the one for the cosmic running ($\nu\propto\nueff$).

The following remarks are in order:
\begin{description}

\item i) for $\nu=0$ there is no cosmic running of the strong
    interaction,

\item ii) for $\nu>0$ the strong coupling $\alpha_s(\mu_R;H)$ is
    ``doubly asymptotically
free''. It decreases for large $\mu_R$ and also for large $H$, whereas
for $\nu<0$ the cosmic evolution drives the running of $\alpha_s$
opposite to the normal QCD running,

\item iii) the velocity of the two runnings is very different, because
    $H$ is slowly varying with time and $|\nu|\ll1$ and $|\nu|\ll
    b_0\lesssim 1$. The cosmic running only operates in the cosmic
    history and is
weighed with a very small $\beta$-function. But it may soon be
measured in the experiments with atomic clocks and through
astrophysical observations.
 \end{description}

We should not overlook the fact that the previous equations describe not
only the leading cosmic evolution of the QCD scale and the proton mass
with the redshift and the expansion rate $H$ of the universe, but they can
account for the redshift evolution of the nuclear masses. For the neutron
we can write approximately: $m_n\simeq c_{\rm QCD}\,\LQCD$. For an atomic
nucleus of current mass $M_A$ {and atomic number $A$} we have
$M_A=Z\,m_p+(A-Z)\,m_n-B_A$, where $Z$ is the number of protons and $A-Z$
the number of {neutrons}, and $B_A$ is the binding energy. {Although
$B_A$} may also change with the cosmic evolution, the shift should be less
significant, since at leading order the binding energy relies on pion
exchange among the nucleons. The pion mass has a softer dependence on
$\LQCD$: $m_{\pi}\sim \sqrt{m_q\,\LQCD}$, due to the chiral symmetry.

In the previous approximations we have neglected the light quark masses
$m_q$. We can assume that the binding energy has a negligible cosmic shift
as compared to the masses of the nucleons. In the limit where we neglect
the proton-neutron mass difference and assume a common nucleon mass
$m_N^0$ at present, the corresponding mass of the atomic nucleus at
redshift $z$ is given at leading order by:
\begin{equation}\label{eq:MAz}
M_A(z)\simeq A\,m_N^0\,\left(1+z\right)^{-3\,(\ODMo/\OMBo)\,\nueff}-B_A\,.
\end{equation}
Although the chemical elements redshift their masses, a disappearance or
overproduction of nuclear mass (depending on the sign of $\nu$) is
compensated by a running of the vacuum energy $\rL$, which is of opposite
in sign, see (\ref{eq:deltaLambda}).

\subsection{Discovering dark matter from the running of vacuum energy}\label{sect:DiscoveringDM}

Above we have described a simplified case, in which the nuclear matter
evolves with the cosmic evolution as a result of the evolution of the
fundamental QCD scale. In this scenario the light quark masses are
neglected, and the DM does not participate in the cosmic time evolution.

Alternatively we can assume that the nuclear matter does not vary with
time, i.e. $\dot\Lambda_{\rm QCD}=0$, and only the DM particles account
for the non-conservation of matter. In general we expect a mixed
situation, in which the temporal rates of change for nuclear matter and
for DM particles are different:
\begin{equation}\label{eq:rateNandDM}
\frac{\dot{\Lambda}_{\rm QCD}}{\LQCD}=3\,\nuQCD\,H\,,\ \ \ \ \ \ \ \ \frac{\dot{m}_{X}}{m_X}=3\,\nuX\,H\,.
\end{equation}
We have defined the QCD time variation index, which is characteristic of
the redshift rate of the QCD scale, while $\nuX$ is the corresponding one
for the DM. In this more general case we find:
\begin{equation}\label{eq:LQCDmxDz}
\LQCD(z)=\LQCD^0\,\left(1+z\right)^{-3\,\nuQCD}\,,\ \ \ \ \ \ \ \ m_X(z)=m_X^0\,\left(1+z\right)^{-3\,\,\nuX}\,.
\end{equation}
We introduce the effective baryonic redshift index $\nuB$:
\begin{equation}\label{eq:nuB}
\nuB=\frac{\OMB}{\ODM}\,\nuQCD\,.
\end{equation}
The equations (\ref{eq:LQCDmxDz}) satisfy the relation
(\ref{eq:reltimeMdensityUniv3}), provided the coefficients $\nuB$ and
$\nuX$ are related by
\begin{equation}\label{eq:nunuQCDnuX}
\nueff=\nuB+\nuX\,.
\end{equation}
$\nuQCD$ is the intrinsic cosmic rate of variation of the strongly
interacting particles. The effective index $\nu_B$ weighs the redshift
rate of these particles taking into account their relative abundance with
respect to the DM particles. Even if the intrinsic cosmic rate of
variation of $\LQCD$ would be similar to the DM index  (i.e. if
$\nuQCD\gtrsim \nuX)$, the baryonic index (\ref{eq:nuB}) would still be
suppressed with respect to $\nuX$, because the total amount of baryon
matter in the universe is much smaller than the total amount of DM.

In this mixed scenario the mass redshift of the dark matter particles
follows a similar law as in the case of protons (\ref{eq:mN}), except  now
we have $\nueff\rightarrow\nuB$. The proton would have the index $\nuQCD$
characteristic of the free (and bound) stable strongly interacting matter:
\begin{equation}\label{eq:mN2}
m_p(z)=m_p^0\,\left(1+z\right)^{-3\,(\ODM/\OMB)\,\nuB}=m_p^0\,\left(1+z\right)^{-3\nuQCD}\,.
\end{equation}
The DM particles have another independent index $\nuX$. The sum
(\ref{eq:nunuQCDnuX}) must reproduce the original index
$\nueff\propto\nu$, which we associated with the non-conservation of
matter.

Finally we consider the possible quantitative contribution to the matter
density anomaly from the dark matter. The global mass defect (or surplus)
is regulated by the value of the $\nu$ parameter, but the contribution of
each part (baryonic matter and DM) depends on the values of the individual
components $\nuB$ and $\nuX$. We can obtain a numerical estimate of these
parameters by setting the expression (\ref{eq:reltimeMdensityUniv}) equal
to (\ref{eq:deltadotrho2}). The latter refers to the time variation of the
matter density $\rM$ without tracking the particular way in which the
cosmic evolution can generate an anomaly in the matter conservation. The
former does assume that this anomaly is entirely due to a cosmic shift in
the masses of the stable particles. Taking the absolute values, we obtain:
\begin{equation}\label{eq:equatetwodensit}
3|\nueff|\,H\simeq \left|\frac{4}{23}\,\frac{\dot{\Lambda}_{\rm QCD}}{\LQCD}+
\frac{\dot{m}_X}{m_X}\right|< \frac{4}{23}\times 10^{-14}\,{\rm yr}^{-1}+\left|\frac{\dot{m}_X}{m_X}\right|\,.
\end{equation}
Here we have used the experimental bound (\ref{eq:timeLambda2}) on the
time variation of $\LQCD$.

Several cases can be considered, depending on the relation between the
intrinsic cosmic rates of variation of the strongly interacting particles
and DM particles, $\nuQCD$ and $\nuX$. Since these indices can have either
sign, we shall compare their absolute values:

\begin{itemize}

\item 1) $|\nuX|\ll|\nuB|$:

This condition implies $|\nuX|\ll|\nuQCD|$. By demanding the stronger
condition $|\nuX|\ll|\nuB|$, we insure that the intrinsic QCD cosmic
rate $|\nuQCD|$ is much larger than the corresponding DM rate
$|\nuX|$. We can neglect the $\dot{m}_X/m_X$ term on the
\textit{r.h.s.} of (\ref{eq:equatetwodensit}), and we recover the
equations (\ref{eq:LQCDz})-(\ref{eq:alphasH}) with $\nueff\simeq\nuB$.
Using $H_0\simeq 7\times 10^{-11}\,{\rm yr}^{-1}$\,, we find:
\begin{equation}\label{eq:scenario1a}
|\nuX|\simeq 0\,,\ \ \ \ \ |\nueff|\simeq|\nuB|< 10^{-5}\,,\ \ \ \ |\nuQCD|<5\times 10^{-5}\,.
\end{equation}

\item 2) $|\nuX|\simeq|\nuB|$:

Here we still have $|\nuX|$ smaller than $|\nuQCD|$, but the
requirement is weaker. It follows: $|\nueff|\simeq2|\nuX|\simeq
2|\nuB|=2(\OMB/\ODM)\,|\nuQCD|$, and we find
\begin{equation}\label{eq:scenario2}
|\nueff|<2\times 10^{-5}\,,\ \ \ \ \ |\nuX|\simeq |\nuB|<10^{-5}\,,\ \ \ \ \ \ \ \ |\nuQCD|<5\times 10^{-5}\,.
\end{equation}
\item 3) $|\nuX|\simeq|\nuQCD|$:

The two intrinsic cosmic rates for strongly interacting and DM
    particles are similar, i.e. $\dot{\Lambda}_{\rm QCD}/\LQCD$ and
    $\dot{m}_X/m_X$ do not differ significantly.  In this case
 Eq.\,(\ref{eq:equatetwodensit}) leads to
\begin{equation}\label{eq:scenario2b}
3|\nueff|\,H<\left(\frac{4}{23}+ 1\right)\times 10^{-14}\,{\rm yr}^{-1}\,.
\end{equation}
There are two sign possibilities ($\nuQCD=\pm\nuX$), and we take the
absolute value:
\begin{equation}\label{eq:nuQCDeqnuX}
|\nueff|\lesssim\left(\frac{\OMB}{\ODM}+ 1\right)\,|\nuQCD|\simeq |\nuQCD|\,.
\end{equation}
We find:
\begin{equation}\label{eq:scenario2c}
|\nueff|\lesssim|\nuQCD|\simeq |\nuX|<5\times 10^{-5}\,.
\end{equation}
\item 4) $|\nuQCD|\ll|\nuX|$:

Here the nuclear part is frozen. The non-conservation of matter is
entirely due to the time variation of the DM particles.
Eq.\,(\ref{eq:equatetwodensit}) gives:
\begin{equation}\label{eq:scenario3}
3\nu\,H\simeq \frac{\dot{m}_X}{m_X}\,\left(1-\frac{\OMB}{\ODM}\right)\,.
\end{equation}
\begin{table*}[t]
\tabcolsep 3pt \vspace {0.2cm}
\begin{center}
\begin{tabular}{|c||l|c|c|}
  \hline
   &\phantom{XXXXX} $|\nu|^{\rm cosm}$ & $|\nu|^{\rm lab}=|\nuB|$ & $|\nuX|^{\rm cosm}$ \\ \hline\hline
  {\rm Model II} & $10^{-3}$ (SNIa+BAO+CMB)& $10^{-5}$ (Atomic clocks+Astrophys.) & $10^{-3}$ \\ \hline
   {\rm Model III} & $10^{-3}$ (BBN)& $10^{-5}$ (Atomic clocks+Astrophys.) & $10^{-3}$ \\ \hline
   {\rm Model IV} & $10^{-3}$ (SNIa+BAO+CMB)+BBN& 0 & 0 \\
  \hline
\end{tabular}
\caption[]{Upper bounds on the basic parameter $|\nu|$ for the various
models defined in sect.\,\ref{sect:DifferentScenarios} -- cf.
\,\cite{FritzschSola2012}. Only for Models II and III a non-vanishing
value of $|\nu|$ is related to non-conservation of matter and a
corresponding time evolution of $\rL$ and $G$, respectively. For these
models, the baryonic part of $\nu$ (denoted $\nuB$) can be accessible
to accurate lab experiments -- cf. Eq.\,(\ref{eq:timeLambda2}) --
whereas the DM part ($\nuX$) can only be bound indirectly from
cosmological observations (same cosmological bound as for the overall
$\nu$). For Model IV matter is conserved, and a non-vanishing value of
$|\nu|$ (only accessible from pure cosmological observations) is
associated to a simultaneous time evolution of $\rL$ and $G$ -- with
no microphysical implications.}
\end{center}
\end{table*}
We have written this expression directly in terms of the original
$\nu$ parameter. In this case we cannot get information from any
laboratory experiment on $\dot{m}_X/m_X$, but we do have independent
experimental information on the original $\nu$ value (irrespective of
the particular contributions form the nuclear and DM components). It
comes from the cosmological data on type Ia supernovae, BAO, CMB and
structure formation. The analysis of this
data\,\cite{BPS09,BasPolarSola12} leads to the bound
(\ref{eq:numodeliicosmology}), which applies to all models, in which
matter follows the generic non-conservation law (\ref{mRG2}) and the
running vacuum law (\ref{RGlaw2}) --- or the same matter
non-conservation law and the running gravitational coupling law
(\ref{GzfixedLsmallnu}), as shown in Eq.\,(\ref{eq:nuboundGvariable}).
Since it depends on the cosmological effects from all forms of matter,
it applies to the DM particles in particular. We find:
\begin{equation}\label{eq:scenario3b}
|\nuX|^{\rm cosm}\lesssim 10^{-3}\,.
\end{equation}
This bound is significantly weaker than any of the bounds found for
the previous scenarios in which the nuclear matter participated of the
cosmic time variation. It cannot be excluded that the matter
non-conservation and corresponding running of the vacuum energy in the
universe is mainly caused by the general redshift of the DM particles.
In this case only cosmological experiments could be used to check this
possibility. If the nuclear matter also participates in a significant
way, it could be analyzed with the help of experiments in the
laboratory. For a summary of the bounds, see Table 1.

\end{itemize}

If in the future we could obtain a tight cosmological bound on the
effective $\nueff$-parameter (\ref{eq:nunuQCDnuX}), using the
astrophysical data, and an accurate laboratory (and/or astrophysical)
bound on the baryonic matter part $\nuB$, we could compare them and derive
the value of the DM component $\nuX$. If $\nueff$ and $\nuB$ would be
about equal, we should conclude that the DM particles do not appreciably
shift their masses with the cosmic evolution, or that they do not exist.
If, in contrast, the fractional difference $|\,(\nueff-\nuB)/\nueff\,|$
would be significant, the DM particles should exist to compensate for it.

\newpage

\section{Final reflections, a few remarks and some conclusions}

In this summarized review, we have discussed a few old and new aspects of
the $\CC$-term in Einstein's equations and its relation with the vacuum
energy. After some brief historical remarks, we have dealt with the
cosmological constant (CC) problem and assessed its significance and
possible phenomenological implications. In doing so we have focused on the
notion of vacuum energy in quantum field theory (QFT) both in flat and
curved spacetime, and we have dwelled upon the fact that the term which is
usually interpreted as the vacuum energy density (originating from the
quantum vacuum fluctuations of the matter fields) is ostensibly the same
both in flat and curved spacetime. This is a well-known fact, but this
does not make it less disquieting, since it implies that particles with
mass $m$ provide huge quartic contributions to the zero-point energy
(ZPE), $\ZPE\sim m^4$, and these do not seem to be acceptable from the
phenomenological point of view.

\subsection{ZPE and SSB: two sources of vacuum energy}

In the SM of particle physics there are other, no less preoccupying,
contributions to the vacuum energy that come from the spontaneous symmetry
breaking (SSB) of the electroweak gauge symmetry, specifically from the
Higgs mechanism. These contributions are associated to the vacuum
expectation value of the Higgs potential and increase as $\langle
V\rangle\sim v^2\,M_{\cal H}^2\sim 10^8$ GeV$^4$, where $v={\cal O}(100)$
GeV is the vacuum expectation value that defines the electroweak (EW)
scale, and $M_{\cal H}\simeq 125$ GeV is the presumed physical mass of the
Higgs boson\,\cite{HiggsDiscovery1}. While every single vacuum energy
source alone is already vastly worrisome, the combination of them all
amounts to a devastating fine tuning problem, which is further aggravated
at the quantum level when we consider the many higher loop effects
involved. We have illustrated quite vividly this fact in
Sect.\,\ref{sect:effHiggsFLAT} by considering what are the highest loop
diagrams still contributing to the CC value (and to the awful fine tuning
process).

Of the two main sources of difficulties with the vacuum energy in QFT,
namely the ZPE and the SSB, the reality of the former has been disputed
sometimes by the inconclusive interpretations about the origin of the
Casimir effect as being a pure QFT vacuum effect or something else;
whereas the latter also remained in the limbo of the theoretical ideas as
long as the Higgs mechanism could not be fully substantiated on the
experimental side. Because of this the CC problem could remain dormant for
a long time in the ethereal world of the theoretical conundrums; these are
a kind of problems whose dangerousness and threatening power on the
physical world of mortals is only potential, not yet factual. Such
situation, however, may have given signs of changing quite dramatically in
recent times. Needless to say, not because the physical world changes an
inch every time the human knowledge gives a jerk in its perception of the
reality, but because we might now be reaching a situation where the two
giant paradigms of the fundamental physics knowledge, viz. the SM of
particle physics and the SM of cosmology, have finally been put furiously
face to face for a very serious parley. Indeed, we have recently heard of
the exciting news from CERN about the $\sim 5\sigma$ (and increasing)
evidence on the discovery of a bosonic Higgs-like resonance at the LHC
collider\,\cite{HiggsDiscovery1}. If fully confirmed, this can be
considered as one of the greatest triumphs of particle physics ever. But,
at the same time, we should not overlook the fact that such discovery
should be certifying the very existence of the EW vacuum energy, with all
its woes and potentially distressing consequences for (theoretical)
cosmology.

On the other hand, we should not forget that apart from the share from the
Higgs vacuum energy we have other sorts of vacuum fluctuations in the
electroweak domain that are perfectly ``alive and kicking'' since long
ago. Recall that quantum corrections to high precision EW observables have
reached a level of certainty that is beyond any possible doubt. A simple
example should suffice. Take the famous $\Delta r$ parameter from
electroweak theory. This is the famous parameter that allows to compute
the $W^{\pm}$ gauge boson mass, $M_W$, in the on-shell renormalization
scheme with quantum precision, i.e. including the quantum output from
radiative corrections\,\footnote{The $\Delta r$ parameter corrects the
\textit{r.h.s.} of  Eq. (\ref{eq:GF}) at the quantum
level\,\cite{Hollik95,HollikDetar}. See also Ref.\cite{DavidJoan2013a},
and references therein, for detailed contextual explanations and for a
comprehensive and updated study of that important electroweak parameter
within the SM and beyond.}. We may ask ourselves, to which extent we can
attest that the ``genuine'' EW quantum effects (i.e. those beyond the pure
QED running of $\alpha_{\rm em}$) have been measured when we compare the
theoretical value of $M_W$ and the experimentally measured one. To ask
(and answer!) this question is important since the EW radiative
corrections are a manifestation of the properties of the electroweak
vacuum at the quantum level, namely the same vacuum that is supported by
the likely existence of the Higgs particle. The answer is known, and is
astounding: we know unmistakably that they are there with a confidence
level of $\sim 26\sigma$!\,\cite{Sirlin2012}. The formerly proclaimed
$\sim 5\sigma$ evidence of the Higgs-like particle now pales in
comparison! But the latter kind evidence is direct whereas the former is
indirect, so when we put both the direct and indirect signatures together
they dramatically reinforce the case for the EW vacuum energy; and,
overall, the overwhelming evidence of the electroweak vacuum becomes even
more defiant and challenging for cosmology.

\subsection{Running vacuum and CC problem}

Somehow time has come to try to find a final solution which comes to grips
with the basic notion of vacuum energy in QFT, rather than constantly
eschewing the issue in an almost non-denumerable number of ways. Although
the problem is huge, and far from being solved, some avenues for its
eventual solution might be looming in the horizon. We have discussed a
possible reinterpretation of the results obtained in the calculation of
the vacuum energy density $\rL$  in QFT in curved spacetime, and suggested
that although the resolution of the CC problem cannot be addressed at
present from a rigorous computation of $\rL$  in an expanding FLRW
universe, at least some consistency relations seem to hint at the possible
form of the correct dynamical dependence of that important quantity as a
function of the Hubble rate. For example, if both the vacuum energy and
the gravitational coupling are evolving with the Hubble rate, $H$, the
Bianchi identity leads to the possible form for the running of the vacuum
density $\rL=\rL(H)$, given the logarithmic running suggested for $G=G(H)$
in the expanding spacetime. As we have discussed in
Sect.\,\ref{sect:VacEnerCurvature}, one is led to the following behavior
for the low energy regime: $\rL(H)=c_0+\beta\,M_P^2\,H^2$, where $M_P$ is
the Planck mass, $c_0$ a constant (close to the current CC density value
$\rLo$) and $\beta$ is a dimensionless coefficient parameterizing the CC
running. This form for the evolving vacuum energy density is perfectly
tenable and has been profusely tested against the latest cosmological
data\,\cite{BPS09,GSBP11,BasPolarSola12} -- cf. also
\cite{SolaReview2011,SolaReview05}.  This could eventually provide a
phenomenological evidence in favor of the vacuum energy as being a serious
candidate for dynamical dark energy (DE). Of special significance for
testing these ideas is the study of the effective equation of state of the
DE, and its possible time variation\,\cite{BasSola13a,SS05}.

The idea of a running vacuum energy in an expanding universe, i.e.
$\rL=\rL(H)$, appears as a most natural one. It would be very
difficult to admit the existence of a tiny and immutable constant
from the early times to the present days, maybe even ruling the
entire future evolution of the universe. Nonetheless, the ultimate
value that $\rL(H)$ takes at present, i.e. $\rLo$, cannot be
predicted within these models and hence can only be extracted from
observations. Notice that if we would have the ability to predict
this value it would be tantamount to solve the old CC
problem\,\cite{CCPWeinberg}. This is of course the toughest part of
the job. In our discussion, however, the running vacuum paradigm
ascribes a new look to the problem, one that could perhaps make it
more amenable for an eventual solution; namely it conceives the
cosmological term as a time evolving variable that underwent a
dramatic reduction from the inflationary time till the present days.

\subsection{Running vacuum and inflation}

A generic model that implements this idea is the following:
$\rL(H)=c_0+c_2\,H^2+c_4\, H^4$ -- cf. Sect.\ref{sect:EarlyUniverse} -- in
which the highest power of the Hubble rate, $H^4$, would only be relevant
for the early universe, i.e. for values of $H$ of order of the
inflationary expansion rate $H_I$. Models of this kind can be motivated by
the general running vacuum framework that we have described here, and they
should have a real chance to provide a complete description of the cosmic
history\,\cite{LimaBasSola13a,HMention2013,Perico2013a}. In contrast to
the standard $\CC$CDM model, in which the two opposite poles of the cosmic
history (inflation and DE) are completely disconnected, the running vacuum
models offer a clue for interconnecting them and let the present DE appear
as a ``quantum fossil'' from the inflationary universe \cite{Fossil07}. In
this way the two de Sitter epochs, viz. the primordial inflationary one
and the late DE epoch, can be thought of as two vacuum dominated stages of
the cosmic evolution smoothly interpolated (within a single unified model)
by the radiation and matter dominated epochs.

The aforesaid power series model for $\rL(H)$ is attractive in that the
vacuum undergoes a significant dynamical evolution together with matter.
It first triggers inflation through $\sim H^4$, while $H$ is comparable to
$H_I$ (the inflationary expansion rate, which is a solution of the
equations); for $H<H_I$ it transits automatically (i.e. ``gracefully
exits'') into radiation and remains subdominant, thus unharmful to
primordial nucleosynthesis and to any other momentous stage of the cosmic
evolution.  Beyond this point, it behaves as a mild ``affine'' quadratic
function of the Hubble rate: $\rL(H)=c_0+c_2\,H^2$. The vacuum energy
density keeps falling (staying below the matter density), until the final
de Sitter phase is achieved. This phase is triggered by the $c_0$ term,
which appears in a natural way in the renormalization group (RG)
formulation.

Obviously this picture gives us an insight into why the present vacuum
energy density has to be small as compared to its primordial value at the
inflationary scale, i.e. $\rLo\ll\rL(H_I)$. It also suggests that there is
no a priori reason for being able to predict the value of $c_0$, and hence
of the current CC value either. In the RG framework we have considered
here, this parameter should rather be viewed as an experimental input tied
to the renormalization point $\mu=H_0$, much in the same way as it is done
with any other parameter in renormalization theory. After we remove the
flat spacetime vacuum energy in order to insure the consistency of the
Minkowskian metric with Einstein's equations in vacuo, the additive term
$c_0$ still enters the low energy expression $\rL(H)=c_0+\beta M_P^2\,H^2$
on account of the arbitrariness of the renormalization procedure when
performing the calculations in curved space. In the RG approach, the
existence of the $c_0$ term is mandatory; it is the natural result of
integrating the renormalization group equation. That term is actually
indispensable for the good phenomenological status of the model. As we
have emphasized, in its absence a vacuum energy density $\rL(H)$ of the
above form  could not predict a cosmic transition from deceleration into
acceleration.

\subsection{Predicting $\CC$ from first principles?}

We may ask ourselves why the mass scale $m_{\CC}\equiv({\rLo})^{1/4}$
associated to the cosmological term is of order of $m_{\CC}={\cal
O}(10^{-3})$ eV rather than, say, ten times or a hundred times bigger.
While such values cannot be admitted, as they would obviously be
incompatible with the observations, here we are addressing a matter of
principle, i.e. we are asking: is there a fundamental physical theory
which can explain the value of the vacuum energy that we measure at
present? Put another way: should the millielectronvolt energy scale
$m_{\CC}$ be ultimately predictable, for example from the value of the
Planck scale $M_P$? This is a very interesting question, but is not at all
an obvious one -- cf.\,\cite{Paddy1213},\cite{BFLWard}.

When we face the possibility to explain the mass scale $m_{\CC}$ in our
universe, we should note that in particle physics essentially all physical
scales remain still unexplained! For example, we cannot explain why the
value of the electron mass is $m_e\simeq 0.511$ MeV in our universe, since
we do not understand why its Yukawa coupling, $Y_e$, takes on the value it
takes in order to yield the precise value of the electron mass from its
product with the vacuum expectation value of the Higgs doublet:
$m_e=Y_e\,(v/\sqrt{2})$. Similarly, we do not understand why we have so
many fermion flavors and with widely different mass scales (or family of
Yukawa couplings). In particular, why the masses of the neutrinos are much
lighter, $m_{\nu}/v<10^{-11}$, and why one species of neutrinos is
possibly not far away from the same $\sim$ meV scale $m_{\CC}$ associated
to  the cosmological constant -- in this case, $m_{\nu}/v<10^{-14}$!

Whether we can ultimately predict or not the scale of the vacuum energy in
the present universe can be a debatable question, but what is perhaps less
debatable is that there is, in principle, no reason why we should be able
to bypass all the foreseeable difficulties well before we can minimally
understand the origin of the more down-to-earth scales of the fermion and
boson masses in the SM of particle physics, and of course the values of
the two basic ``order parameters'' that set the fundamental scales of the
model, to wit: i) the EW scale $v\simeq G_F^{-1/2}={\cal O}(100)$ GeV
$\sim M_W$ (associated to the Higgs mechanism and linked to Fermi's
constant); and ii) the value of the QCD scale $\LQCD={\cal O}(100)$ MeV
(associated to the strong interactions, being determined by the
non-perturbative confinement dynamics of QCD). In both cases an
experimental input is needed to make contact with physics. For the EW
sector (which provides the dominant contribution to the vacuum energy of
the SM) we need to perform a precise measurement of Fermi's constant from
muon decay. Similarly, in the domain of strong interactions we have to
measure $\LQCD$ to account for the QCD vacuum contribution. It is
fascinating to entertain that the possible cosmic evolution of $\LQCD$,
i.e. $\LQCD=\LQCD(H)$ -- and with it the masses of all nucleons and nuclei
of the universe! -- could also play a decisive role in elucidating the
nature of the CC problem, as we have amply emphasized in
Sect.\,\ref{sect:DynamicalVacFundConst}.

Owing to its especial position in the realm of the physical quantities,
and because of the general covariance of Einstein's equations, we should
expect a tight connection of the CC with the remaining scales of the
universe. For this reason the help of the phenomenological input appears
as indispensable. Is this not a most fundamental physical requirement,
even for the glory of the CC problem? In this sense, if the vacuum energy
density is treated as a running quantity in an expanding universe, it
would be natural to input its value at a given cosmic time (say, now) and
then focus our efforts on finding its past and future evolution. After all
it is not granted that we can reach a purely theoretical solution, unless
e.g. all scales of the universe should come from a single one, say the
Planck mass $M_P$, and all the others be referred to it through
dimensionless (and computable) ratios. But it is far from obvious that we
can happily make such an aprioristic assumption. For the time being
quantum gravity is not a consistent theory, and we {\em cannot} know for
certain if $M_P$ -- which is nothing but a shorthand for
$\left(\hbar\,c/G\right)^{1/2}$ ($G$ being Newton's constant) -- is truly
a physical scale or a theoretical construct!

\subsection{Analogies with the CC: Casimir effect and light-by-light scattering}

The bare truth is that we still need the help from the phenomenological
input so as to set the scale for the vacuum energy in our low energy
universe, and there is no foreseeable change in this situation. But this
does not mean we cannot hope for some progress. Take for example the RG
approach we have discussed here; after we input $m_{\CC}$ (or,
equivalently, $c_0$) what remains of the CC problem is still a hard enough
challenge for our intellect! The CC problem is then formulated as the
problem of explaining why the dynamical term in the low energy expression
$\rL(H)=c_0+\beta M_P^2\,H^2$ is just the soft (and completely harmless)
$\sim H^2$ contribution in the context of QFT in an expanding spacetime.
This term can be thought of as being caused by the ``differential
vibrational modes'' of the vacuum when we remove the flat spacetime, in
analogy to the physical effect that remains when we remove the ``absolute
ZPE'' contribution (viz. the total output from the pure vacuum-to-vacuum
effects)  without the plates in the Casimir effect.

We have more useful analogies. For example, in the same way as we cannot
identify the mass of a particle, say the electron or the $W^{\pm}$ gauge
boson, with the (UV-divergent) quantum fluctuations associated to their
radiative corrections, and we need to perform a subtraction (``mass
renormalization'') to achieve a physical result,  the ``absolute ZPE''
should not be considered as a physical quantity either, namely a form of
tangible energy endowed with a gravitational mass (and corresponding
inertia). For this reason the ZPE (and, for that matter, any additional,
contribution to the vacuum energy, say from spontaneous symmetry breaking)
should be renormalized to exactly zero in Minkowskian spacetime. Only in
this way Einstein's equations in vacuo can be satisfied for this metric.
In fact, this ought to be considered the physical requirement or
``renormalization condition'' for the vacuum energy density in flat
spacetime. Similarly, when we compute the corresponding energy density of
the vacuum in a curved spacetime we should subtract the flat spacetime
result obtained in the same renormalization scheme, thus we should
subtract the $\sim m^4$ quantity (\ref{renormZPEoneloop2}). The net or
``distinctive vibrational'' vacuum modes that remain at low energies after
this operation, should be (as in the Casimir effect) the only physically
measurable quantity, namely the expression $\rL(H)=c_0+\beta M_P^2\,H^2$.

It may be worth further prolonging the analogy we have devised in
Sect.\,\ref{sect:oddH} with the effective Heisenberg-Euler-Weisskopf
Lagrangian for photon self-interactions at low energy, when the electrons
are integrated out of the theory. Quite in a similar manner as we subtract
from the vacuum energy density the flat spacetime contribution
(represented by the ``bald'' blob in Fig.\ref{Fig1:blobsHair}), and
thereby the $\sim m^4$ quantum effects are disposed off, for the case of
the low energy light-by-light scattering we must equally subtract the
first diagram of the expansion, namely the (UV-divergent) loop graph with
only two photons. Although it is not a ``bald'' diagram as in the vacuum
case, it is the one with the minimum possible number of photon legs, and
in this sense is the analog of the vacuum-to-vacuum diagram with zero legs
in Fig.\ref{Fig1:blobsHair}. Now, it is well-known that the loop graph
with two photons contributes to charge renormalization. Therefore, upon
implementing standard charge renormalization, which entails the
subtraction of the UV divergence from the first graph of the effective QED
theory, the remaining multiphoton scattering amplitudes can be computed in
a well-defined, UV-finite, theory in which the IR effects can also be
properly accounted for. Needless to say, at high energies the
electromagnetic charge will be running and in this sense finite effects
from that diagram will appear in the ``UV regime'', but for the effective
low energy QED theory at stake here charge renormalization virtually
implies the complete removal of the first diagram of the expansion. So the
situation is entirely similar to the vacuum energy case under discussion,
as the subtraction of the bald diagram is also associated to a
renormalization: the (vanishing) value of the CC in flat spacetime.

\subsection{The hierarchy problem in particle physics and cosmology}

With some resemblance, although with important differences, we have the
fine tuning situation in the Higgs sector of gauge theories, where a
hierarchy problem is motivated owing to the quadratic corrections that
receive the Higgs boson masses. This fact makes naturalness in the Higgs
sector a pressing issue, especially now that a relatively light Higgs
boson seems to have been found. For example, the Higgs boson mass in the
SM receives quadratic radiative corrections of the form
\begin{equation}\label{eq:HiggsQuadratic}
\delta M_{\cal H}^2=\frac{\CUV^2}{(4\pi)^2}\,\beta_{\HB}+{\cal O}\left(M_i^2\ln\frac{\CUV^2}{M^2_{\cal
H}}\right)\,,
\end{equation}
where ${\CUV}$ is the cutoff and $M_i$ a collection of mass parameters of
the electroweak group $SU(2)_L\times U_Y(1)$. The dimensionless
$\beta_{\HB}$-coefficient depends on the gauge couplings $g,g'$ of
$SU(2)_L$ and $U_Y(1)$ respectively, and of the Higgs self-coupling
$\lambda$ defined in Sect.\,\ref{sect:HiggsVacuum}. At one loop it reads
\begin{equation}\label{eq:couplingsmasses}
\beta_{\HB}=\frac92\,g^2 +\frac32\,g'^2+2\lambda-12 Y_t^2\,,
\end{equation}
where $Y_t=\sqrt{2}\,m_t/v$ is the top quark Yukawa coupling (the
contributions from the remaining fermions are neglected). Note that
$\beta_{\HB}$ can be written as a combination of ratios of mass squares of
the weak gauge bosons, the Higgs boson and the top quark mass, with the
Higgs VEV $v$ defined in (\ref{5N}). Using the relations given in
Sect.\,\ref{sect:HiggsVacuum} we immediately find
\begin{equation}\label{eq:couplingsmasses2}
\beta_{\HB}=\frac{6}{v^2}\,\left(M_Z^2+2\,M_W^2+M_{\cal
H}^2-4\,m_t^2\right)\,.
\end{equation}
By requiring that $\delta M_{\cal H}^2/M_{\cal H}^2<1$
Eq.\,(\ref{eq:HiggsQuadratic}) entails that ${\CUV}$ is of order 1 TeV at
most. But as we know the regularization of quadratic divergences with a
cutoff leads to some arbitrariness, i.e. they depend on the regularization
procedure, and in fact long ago it was proposed  that the coefficient of
$\CUV^2$ in the above formula should vanish\,\cite{Veltman81}. Using the
form (\ref{eq:couplingsmasses2}), this condition leads to a strong fine
tuning between the electroweak masses in the one-loop relation
(\ref{eq:couplingsmasses2}), and also to an estimate of the Higgs mass:
$M_{\cal H}\simeq 315$ GeV. The latter amply overshoots the presently
known value ($\sim 125$ GeV) and hence that argument is ruled out.
Moreover, at two loops\,\cite{MHQuadraticUV} it gives a different result,
$M_{\cal H}\simeq 280$ GeV, which is also excluded.  In contrast, if we
consider a more amenable renormalization framework, viz. one focusing on
the logarithmic divergences (as e.g. dimensional regularization), then
only the subleading terms indicated on the \textit{r.h.s.} of
Eq.\,(\ref{eq:HiggsQuadratic}) are projected. The log terms do not present
the ambiguities of the quadratic ones. In this way one finds a gentler
expression for the one-loop result:
\begin{equation}\label{eq:HiggsQuadraticGeneral}
\delta M_{{\cal
H}_i}^2=\frac{1}{(4\pi)^2}\,\sum_i\xi_i\,g_i^2\,M_i^2\,\ln\frac{\CUV^2}{M^2_{{\cal H}_i}}\,,
\end{equation}
where $\xi_i$ are some coefficients, and $g_i$ and $M_i$ are a collection
of couplings and mass dimension parameters in the given model. Specific
realizations of this formula will depend on the model itself of course.
For example, in SUSY theories\,\cite{SUSYTheories} the cancelation of
quadratic corrections to the Higgs boson masses is automatic. In fact this
is actually one of the traditional motivations for SUSY. One obtains an
expression of the form
(\ref{eq:HiggsQuadraticGeneral})\,\cite{KoldaMurayama00}. Whatever it be
the renormalization framework where the corrections can be arranged as in
\ref{eq:HiggsQuadraticGeneral} they will remain small as compared to the
Higgs boson mass, even if ${\CUV}$ (playing here the role of scaling
parameter) moves from  the EW scale to a high GUT scale below the Planck
mass\,\cite{Feng2013}.

The above hierarchy problem in particle physics can be formulated on RG
grounds by considering the general form of the renormalization group
equation satisfied by the Higgs boson mass in usual gauge theories. Using
a regulator that projects the log terms only, or a supersymmetric theory
where the quadratic divergences are actually canceled above some soft
SUSY-breaking scale $M_{\rm SUSY}$,  the RGE's for the Higgs boson masses
take on the form
\begin{equation}\label{eq:RGEHiggs}
\frac{d M_{\HB}^2}{d\ln\mu^2}=\frac{1}{(4\pi)^2}\sum_i \beta_i\,M_i^2\,,
\end{equation}
where $M_i$ and $\beta_i$ is a collection of mass parameters and
dimensionless coefficients associated to the given model, respectively. In
SUSY theories, the $M_i$ is a set of soft SUSY-breaking mass parameters,
all of them of order $M_{\rm SUSY}$. For instance, in the absence of
flavor mixing, the RGE for the Higgs mass parameter $M^2_{\HB_u}$ (the one
sensitive to the top quark loop corrections) reads
\begin{equation}\label{eq:RGEMSSMHiggs}
\frac{d M^2_{\HB_u}}{d\ln\mu^2}=-\frac{3\,Y_t^2}{(4\pi)^2}\left(M^2_{\tilde{t}_{L}}+M^2_{\tilde{t}_{R}}+|A_t|^2\right)\,,
\end{equation}
where $Y_t$ is the top quark Yukawa coupling, $M^2_{\tilde{t}_{L,R}}$ are
the soft SUSY-breaking diagonal mass squared terms in the stop mass
matrix, and $A_t$ is the stop trilinear soft SUSY-breaking parameter.
Integration of Eq.\,(\ref{eq:RGEMSSMHiggs}) obviously leads to a
logarithmic mass correction of the form (\ref{eq:HiggsQuadraticGeneral}),
for $\mu$ up to some large scale $\CUV$\,\cite{Blanke3}.

Of course the great hope of SUSY theories, in particular of the
MSSM\,\cite{SUSYTheories}, is that $M_{\rm SUSY}={\cal O}$(TeV) --- this
being one of the reasons for having constructed the LHC collider!  (the
other, the quest for the Higgs, seems to have been fulfilled to some
extent!). The upshot is that, in SUSY theories, the Higgs boson masses are
protected from quadratic divergences up to the scales of soft
SUSY-breaking, so if this scale is not too high ($M_{\rm SUSY}\sim 1-10$
TeV) the hierarchy problem is under control and there is a hope to find
some of the SUSY particles at the LHC -- see \cite{Jegerlehner13} for
alternative points of view. Of course this is what SUSY people should now
expect after we discovered what seems to be a relatively light Higgs boson
particle of ``only'' $\sim 125$ GeV\,\footnote{The lightest $\CP$-even
SUSY Higgs particle of the MSSM cannot be heavier than $\sim 130$
GeV\,\cite{MSSMHiggsmasses1,MSSMHiggsmasses2} for any set of values of the
remaining parameters. However, we cannot tolerate arbitrarily large values
for these parameters because otherwise this would lead to fine tuning.
Thus, if the found Higgs is the lightest MSSM one we should expect some
news from the rest of the SUSY spectrum during the LHC lifetime!}.

We may now briefly confront the hierarchy problem in the Higgs boson case
with the hierarchy problem in cosmology. To this end let us compare the
above RG equation (\ref{eq:RGEHiggs}) with the proposed RGE for the CC
term, Eq.\,(\ref{seriesRLH}), which we write down here in the following
simplified form:
\begin{eqnarray}\label{seriesRLH2}
\frac{d\rL}{d\ln
\mu^2}=\frac{1}{(4\pi)^2}\sum_{i}\,a_{i}M_{i}^{2}\,\mu^{2}
+{\cal O}\left(\mu^{4}\right)\,,
\end{eqnarray}
in which $\mu$ is associated with $H$. The role played by the quadratic
terms in the Higgs case should apparently be taken here by the quartic
contributions $\sim M_i^4$, which are missing in (\ref{seriesRLH2}). Here
one might naively dream of using also SUSY to kill these quartic terms
because of the $(-1)^{2J}$ factor in the CC $\beta$-function
Eq.\,(\ref{eq:BetaFunctionSpinJ}), since this factor alternates sign
between bosons and fermions. However, the cancelation between boson and
fermion contributions  would be valid only above the scale $M_{\rm SUSY}$
of soft SUSY-breaking. As a result the remnant contribution to the CC
would still be of order $\rL\sim M_{\rm SUSY}^4$. Being, however, $M_{\rm
SUSY}={\cal O} (1)$ such ``residue'' of SUSY breaking is at least as huge
as the SM contribution itself! (cf. Sect.\,\ref{sect:HiggsVacuum}).
Therefore, the partial cancelation is of no practical help at all.

But in fact we do not expect that the absence of those terms comes from
any miraculous cancelation, not even from SUSY.  As we said, the very
reason why the $\sim M_i^4$ contributions are not there is quite another;
it stems from the impossibility of the Minkowski metric from being a
solution of Einstein's equations in vacuo. So we established it as a
renormalization condition, Eq.\,(\ref{eq:zeroVacEnergFlatSpacetime}),
which eliminates these terms. Furthermore, as explained by the end of
Sect.\,\ref{sect:VacEnerCurvature} -- cf. Eq.\,(\ref{eq:activeMi}) -- such
condition is perfectly consistent with the RG since the average SM
particle with mass $m$ cannot satisfy $H>m$ at any time of the
cosmological evolution, so the usual SM particles cannot be active degrees
of freedom for the RGE (\ref{seriesRLH2}). So the first term in its
\textit{r.h.s.} cannot be $\sim M_i^4$ but $\sim M_i^2\,H^2$. We will
further discuss in the next section the naturalness of that
renormalization condition.

The difference between the hierarchy problems in particle physics and
cosmology is now quite evident. In contrast to the RGE for the Higgs mass,
Eq.\, (\ref{eq:RGEHiggs}), the RGE for the CC term, Eq.\,
(\ref{seriesRLH}), does not lead to a log behavior of the CC versus
$\mu=H$, but a quadratic one.  This is possible because $\rL$ has
dimension $4$ in natural units whereas $M_{\HB}^2$ has dimension $2$. Such
important dimensional difference traces a parallelism between the
renormalization of the Higgs boson mass squared and the CC term, to wit:
while in the former case the first consistent contribution is only
logarithmic rather than quadratic, in the latter the first admissible
effect is the quadratic one rather than the outrageous quartic.

The lowering of powers by two units in both cases acts in a different
manner for each case: i) on the one hand, in particle physics, it solves
the hierarchy problem (which means that the Higgs sector can then be
\textit{natural}, i.e. no longer subdued by its radiative instability when
the SM is embedded in a GUT); ii) and, on the other, i.e. in the
cosmological case, the remaining two powers imply a ``soft decoupling'' of
the CC term. This feature is actually crucial and is what enables us to
``trade'' the fine tuning problem for simple renormalization, as explained
in the next section.

\subsection{Fine tuning or renormalization?}

We close our reflections with what is usually perceived as an excruciating
fine tuning adjustment associated to the CC problem. We have presented it
in all its severity and rawness in Sect.\,\ref{sect:FineTuningMother}, and
indeed it looks completely hopeless. On the face of that glooming and
unearthly panorama we may still dare asking, in an almost state of shock:
is this harrowing description of the situation truly real? One may still
hope for a miracle, a fortuitous cancelation or some unexpected symmetry
etc. But we do not think there is anything like that. In
Sect.\,\ref{sect:Casimir}, we have suggested that it could be viewed as
just nominal renormalization, what would of course be harder to swallow
for a strictly constant $\CC$. Notwithstanding, for a dynamical CC term
the picture can be quite different. Please notice that the fine-tuning
associated to the CC problem is looked upon as extremely bizarre only
because we live in a very low energy universe and we believe (following
the standard $\CC$CDM paradigm) that the CC remained constant all the time
(at least after inflation).

The ``mirage'' of the CC problem appears only in the late universe, when
$H$ is very small and both terms of the sum $\rL(H)=c_0+\beta M_P^2\,H^2$
are comparable, both being much smaller than the average (naively
``expected'') $\sim m^4$ effect. In point of fact, the mirage cannot occur
with any of the usual parameters in QFT. All fermion masses and couplings
scale logarithmically with the energy, and for SUSY theories the Higgs
boson mass itself also scales logarithmically (as we have seen in the
previous section). In the gravity sector the logarithmic scaling includes
also Newton's gravitational constant as well -- cf.
Sect.\,\ref{sect:RGequations}. After the quartic $\sim M_i^4$ effects have
been removed, general covariance arguments imply that $\rL$ scales
quadratically with $\mu=H$.  We called this feature the ``soft
decoupling'' behavior of $\rL$ because it implies that $\rL$ increases in
the subleading form $\sim M_i^2\,H^2$ rather than as the naive $\sim
M_i^4$ one. There is indeed a softening of the increasing behavior, for we
have $M_i^2\,H^2\ll M_i^4$ in the entire matter dominated epoch. Moreover,
in the radiation epoch at temperature $T$, we can integrate
(\ref{seriesRLH2}) using $\mu=H$ for the RG scale, and we find the
following expression for the vacuum energy density as a function of the
temperature (neglecting the constant term $c_0$ here):
\begin{equation}\label{eq:radiatdensity}
\rL(T)\simeq\frac{1}{(4\pi)^2}\sum_i\,a_i\,M_i^2\,H^2(T)=\beta\,M_P^2\,H^2(T)=\nu\,\rR(T)\,,
\end{equation}
where we have used (\ref{eq:nuloopcoeff2}) and $\beta=3\nu/8\pi$. Let us
recall that $M_i$ is the collection of \textit{all} possible masses below
the Planck scale, but in practice only the largest masses available
(typically those associated to the heaviest degrees of freedom of some GUT
near $M_P$) will contribute significantly. Furthermore, in the last
equality of the previous equation we have used
\begin{equation}\label{eq:HfunctionT}
H^2(T)=\frac{8\pi\,G}{3}\,\rR(T)=\frac{8\pi}{3\,M_P^2}\,\left(g_*\,\frac{\pi^2}{30}\,T^4\right)\,,
\end{equation}
where $\rR(T)$ is the radiation density at temperature $T$, say at the EW
scale $T\sim v={\cal O}(100)$ GeV; and $g_{*}$ is the effective number of
relativistic species at this temperature, typically $g_{*}={\cal O}(100)$.
Now, being $|\nu|\lesssim{\cal O}\left(10^{-3}\right)$ it follows from
(\ref{eq:radiatdensity}) that the vacuum energy density associated to the
curvature terms $M_i^2\,H^2$ never dominates, neither in the matter nor in
the radiation epoch. This is of course a welcome feature, as it insures
that primordial BBN is perfectly safe. However, even if it does not
dominate in the radiation epoch, the vacuum energy density is quite
sizeable: it is only a factor $|\nu|\sim 10^{-3}$ below the electroweak
density (at the natural EW scale), not 55 orders of magnitude smaller (as
it would be the case if $\rL$ would be thought of as strictly constant!).
At this point we impose the renormalization condition according to which
the CC is zero in Minkowski space -- i.e. we invoke
Eq.\,(\ref{eq:zeroVacEnergFlatSpacetime}) -- and with it we remove all the
$\sim M_i^4$ contributions.

Finally, when we look at the purported true value (\ref{eq:radiatdensity})
of the vacuum energy density induced in curved spacetime (as a
``distinctive effect'' with respect to the Minkowski vacuum), we find that
the radiative corrections involved in the previous renormalization
condition are indeed small as compared to that effective vacuum energy.
Comparing equations (\ref{eq:radiatdensity}) and (\ref{nloops}) with
$T\sim v$ and using the aforementioned values of these parameters,  we
find
\begin{equation}\label{eq:comparison1}
|\nu|\,g_{*}\,\frac{\pi^2}{30}\gtrsim\left(\frac{g^2}{16\pi^2}\right)^n\,\ \ \forall n\,,
\end{equation}
in contrast to the old (and highly unnatural) situation:
\begin{equation}\label{eq:comparison2}
\rLo\lll \left(\frac{g^2}{16\pi^2}\right)^n\,v^4\,\ \ \ {\rm for}\ n=1,2,3....,\sim20\,,
\end{equation}
where $n$ counts the order of the electroweak loop diagrams beyond the
first one (which has no coupling $g$ of any kind since there is no vertex
at lowest order for the ZPE). From the foregoing it is evident that the
aforesaid renormalization condition looks perfectly fine when formulated
at the EW scale, rather than at the current (extremely low energy)
universe. The naturalness of equation (\ref{eq:comparison1}) can be
fulfilled because $\nu$ receives contributions from the masses $M_i$ of
\textit{all} fields below the Planck scale, see
Eq.\,(\ref{eq:nuloopcoeff2}), in particular from fields of GUT's defined
at a nearby $M_X\lesssim M_P$. Thanks to this fact $\nu$ is not extremely
small and can stay $|\nu|\lesssim{\cal O}\left(10^{-3}\right)$, what is
consistent with all observational
data\,\,\cite{BPS09,GSBP11,BasPolarSola12,BasSola13a}.

From the above story we learn there is a big difference between the
parameter $\rL$ and the other parameters, such as couplings and masses. By
virtue of the higher dimensionality of $\rL$ (that is to say, dimension
$4$) the softening it undergoes is not enough to enforce the subleading
effects on $\rL$ to follow the traditional (tamed) form of the decoupling
effects -- as dictated by the Appelquist-Carazzone
theorem\,\cite{AppCarazz74}. In other words, owing to the remnant
quadratic scaling the dynamical part of the vacuum energy density is still
dominated by the highest masses, $\rL(H)\sim \sum _i\,M_i^2\,H^2\sim\beta
M_P^2 H^2$, and this is tantamount to say that $\rL$ can follow (from
below) the evolution of the radiation density through $\rL(T)
=\nu\,\rR(T)\sim \nu\,T^4$ at any epoch of the cosmic evolution where
$\rR(T)$ is defined. As a result $\rL(H)$ can slide from the highest
energy scales (from where the renormalization condition can be formulated
naturally)  to the tiny value it has today (the point where that condition
looks unnatural). This is the reason why we have called it ``soft
decoupling'', because while there is indeed softening there is no real
decoupling. This feature does not occur for the lower dimensionality
parameters (masses and dimensionless couplings), and in this sense $\rL$
is special as a running parameter.

The above considerations would be impossible, or extremely contrived, for
a strictly constant CC term, which is why a rigid vacuum energy density
causes the excruciating fine tuning conundrum. Obviously, it is thanks to
that quadratic scaling of $\rL$ that we can place our renormalization
settings at that vantage point in the remote past of our cosmic history,
i.e. the electroweak scale (or any other relevant scale in the early
universe which we think causes a fine tuning problem), and from there we
can just dispatch the CC fine tuning issues following the conventional
wisdom of renormalization theory. After that, when we move back to our
cold universe, the term $M_P^2\,H^2$ is not only much smaller than $\sim
m^4$ for the average SM particle, but is already as small as the additive
term $c_0$ in the final expression
\begin{equation}\label{eq:rLotoday}
\rLo=c_0+\beta M_P^2\,H_0^2\sim 10^{-47}\, {\rm GeV}^4
\end{equation}
for the current vacuum energy density. The dynamical part is in fact
waning more and more in our old universe and is becoming dominated by the
constant $c_0$. It is this constant that finally determines the ultimate
de Sitter stage of the cosmic evolution.

A the end of the day, we see that if we accept such dynamical picture of
the vacuum energy, the ``Mother'' of all the fine tunings, as we have
called it in Sect.\,\ref{sect:FineTuningMother}, could well tumble from
her arrogant position and perhaps be demoted to just the ordinary status
of any parameter undergoing a process of standard QFT renormalization at
its natural scale.

\subsection{Dynamical vacuum energy as dark energy}

We have argued that there is no a priori reason for being able to predict
the value of $c_0$ in (\ref{eq:rLotoday}) from first principles because,
ultimately, we need a physical input for the running vacuum energy density
at the scale of its measurement, namely at $\mu=H_0$. This is so for any
parameter of QFT, including of course the SM of particle physics. There is
no reason why $\rLo$ should be an exception. If we can accept this, the
real QFT challenge on the pure theoretical side should actually focus on
the rigorous computation of the time dependent disturbance $\sim H^2$,
which endows the vacuum energy density of a mild (albeit non-negligible)
time evolution. It is therefore a very attractive option, in which the
time effect could eventually surface in the form of a dynamical vacuum
energy of the form $\rL=c_0+\beta M_P^2\,H^2$ whose value in the present
universe is (\ref{eq:rLotoday}), but capable of showing some variation
around it when we explore our recent past. This is of course a handle to
test the model. If eventually confirmed, it could arguably be a plausible
QFT explanation for the sought-for dynamical DE. But to check the reality
of this effect is, of course, a task reserved for the future observations.

At the moment it is intriguing that the effective equation of state
parameter of the dark energy has a peculiar tilt in the phantom regime
$\weff=-1.13^{+0.13}_{-0.10}$. This is the situation after the first
release of precise cosmological data from the PLANCK
satellite\,\cite{PLANCK2013}, but such feature is not new at all; it has
been reckoned by the long series of WMAP observations \,\cite{WMAP,Kom11}.
Despite the many attempts to accommodate it in a framework of pure scalar
field dynamics, where the DE is thought of as the result of the crosstalk
between one or more fundamental scalar fields (one of them at least been
phantom-like, i.e. with a negative kinetic term), this is probably not
very much convincing. It is encouraging, however, that a dynamical vacuum
energy framework can mimic the equation of state of quintessence and
phantom DE without involving any such sort of fields at
all\,\cite{BasSola13a}.

We conclude  by saying our credo: we believe that the CC problem can only
be solved as a physical problem, not just as a theoretical conundrum. This
means that only through phenomenological tests it should be possible to
disentangle the most difficult theoretical aspects of the CC problem. In
this respect, another potentially interesting aspect of the dynamical
vacuum models is the fact that they could provide an explanation for a
possible variation of the so-called fundamental constants of
Nature\,\cite{FritzschSola2012}. There is currently plentiful of
experimental activity, both in the lab and from observations in the
astrophysical domain, that will provide sooner or later interesting news
on this field\,\cite{FundConst,Reinhold06}. Future data from these
experiments should be very helpful for effectively testing the proposed
vacuum ideas in the near future. For this reason we have discussed this
issue at length in Sect.\,\ref{sect:DynamicalVacFundConst}.

On the theoretical side, it is probably no exaggeration to assert that the
investigations on the dynamical behavior of the vacuum energy in QFT in
curved spacetime are at the heart of one of the most important endeavors
of fundamental cosmology in the years to come. While we have given here
some indirect clues on how to effectively achieve an appealing scenario
for the desired expression for $\rL(H)$ in that context, much more work
(and thought!) is of course needed to tackle the most challenging and
fierce angles of the CC problem, face to face.

\newpage

%\vspace{1.5cm}

\acknowledgments

\noindent I have been partially supported by DIUE/CUR Generalitat de
Catalunya under project 2009SGR502, by the research Grant FPA 2010-20807,
and by the Consolider CPAN project SD2007-00042. It is my pleasure to
thank  S. Basilakos, F. Bauer, N. Bili{\'c}, J.C. Fabris, H. Fritzsch, J.
Grande, B. Guberina, J.A.S. Lima, D. L\'opez-Val, R. Opher, A. Pelinson,
M. Plionis, D. Polarski, I.L. Shapiro and H. \v{S}tefan\v{c}i\'{c} for
sharing their knowledge on these matters, sometimes over the years, and
for the direct collaboration I maintained with most of them recently or in
the past on some of the topics discussed here. Interesting observations on
the first version of this manuscript by J.D. Barrow and A. R. Zhitnitsky
are also gratefully acknowledged. I felt encouraged to include additional
(hopefully useful) discussions in this new version. Last but not least, I
am indebted to the organizers of NEB 15, in particular to E. Vagenas, for
the invitation to review the cosmological constant problem in the
incomparable scenario of the Hellenic Society conference on Recent
Developments in Gravity, held in the town of Chania, in the beautiful
island of Crete, Greece.

\bibliographystyle{JHEP}
%\bibliography{aspic}

%%%%%%%%%%%%%%%%%%%%%%%%%%%%%%%%%%%%%%%%%%%%%%%%%%%%%%%%%%%%%%%%%%%%%%%%%
%\newcommand{\JHEP}[3]{{ J. of High Energy Physics } {JHEP} {#1} (#2)  {#3}}
\newcommand{\JHEP}[3]{ {JHEP} {#1} (#2)  {#3}}
\newcommand{\NPB}[3]{{ Nucl. Phys. } {\bf B#1} (#2)  {#3}}
\newcommand{\NPPS}[3]{{ Nucl. Phys. Proc. Supp. } {\bf #1} (#2)  {#3}}
\newcommand{\PRD}[3]{{ Phys. Rev. } {\bf D#1} (#2)   {#3}}
\newcommand{\PLB}[3]{{ Phys. Lett. } {\bf B#1} (#2)  {#3}}
\newcommand{\EPJ}[3]{{ Eur. Phys. J } {\bf C#1} (#2)  {#3}}
\newcommand{\PR}[3]{{ Phys. Rep. } {\bf #1} (#2)  {#3}}
\newcommand{\RMP}[3]{{ Rev. Mod. Phys. } {\bf #1} (#2)  {#3}}
\newcommand{\IJMP}[3]{{ Int. J. of Mod. Phys. } {\bf #1} (#2)  {#3}}
\newcommand{\PRL}[3]{{ Phys. Rev. Lett. } {\bf #1} (#2) {#3}}
\newcommand{\ZFP}[3]{{ Zeitsch. f. Physik } {\bf C#1} (#2)  {#3}}
\newcommand{\MPLA}[3]{{ Mod. Phys. Lett. } {\bf A#1} (#2) {#3}}
%%%%%%%%%%%%%%%%%%%%%%%%%%%%%%%%%%%%%%%%%%%%%%%%%%%%%%%%%%%%%%%%%%%%%%%%%
%%%%%%%%%%%%%%%%%%%%%%%%%%%%%%%%%%%%%%%%%%%%%%%%%%%%%%%%%%%%%%%%%%%%%%%%%
\newcommand{\CQG}[3]{{ Class. Quant. Grav. } {\bf #1} (#2) {#3}}
\newcommand{\JCAP}[3]{{ JCAP} {\bf#1} (#2)  {#3}}
\newcommand{\APJ}[3]{{ Astrophys. J. } {\bf #1} (#2)  {#3}}
\newcommand{\AMJ}[3]{{ Astronom. J. } {\bf #1} (#2)  {#3}}
\newcommand{\APP}[3]{{ Astropart. Phys. } {\bf #1} (#2)  {#3}}
\newcommand{\AAP}[3]{{ Astron. Astrophys. } {\bf #1} (#2)  {#3}}
\newcommand{\MNRAS}[3]{{ Mon. Not. Roy. Astron. Soc.} {\bf #1} (#2)  {#3}}
\newcommand{\JPA}[3]{{ J. Phys. A: Math. Theor.} {\bf #1} (#2)  {#3}}
\newcommand{\ProgS}[3]{{ Prog. Theor. Phys. Supp.} {\bf #1} (#2)  {#3}}
\newcommand{\APJS}[3]{{ Astrophys. J. Supl.} {\bf #1} (#2)  {#3}}
%%%%%%%%%%%%%%%%%%%%%%%%%%%%%%%%%%%%%%%%%%%%%%%%%%%%%%%%%%%%%%%%%%%%%%%%%

\newcommand{\Prog}[3]{{ Prog. Theor. Phys.} {\bf #1}  (#2) {#3}}
\newcommand{\IJMPA}[3]{{ Int. J. of Mod. Phys. A} {\bf #1}  {(#2)} {#3}}
\newcommand{\IJMPD}[3]{{ Int. J. of Mod. Phys. D} {\bf #1}  {(#2)} {#3}}
\newcommand{\GRG}[3]{{ Gen. Rel. Grav.} {\bf #1}  {(#2)} {#3}}

%  Example:  \NPB {\bf 20} {1992}  {200}

%%%%%%%%%%%%%%%%%%%%%%%%%%%%%%%%%%%%%%%%%%%%%%%%%%%%%%%%%%%%%%%%%%%%%%%%%
%\newpage
%%%%%%%%%%%%%%%%%%%%%%%%%%%%%%%%%%%%%%%%%%%%%%%%%%%%%%

%\vspace{1cm}

%\newpage

\end{document}